\documentclass[11pt]{article}
\usepackage{aaspp4,lscape,astrobib}

\let\oldfootsep=\footnotesep
\def\spose#1{\hbox to 0pt{#1\hss}}
\def\simlt{\mathrel{\spose{\lower 3pt\hbox{$\mathchar"218$}}
     \raise 2.0pt\hbox{$\mathchar"13C$}}}
\def\simgt{\mathrel{\spose{\lower 3pt\hbox{$\mathchar"218$}}
     \raise 2.0pt\hbox{$\mathchar"13E$}}}

\def\kmskpc{\,{\rm km \, s^{-1} \, kpc^{-1}}}

\def\msun{{\rm M}_\odot} 
\def\rsun{{\rm R}_\odot}
\def\Re{R_{\rm E}}

\def\Amax{A_{\rm max}} 
\def\umin{u_{\rm min}}
\def\rstar{R_{\rm *}}

\def\vperp{v_{\rm \perp}}
\def\umin{u_{\rm min}}
\def\t0{t_{\rm 0}}

\def\tstar{t_{\rm *}}

\def\delchi{\Delta\chi^2}

\def\ie{{\it i.e. }}
\def\mlens {M} 
\def\mtot {M_{\rm tot}} 
\def\m1 {M_1} 
\def\m2 {M_2} 
\def\dlens {D_{ol}} 
\def\teff{T_{\rm eff}}
\def\mbol{m_{\rm bol}}

\def\Av{A_{V}}
\def\Evr{E(V-R)}

\def\vhat{\widehat{v}} 
\def\that {\widehat{t}}
 
\def\kms {\,{\rm km \, s^{-1} }}
\def\kpc {\, {\rm kpc}} 

\begin{document}

\title{Binary Microlensing Events from the MACHO Project}
\author{
  C.~Alcock\altaffilmark{1,2},           
  R.A.~Allsman\altaffilmark{3},          
  D.~Alves\altaffilmark{1,4},            
  T.S.~Axelrod\altaffilmark{5},          
  D.~Baines\altaffilmark{6},
  A.C.~Becker\altaffilmark{2,7},         
  D.P.~Bennett\altaffilmark{1,2,8},      
  A.~Bourke\altaffilmark{9},      
  A.~Brakel\altaffilmark{6},
  K.H.~Cook\altaffilmark{1},             
  B.~Crook\altaffilmark{6},
  A.~Crouch\altaffilmark{9},
  J.~Dan\altaffilmark{11},
  A.J.~Drake\altaffilmark{5},          
  P.C.~Fragile\altaffilmark{8},
  K.C.~Freeman\altaffilmark{5},          
  A.~Gal-Yam\altaffilmark{11},
  M.~Geha\altaffilmark{1,12},
  J.~Gray\altaffilmark{9},
  K.~Griest\altaffilmark{2,10},           
  A.~Gurtierrez\altaffilmark{6},
  A.~Heller\altaffilmark{11},
  J.~Howard\altaffilmark{6},
  B.R.~Johnson\altaffilmark{13},
  S.~Kaspi\altaffilmark{11},
  M.~Keane\altaffilmark{14},
  O.~Kovo\altaffilmark{11},
  C.~Leach\altaffilmark{6},
  T.~Leach\altaffilmark{6},
  E.M.~Leibowitz\altaffilmark{11},
  M.J.~Lehner\altaffilmark{15},         
  Y.~Lipkin\altaffilmark{11},
  D.~Maoz\altaffilmark{11},
  S.L.~Marshall\altaffilmark{1},         
  D.~McDowell\altaffilmark{6},
  S.~McKeown\altaffilmark{6},
  H.~Mendelson\altaffilmark{11},
  B.~Messenger\altaffilmark{9},
  D.~Minniti\altaffilmark{1,16},
  C.~Nelson\altaffilmark{1,17},
  B.A.~Peterson\altaffilmark{5},         
  P.~Popowski\altaffilmark{1},
  E.~Pozza\altaffilmark{6},
  P.~Purcell\altaffilmark{6},
  M.R.~Pratt\altaffilmark{2,18},          
  J.~Quinn\altaffilmark{8},
  P.J.~Quinn\altaffilmark{19},           
  S.H.~Rhie\altaffilmark{8},             
  A.W.~Rodgers\altaffilmark{5},          
  A.~Salmon\altaffilmark{6},
  O.~Shemer\altaffilmark{11},
  P.~Stetson\altaffilmark{20},
  C.W.~Stubbs\altaffilmark{2,7},         
  W.~Sutherland\altaffilmark{21},        
  S.~Thomson\altaffilmark{9},        
  A.~Tomaney\altaffilmark{7},            
  T.~Vandehei\altaffilmark{2,10},         
  A.~Walker\altaffilmark{14},
  K.~Ward\altaffilmark{6},
  G.~Wyper\altaffilmark{6}
\begin{center}
{\bf (The MACHO/GMAN Collaboration) }\\
\end{center}
}

\altaffiltext{1}{Lawrence Livermore National Laboratory, Livermore, CA 94550
}

\altaffiltext{2}{Center for Particle Astrophysics,
  University of California, Berkeley, CA 94720
}

\altaffiltext{3}{Supercomputing Facility, Australian National University,
  Canberra, ACT 0200, Australia 
}
 
\altaffiltext{4}{Space Telescope Science Institute, 3700 San Martin Drive, 
  Baltimore, MD 21218
}

\altaffiltext{5}{Mt.~Stromlo and Siding Spring Observatories,
  Australian National University, Weston, ACT 2611, Australia
}

\altaffiltext{6}{Reynolds Amateur Photometry Team, Canberra Astronomical Society,
  Canberra, ACT 0200, Australia
}

\altaffiltext{7}{Departments of Astronomy and Physics,
  University of Washington, Seattle, WA 98195
}
 
\altaffiltext{8}{Department of Physics, University of Notre Dame, Notre Dame, IN 46556
}

\altaffiltext{9}{Dept of Mathematics and Statistics, Monash University,
  Clayton, Victoria, 3168, Australia
}

\altaffiltext{10}{Department of Physics, University of California,
  San Diego, CA 92093
}
 
\altaffiltext{11}{School of Physics \& Astronomy and Wise Observatory,
   Tel-Aviv University, Tel-Aviv 69978, Israel
}

\altaffiltext{12}{Department of Physics, University of California,
  Davis, CA 92093
}

\altaffiltext{13}{Tate Laboratory of Physics, University of Minnesota, 
  Minneapolis, MN 55455
}

\altaffiltext{14}{Cerro Tololo Interamerican Observatory, National
  Optical Astronomy Observatories.
}

\altaffiltext{15}{Department of Physics, University of Sheffield,
Sheffield s3 7RH, UK
}
 
\altaffiltext{16}{Departmento de Astronomia, P. Universidad Cat\'olica, 
Casilla 104, Santiago 22, Chile
} 

\altaffiltext{17}{Department of Physics, University of California,
  Berkeley, CA 92093
}

\altaffiltext{18}{Center for Space Research, MIT,
Cambridge, MA 02139
}
 
\altaffiltext{19}{European Southern Observatory, Karl-Schwarzschild Str. 2, D-85748, Garching, Germany
}

\altaffiltext{20}{National Research Council,
  5071 West Saanich Road, RR 5, Victoria, BC V8X 4M6, Canada
}
 
\altaffiltext{21}{Department of Physics, University of Oxford,
  Oxford OX1 3RH, U.K.
}

\setlength{\footnotesep}{\oldfootsep}
\renewcommand{\topfraction}{1.0}
\renewcommand{\bottomfraction}{1.0}     

\newpage

\vspace{-5mm}
\begin{abstract} 
\rightskip = 0.0in plus 1em

We present the lightcurves of 21 gravitational microlensing events
from the first six years of the MACHO Project gravitational
microlensing survey which are likely examples of lensing by binary
systems.  These events were manually selected from a total sample of
$\sim$ 350 candidate microlensing events which were either detected by
the MACHO Alert System or discovered through retrospective analyses of
the MACHO database.  At least 14 of these 21 events exhibit strong
(caustic) features, and 4 of the events are well fit with lensing by
large mass ratio (brown dwarf or planetary) systems, although these
fits are not necessarily unique.
The total binary event rate is roughly consistent with
predictions based upon our knowledge of the properties of binary
stars, but a precise comparison cannot be made without a determination
of our binary lens event detection efficiency.

Towards the Galactic bulge, we find a ratio of caustic crossing to
non-caustic crossing binary lensing events of 12:4, excluding one
event for which we present 2 fits.  This suggests significant
incompleteness in our ability to detect and characterize non-caustic
crossing binary lensing.  The distribution of mass ratios, N($q$), for
these binary lenses appears relatively flat.  We are also able to
reliably measure source-face crossing times in 4 of the bulge caustic
crossing events, and recover from them a distribution of lens proper
motions, masses, and distances consistent with a population of
Galactic bulge lenses at a distance of $7 \pm 1 \kpc$.  This analysis
yields 2 systems with companions of $\sim 0.05 \msun$.

\end{abstract}
\vspace{-5mm}

\newpage
\section{Introduction}
\label{sec-intro}

The search for gravitational microlensing has been very successful
since first envisioned by \citeN{paczynski86}.  Several teams have
reported observations of stellar brightenings that are consistent with
the microlensing interpretation
(\citeNP{macho-nat93,eros-nat93,ogle-pac94a,duo95a}).  The total
number of reported candidate events now exceeds 400.  With second
generation surveys now operating, we can expect a similar number of
events in the coming years (\citeNP{ogle2,palanque-thesis}).

By comparing the (efficiency-corrected) observed event rate with model
predictions, microlensing surveys have ruled out a large class of dark
matter candidates (\citeNP{macho-eros,macho-lmc2,eros-lmc}), and
appear to have detected a previously unknown component of the
Galactic-Magellanic Cloud system in lensing objects \cite{macho-lmc2}.
The extensive microlensing databases have also allowed important
constraints to be placed on Galactic structure
(\citeNP{macho-bulgerrl,ogle-98bulgedist,macho-zpt,macho-nodwarf,zhao-96bar,stanek-map,ogle-94bar})

The gravitational lensing technique uses astronomical sources to
backlight a foreground mass distribution.  Lensing occurs due to the
deflection of light rays by gravity, and most events are well
described by the weak field, small angle limit of the gravitational
scattering of photons.  The equation describing the image positions in
the lensing plane, for a single point lens, is

\begin{equation}
\label{single-lens}
\vec s = \vec i - \Re^2 {\vec i - \vec x \over (\vec i - \vec x )^2} \ ,
\end{equation}
where $\vec s$, $\vec i$, and $\vec x$ are the positions of the
source, image, and lens, projected along the line of sight into the
lens plane. $\Re$ refers to the Einstein ring radius given by

\begin{equation}
\Re^2 \equiv  4GM\left({ D_{ol}D_{ls}\over D_{ol} + D_{ls} } \right) \ ,
\end{equation}
where $D_{ol}$ is the distance to the lens, $D_{ls}$ is the
lens-source distance, $M$ the mass of the lensing object, and $G$ is
the Newtonian gravitational constant.  If we define the origin of the
lens plane as the position of the lens ($\vec x = 0$), then it is
obvious from Eqn. \ref{single-lens} that $\vec i ~||~ \vec s$.  This
symmetry provides two images during a lensing event.  (There is
actually a third image of negligible magnification that can be found
if we give the lens a finite extent, or if we depart from the weak
field approximation.)

The magnification of an image is given by the inverse of the
determinant of the lens mapping from image to source
(Eqn.~\ref{single-lens}), evaluated at the position of the image.  The
sum of the two image magnifications, which, if the images are
unresolved, is the observed magnification of the source, is

\begin{eqnarray}
\label{eq-amp}
 & A(t) = \frac{u^2 + 2}{u\sqrt{u^2 + 4}}, & \\
 {\rm where} & u(t) = \sqrt{u_{min}^2 + (~2(t-t_0)/\that~)^2}  \nonumber & \\
 {\rm and} & \that = 2 \Re / \vperp . \nonumber &
\end{eqnarray}

The lens passes closest to the source at an impact parameter (scaled
by $\Re$) of $\umin$, at the time $\t0$.  The variable $\that$
represents the time it takes the lens to cross its own Einstein
diameter (a function of $\mlens$ and $D_{ol}/D_{os}$), given its
unknown transverse velocity $\vperp$.  This simple form of the event
``lightcurve'' is based on point--source and point--lens
approximations, and an assumption of uniform linear motion between
observer, source and lens.  As a practical consideration, one must
also model additional (unlensed) light which contributes to the source
object's baseline brightness, and modifies the observed shape of the
lightcurve.  This may come from unresolved neighboring stars, or
possibly from the lens itself.  We therefore include a blending
parameter for each passband, $f$, which represents the fraction of the
photometered object's flux which was lensed.

\section{Binary Microlensing}
\label{sec-bin}

Binary lensing is described by the generalization of
Eqn.~\ref{single-lens} to the case of two deflectors:

\begin{equation}
\label{eq-binlens}
\vec s = \vec i - \Re^2 \left( \epsilon_1 {\vec i - \vec x_1 \over (\vec i - \vec x_1 )^2} +
  \epsilon_2 {\vec i - \vec x_2 \over (\vec i - \vec x_2 )^2} \right) \ ,
\end{equation}
where $\epsilon_1 + \epsilon_2 \equiv 1$, and refer to the mass
fraction of each lens.  $\Re$ now refers to the Einstein ring radius
of the complete lens system ($M_1 + M_2$).

As with the single lens, Eqn.~\ref{eq-binlens} represents a mapping
from the image plane to the source plane.  This mapping is well
behaved, but the inverse mapping from the source plane to the image
plane is multi-valued, with each point on the source plane mapping
into either 3 or 5 points on the image plane.  The boundaries between
the regions of 3 and 5 images in the image plane are referred to as
critical curves.  These critical curves map into closed caustic curves
(or caustics) in the source plane.  The shape of a caustic curve is
characterized by inwardly curved segments joined at outwardly spiked
cusps.

A binary lens lightcurve can be described by 7 parameters, if the
orbital motion of the lensing objects is neglected. These parameters
include the 3 parameters for a single lens fit $\that$, $\umin$, and
$\t0$, where $\umin$ and $\t0$ are now measured with respect to the
lensing system's center of mass.  The 3 intrinsic binary parameters
are the binary lens separation (scaled by $\Re$), $a$, the mass
fraction, $\epsilon_1$, and the angle between the lens axis and the
source trajectory, $\theta$.  A final parameter may be added if the
source star crosses a caustic, the source radius crossing time
$\tstar$.  This represents the time it takes the lens to move relative
to the source by an angle equal to the source angular radius --
typically an hour to a day.  This may be constrained by observing the
caustic crossing duration $2 \tstar / \sin \phi$, where $\phi$ is the
angle between the relative motion vector and the caustic line.  An
additional blending parameter $f$ must also be included for each
passband.  An excellent treatment of binary microlensing can be found
in \citeN{dominik98a}.

When a source crosses near or over a caustic, the event lightcurve
can differ quite significantly from the standard microlensing
lightcurve.  During a caustic crossing, the determinant of the lens
mapping in Eqn.~\ref{eq-binlens} goes to zero.  This divergence of the
magnification factor implies that a point source located on a caustic
curve becomes infinitely bright.  In this situation, two of the images
either merge and disappear, or are created, such that the number of
images is greater within the region bounded by the caustic surface.
For a realistic source brightness profile, the observed magnification
is the weighted mean of the magnification factor over the source,
which always leads to finite values.

If one is able to observe a caustic crossing, the temporal width of
the transit provides a measurement of $\tstar$, if $\phi$ is
sufficiently constrained by the global microlensing fit.  An estimate
of the true angular source size then yields the proper motion of the
lensing system with respect to the source star.  This provides an
especially important constraint in binary lensing events seen towards
the Magellanic Clouds, where there is a distinct difference between
the expected relative proper motion of a binary lens in the Galactic
dark halo ($\mu \sim 30 \kmskpc$), and of a binary lens in the
Magellanic Clouds, where the relative proper motion is a function of
the velocity dispersion in the Cloud ($\mu \sim 1 \kmskpc$).  In
general, once a caustic crossing event is detected (the first caustic
crossing is unlikely to be resolved due to its short duration, but may
be inferred from the enhanced magnification between caustics), one is
guaranteed a second caustic crossing and, with dense enough sampling
through the crossing, an estimate of the lens proper motion.

\subsection{Estimates of the Binary Lensing Rate}
\label{subsec-rate}

The phenomenology of lensing by two point masses is well known (e.g.
\citeNP{schneider86}), and the lightcurve morphology expected in the
Galactic microlensing limit well established (e.g. \citeNP{mao91},
\citeNP{mao-distefano}, and \citeNP{distefano97}).  \citeN{mao91}
predicted that $\sim 7 \%$ of all events seen towards the Galactic
bulge should show strong evidence of lens binarity, and that as many
as $\sim 3 \%$ of bulge events might show evidence of a high mass
ratio (planetary) binary system, depending upon the abundance of giant
planets.  The prospect of detecting extra-solar planets through lens
binarity has led to much investigation
(\citeNP{gould-loeb,bolatto94,bennettrhie96,peale97,gaudi-naber}), and
distinct methods and detection rates for multiple classes of planetary
lensing (\citeNP{griest-neda,dis-scalzo1,dis-scalzo2}).

The question of the fractional binary lensing rate has recently been
analyzed in more detail by \citeN{distefano99}, who argues that, in
the extreme case where {\it all} lenses are binaries, $\sim 6$\% of
all events should exhibit caustic crossings.  This fraction will
obviously decrease if some lenses are single.  Thus the majority of
binary lensing events should result in multiply-peaked lightcurves or
low amplitude deviations from single lens lightcurves, without caustic
features.
However, \citeN{distefano99} appears to underestimate the
observational bias towards low impact parameter (small $\umin$)
events.  \citeN{distefano99} considers all events with $\umin \leq 1$
to be detectable if they are sampled frequently, but in fact the MACHO
analysis and Alert System trigger both explicitly exclude lensing
events with $\umin \simgt 0.75$.  Additional cuts on the signal to
noise of the stellar brightening also discriminate against the large
$\umin$ events, which may lie within the relatively large error bars
($\sim 7$\%) of survey system photometry.  Furthermore, most of the
events seen towards both the Galactic bulge and the Magellanic Clouds
are significantly blended, such that the true $\umin$ value is
generally smaller than the observed value used for event selection.
Fig.~2 of \citeN{distefano99} indicates that the fraction of caustic
crossing events would be substantially higher if a smaller $\umin$ cut
(\ie $\umin \simlt 0.3$) were used to better reflect the survey team's
event selection criteria.

The current microlensing follow-up systems, including the Global
Microlensing Alert Network (GMAN) \cite{macho-apj97e}, Microlensing
Planet Search (MPS) team \cite{mps-98smc1}, and Probing Lensing
Anomalies NETwork (PLANET) \cite{planet95}, avoid many of these
difficulties with observing strategies optimized to detect short
timescale, low level deviations in on-going events, with photometry at
the $\sim 1 - 2\%$ level.

\subsection{Previous Binary Lens Observations}

The first unambiguous case of binary microlensing was reported by the
OGLE collaboration (OGLE-7, \citeNP{udalski94c}), and later
corroborated with data from the MACHO database
\cite{macho-bennett95,macho-mao94}.  Subsequent binary lens event
publications include DUO 2 \cite{duo-binary}, MACHO 95-BLG-12
(\citeNP{macho-pratt96a,albrow96}), MACHO LMC-1
\cite{rhie96a,dominik96}, MACHO LMC-9 \cite{macho-bennett96b}, MACHO
97-BLG-28 \cite{planet-97blg28}, and MACHO 98-SMC-1
(\citeNP{macho-98smc1,joint-98smc1}).  In addition, there have been a
number of binary lensing events discovered in progress and announced
on the web sites of the various microlensing survey and follow-up
programs\footnote[1]{
EROS~~~~{\bf~http://www-dapnia.cea.fr/Spp/Experiences/EROS/alertes.html}

MACHO~~{\bf~http://darkstar.astro.washington.edu}

MPS~~~~~~~~{\bf~http://bustard.phys.nd.edu/MPS/}

OGLE~~~~~{\bf~http://www.astrouw.edu.pl/$\sim$ftp/ogle/ogle2/ews/ews.html}

PLANET~{\bf~http://www.astro.rug.nl/$\sim$planet/index.html} }.

\section{Observations}
\label{sec-observations}

The purpose of this paper is to present instances of binary lens type
lightcurves that we have detected in our gravitational microlensing
survey, using the 50--inch telescope at the Mount Stromlo Observatory
\cite{macho-lmc2,macho-hart96a,macho-marshall94}.  These include
events towards the Large Magellanic Cloud (LMC), Small Magellanic
Cloud (SMC), and Galactic bulge.  We have not made a complete
assessment of our binary lens detection efficiency, and the events
presented here were selected by a variety of different mechanisms.
Most were originally discovered with the MACHO Alert System, and were
subsequently observed to deviate from the standard microlensing
lightcurve.  Other events were discovered via our normal analysis
procedure (see \citeNP{macho-lmc2}), difference imaging analysis (DIA)
on a subset of our database \cite{macho-dia}, or during testing of the
Alert System software.

Many of the alert events, and most of the unusual events, from
1995-1998 were also followed with coordinated observations of the
Global Microlensing Alert Network (GMAN) \cite{macho-pratt96a} :

\begin{itemize}

\item These include nightly observations on the Cerro Tololo
Interamerican Observatory (CTIO)~\footnote[2]{Cerro Tololo
Interamerican Observatory, National Optical Astronomy Observatories,
operated by the Association of Universities for Research in Astronomy,
Inc. under cooperative agreement with the National Science
Foundation.} 0.9m telescope at Cerro Tololo, Chile, with extremely
flexible scheduling allowed by the staff to assist in characterizing
the binary events presented here.  The observing strategy was
optimized in 1997 to target LMC/SMC events at highest priority (and
paid off with the real-time detection of binary event MACHO 98-SMC-1
\cite{macho-iauc98a}), and thus the frequency of observations of our
bulge events decreases.

\item We have also coordinated observations on the Wise Observatory
(WISE) 1.0m telescope at Mitzpe Ramon, Israel.  These observations are
generally requested in real-time for a subset of the bulge alert
events which show unusual features.

\item For the observing seasons 1995-1996, we were allotted
approximately $50 \%$ of the bulge time at the University of Toronto
Southern Observatory (UTSO) 0.6m telescope at Las Campanas, Chile for
microlensing follow-up.  Several events were followed at high
frequency, and 3 of these appear here as binary candidates.

\item Starting in 1997, we arranged microlensing follow-up
observations at the Mount Stromlo Observatory 30'' (MSO30) telescope
in Canberra, Australia, with the Reynolds Amateur Photometry Team
(RAPT).  This group of amateur astronomers staffs the telescope
nightly, on a volunteer basis.

\item The Microlensing Planet Search (MPS) team began its pilot season
of observations in 1997, and has obtained use of the MSO 74'' (MSO74)
telescope for rapid, precise monitoring of on-going Galactic bulge
events.

\end{itemize}

The MACHO survey data from the Mt.~Stromlo 50" telescope are generally
reduced within hours after acquisition using the SoDOPHOT photometry
routine \cite{macho-bennett93b}, which was developed from DOPHOT
\cite{schechter93}. The error bars presented here are the standard
SoDOPHOT (or DOPHOT) error estimates with a 1.4\% uncertainty added in
quadrature.  As the 50" data are reduced, the photometry is sent
through the MACHO Alert filter for the purpose of detecting
microlensing events in progress.  Thirteen of the 21 events presented
here were first detected with the MACHO Alert System, although two of
these events were not found while in progress.

All CTIO, MSO30, and UTSO data are reduced in real-time using a suite
of scripts, written in PERL, which make use of the DAOPHOT package
\cite{stetson87}.  After the event has returned to baseline, CTIO,
MSO30, UTSO, and WISE data are reduced {\it en-masse} using the
ALLFRAME package \cite{stetson-allframe}.  To assemble each
lightcurve, the target star's brightness is finally normalized with
respect to several hundred neighboring stars.  The resulting
photometric error bars are multiplied by a factor of 1.5, to account
for additional scatter in the time-series not accounted for by the
formal ALLFRAME error estimates, such as flat fielding and
normalization errors.  The MSO74 data are reduced and normalized using
a slightly modified version of SoDOPHOT.

Finally, we would like to emphasize that for some of these fits there
are probably degeneracies between model parameters.  That is, the fit
presented is not unique, but does represent a local minimum in
parameter space.  Is it likely that for some of these events,
particularly the poorly sampled ones, there are different lens
configurations whose lightcurves, as sampled by our observations, are
similar to the fits presented here (see \citeNP{dominik-bin99}).
There are, however, systematic approaches to determining all physical
solutions for a set of microlensing data (e.g. \citeNP{distefano97}
and \citeNP{planet-binsolns}).

\section{Binary Lensing Events}
\label{sec-events}

Characteristics of the MACHO objects that received the lensed flux are
presented in Table~\ref{tab-sources}.  Each object is given an event
name in addition to the MACHO star ID number which has been previously
assigned.  For each object, we include the transformation to Cousins
$V$ and $V-R$, calibrated using \citeN{macho-calib}.  From the
blending parameters included in the binary microlensing fit, we can
also estimate the true brightness and color of the star which was
lensed.  This information is also included in Table~\ref{tab-sources}.
For all events, we will refer to the (sometimes blended) object
identified in the MACHO database as the 'object', and refer to the
'source' as the actual star in the blend which was lensed.  Fit
parameters for these binary microlensing events are given in
Tables~\ref{tab-params} and~\ref{tab-bparams}.  For each event, we
present a lightcurve of these fits, as well as a schematic of the
critical and caustic curves induced by the binary lens
(Fig.~\ref{fig-lmc1}~--~\ref{fig-17619219978cc}).  The data used in
the fits may be found on the WWW at http://wwwmacho.anu.edu.au/ and
http://wwwmacho.mcmaster.ca/.  Individuals are encouraged to attempt
other binary lens fits to these data.

Fig.~\ref{fig-cmdmc} and Fig.~\ref{fig-cmdblg} display the
distribution of $V$, $V-R$ for $\sim 3,000$ neighboring sources in the
same 5' ``chunk'' (1/64 of MACHO's focal plane) as the lensed object,
for Magellanic Cloud and Galactic bulge sources, respectively.  The
solid circle indicates the location of the MACHO object, and the open
circle represents the location of the lensed source, as determined
from the blending parameters.

In the following, we discuss each of the 21 events individually.


\subsection{LMC-1 (79.5628.1547)}
\label{subsec-lmc1}

This event was the first LMC microlensing event reported by the MACHO
collaboration \cite{macho-nat93}.  The lensed object does not
significantly change color during the event, and its color and
magnitude indicate that it is a slightly reddened red clump giant star
in the LMC.  We thus expect {\it a priori} that a large fraction of
the MACHO object's brightness should be lensed.  A difference image
photometry (DIP, \citeNP{austin-dip96}) analysis of the event images
indicates a very small centroid shift between the photometered object
and lensed source, $0.1 \pm 0.1$'', also suggesting the clump giant
was lensed.

\begin{figure}
\plotone{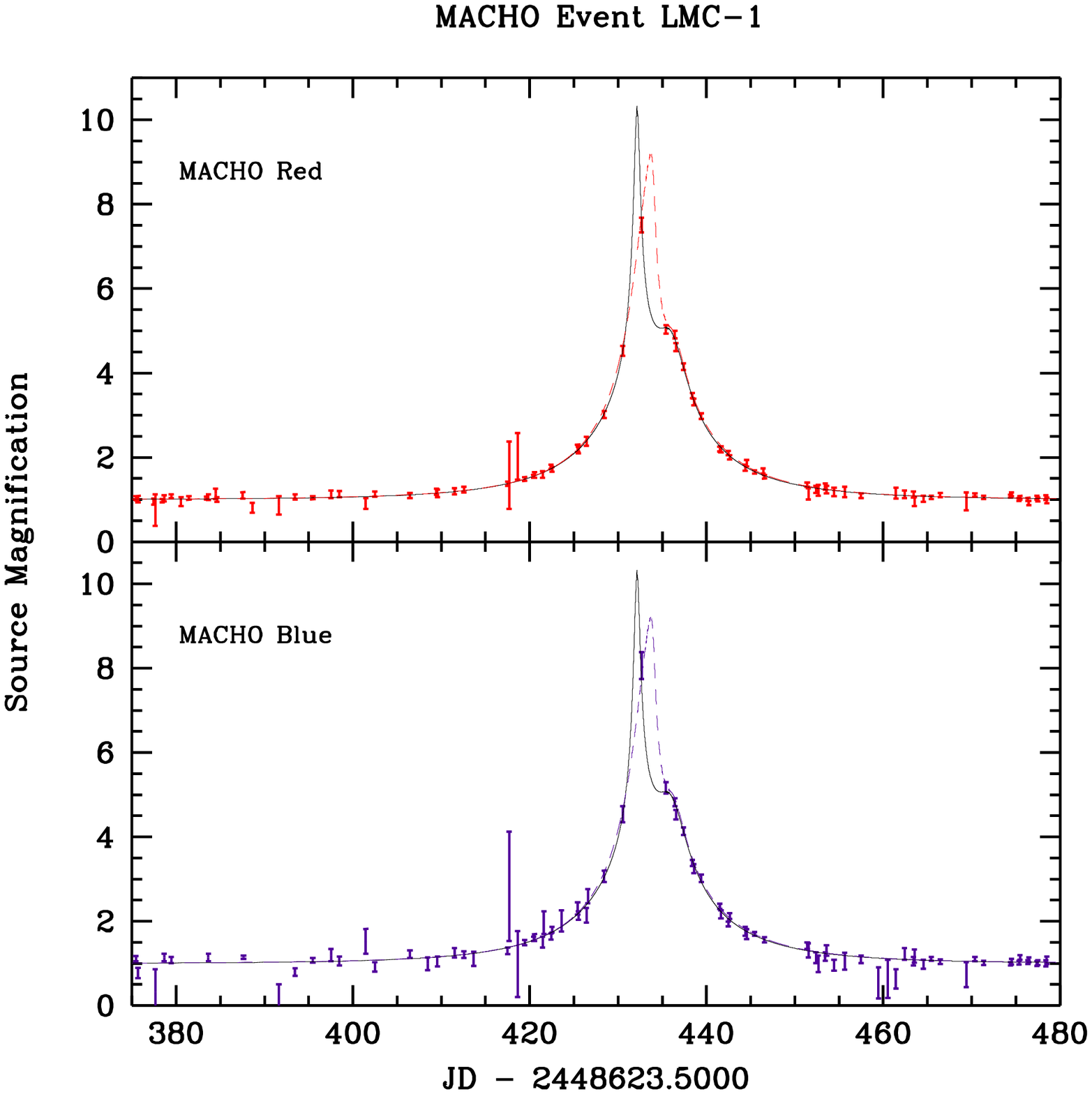}
\figcaption[f1.ps]{\label{fig-lmc1} Lightcurve of MACHO
  event LMC-1, including fits indicating a 'planetary' mass secondary
  lens (-~-~-) and a more standard binary system (-----).  }
\end{figure}

\begin{figure}
\plotone{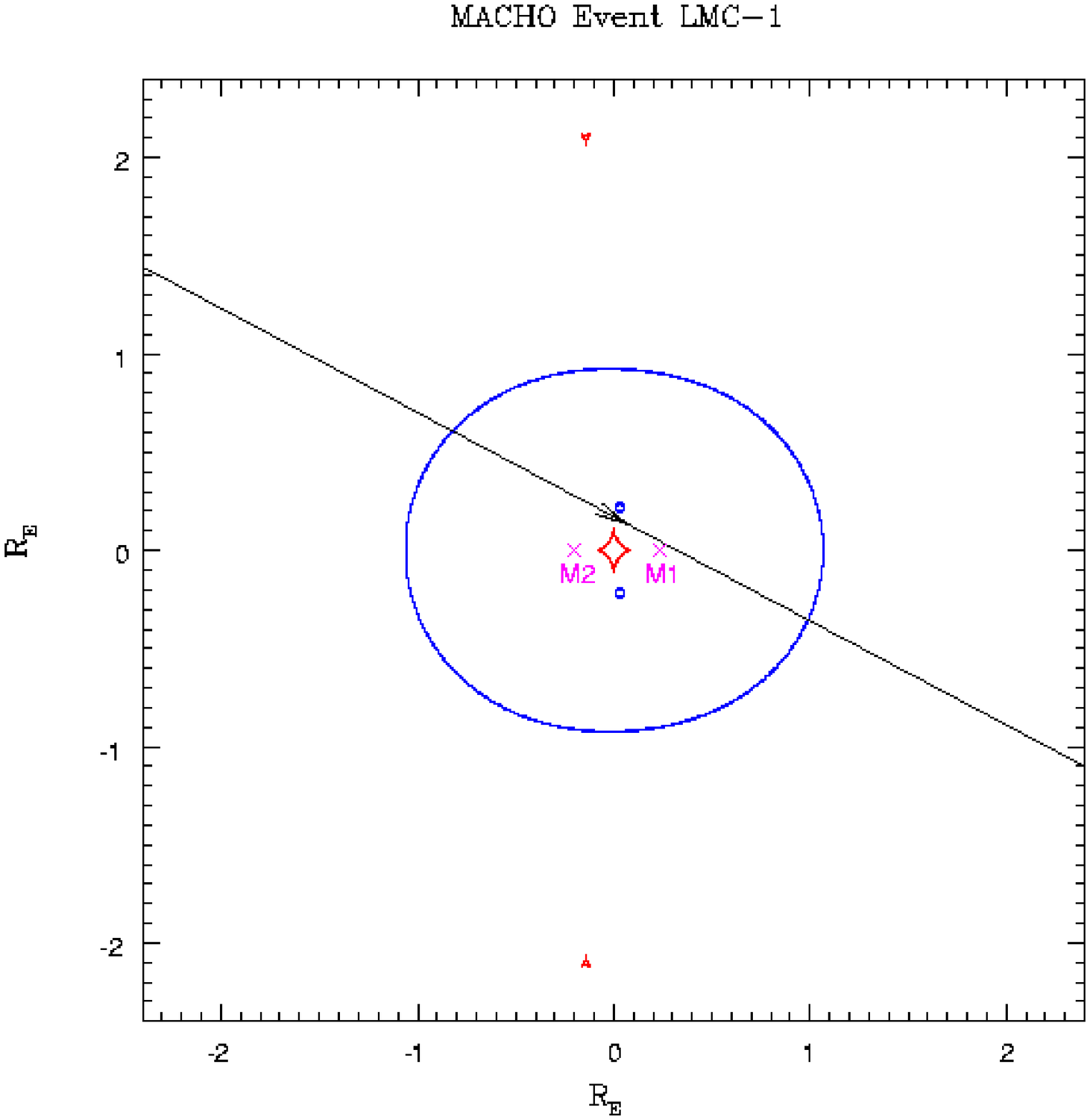}
\figcaption[f2.ps]{\label{fig-lmc1ccn} Location of the
  (red) caustic and (blue) critical curves for the LMC-1 standard binary
  lens fit (-----) presented in Fig.~\ref{fig-lmc1}.  The coordinate
  system, whose origin is at the center of mass, indicates distance in
  units of the system's Einstein ring radius $\Re$.  Also shown are the
  locations of the lensing objects, and the trajectory of the source
  through the caustic structure.  }
\end{figure}

\begin{figure}
\plotone{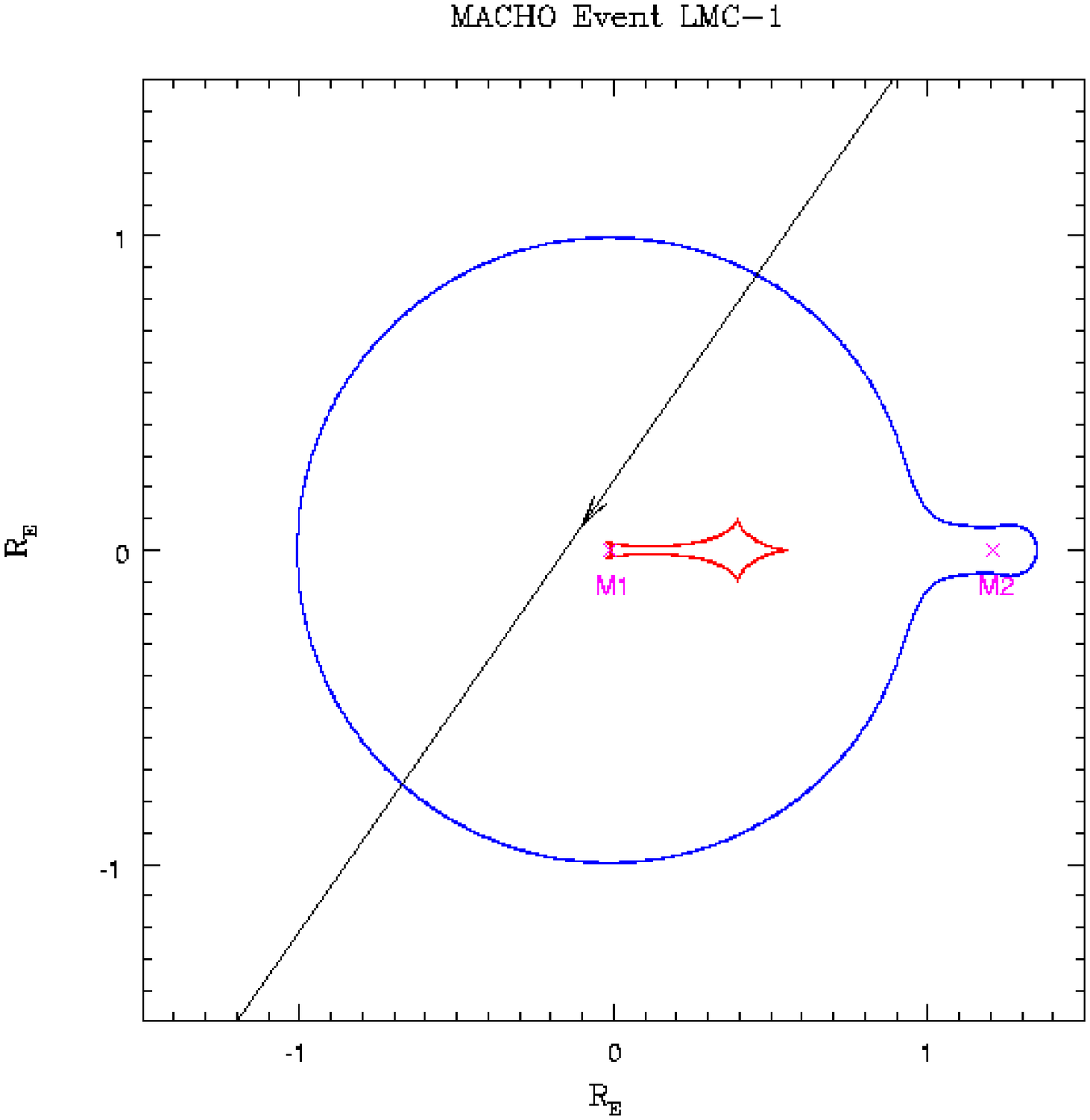}
\figcaption[f3.ps]{\label{fig-lmc1ccp} Location of the
  (red) caustic and (blue) critical curves for the LMC-1 'planetary'
  binary lens fit (-~-~-) presented in Fig.~\ref{fig-lmc1}.  The
  coordinate system, whose origin is at the center of mass, indicates
  distance in units of the system's Einstein ring radius $\Re$.  Also
  shown are the locations of the lensing objects, and the trajectory of
  the source through the caustic structure.  }
\end{figure}

\begin{figure}
\plotone{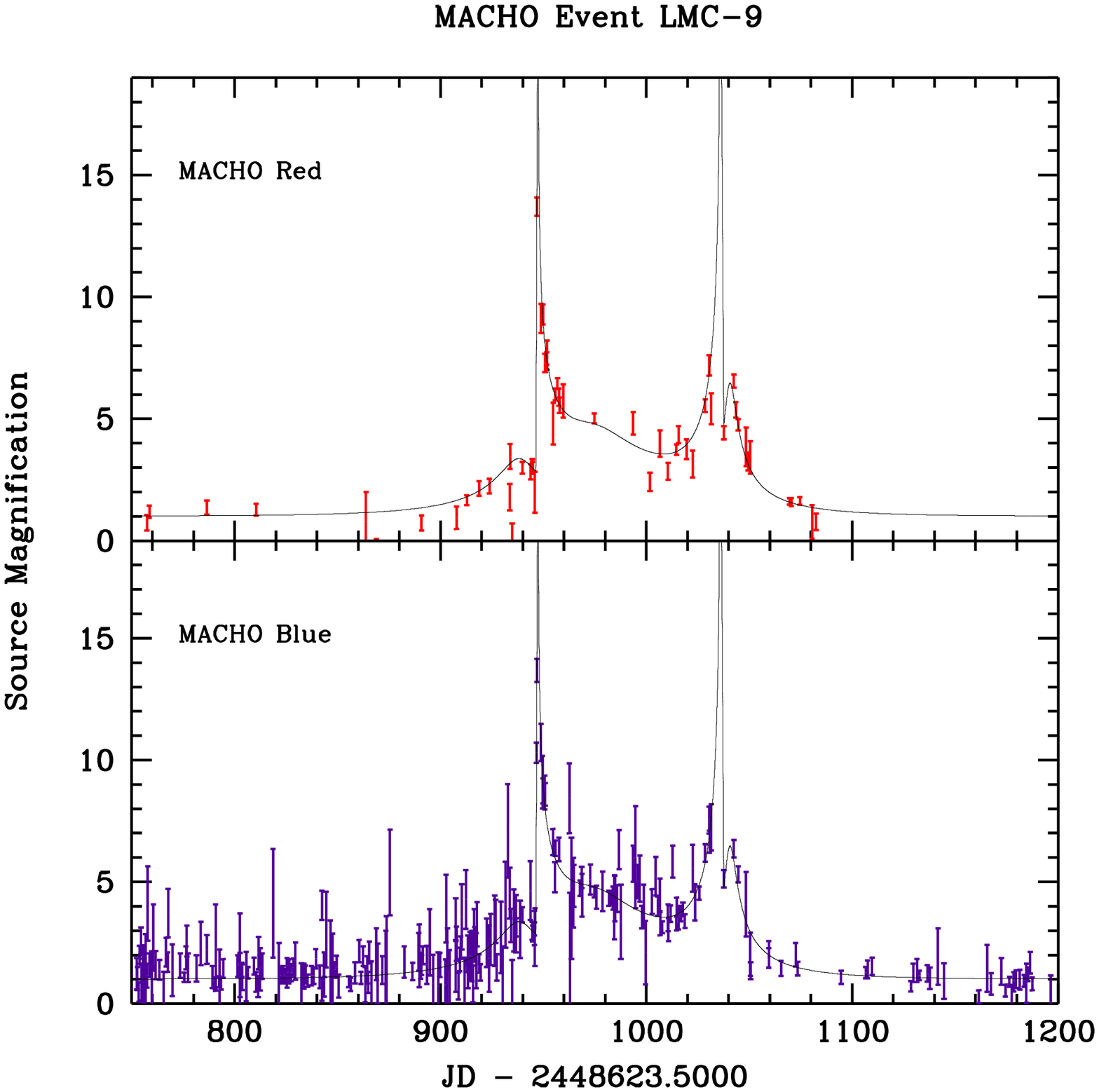}
\figcaption[f4.ps]{\label{fig-lmc9}
  Lightcurve of MACHO event LMC-9, including our fit to binary microlensing.
}
\end{figure}

\begin{figure}
\plotone{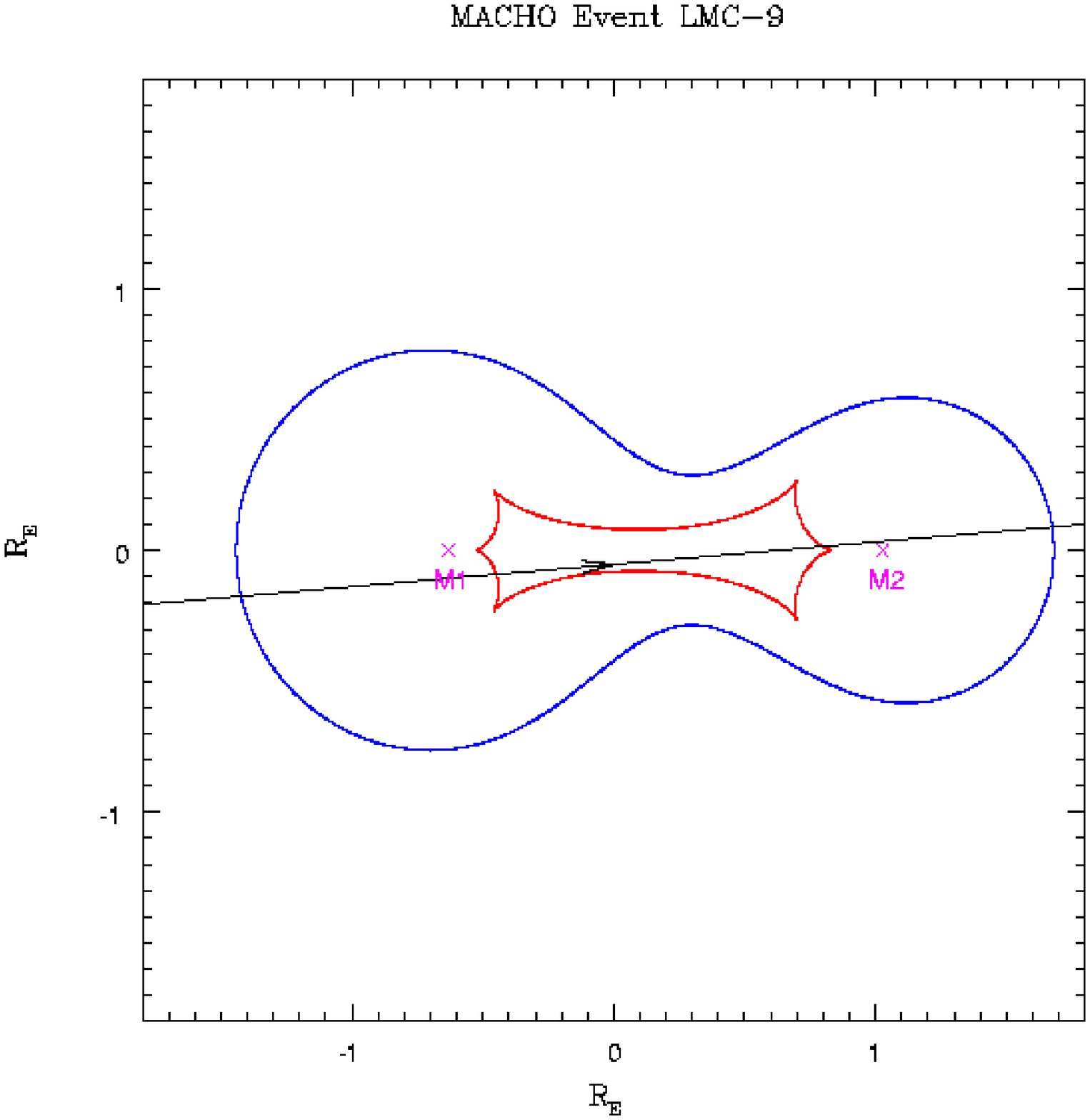}
\figcaption[f5.ps]{\label{fig-lmc9cc} Location of the
  (red) caustic and (blue) critical curves for the LMC-9 binary lens fit
  presented in Fig.~\ref{fig-lmc9}.  The coordinate system, whose origin
  is at the center of mass, indicates distance in units of the system's
  Einstein Ring radius $\Re$.  Also shown are the locations of the
  lensing objects, and the trajectory of the source through the caustic
  structure.  }
\end{figure}

\begin{figure}
\plotone{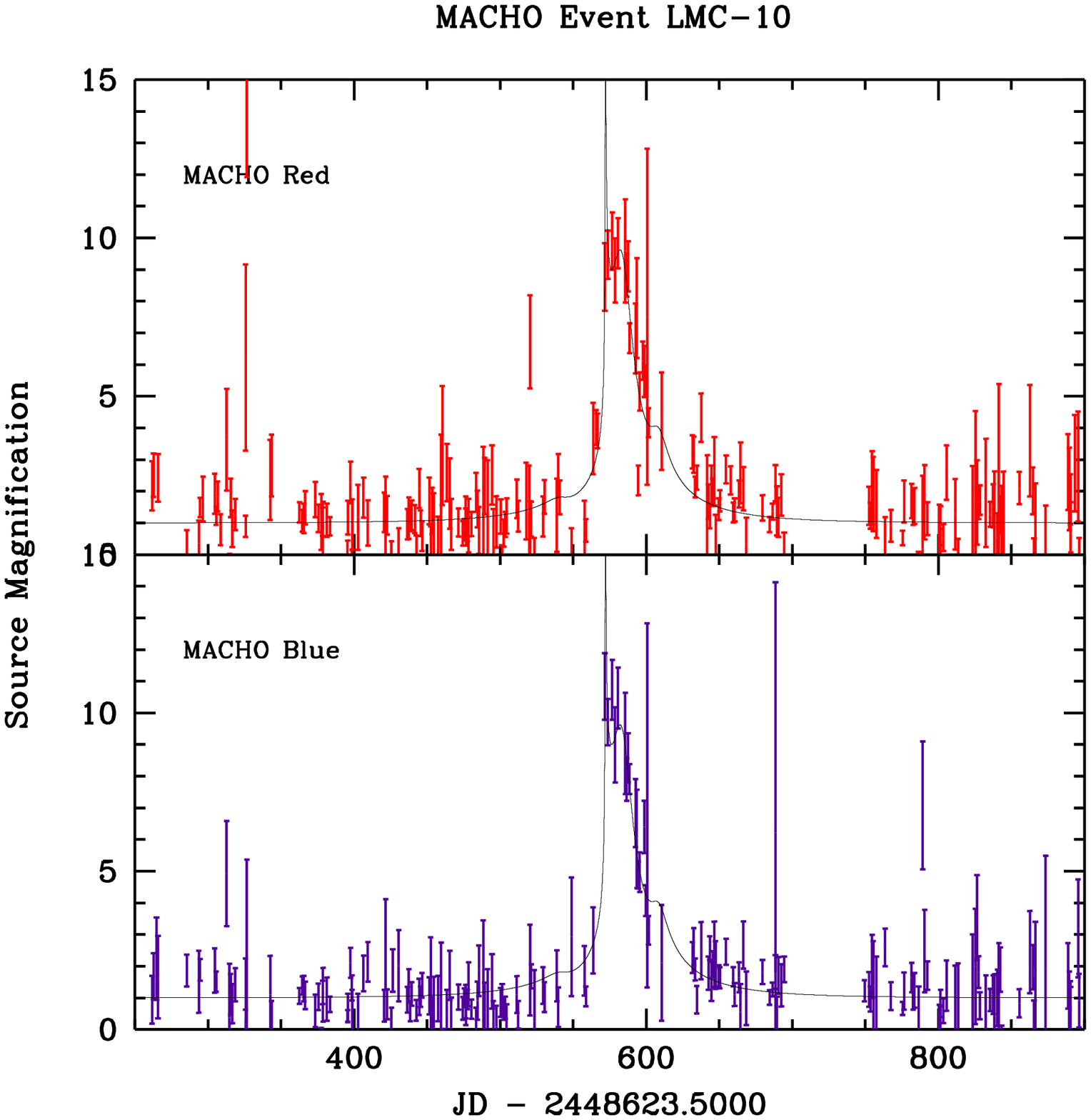}
\figcaption[f6.ps]{\label{fig-lmc10}
  Lightcurve of MACHO event LMC-10, including our fit to binary microlensing.
}
\end{figure}

\begin{figure}
\plotone{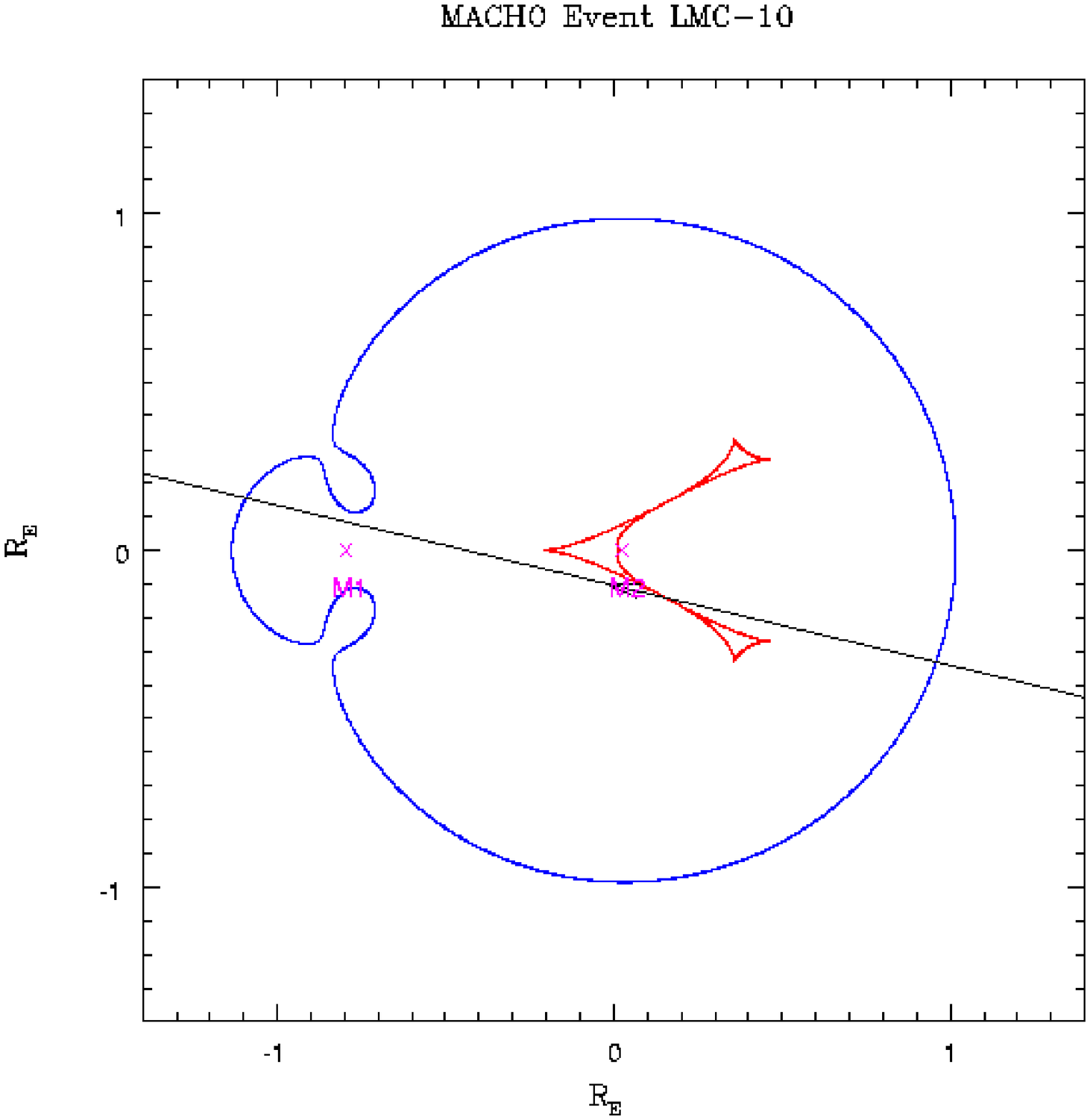}
\figcaption[f7.ps]{\label{fig-lmc10cc} Location of the
  (red) caustic and (blue) critical curves for the LMC-10 binary lens fit
  presented in Fig.~\ref{fig-lmc10}.  The coordinate system, whose origin
  is at the center of mass, indicates distance in units of the system's
  Einstein Ring radius $\Re$.  Also shown are the locations of the
  lensing objects, and the trajectory of the source through the caustic
  structure.  }
\end{figure}

\begin{figure}
\plotone{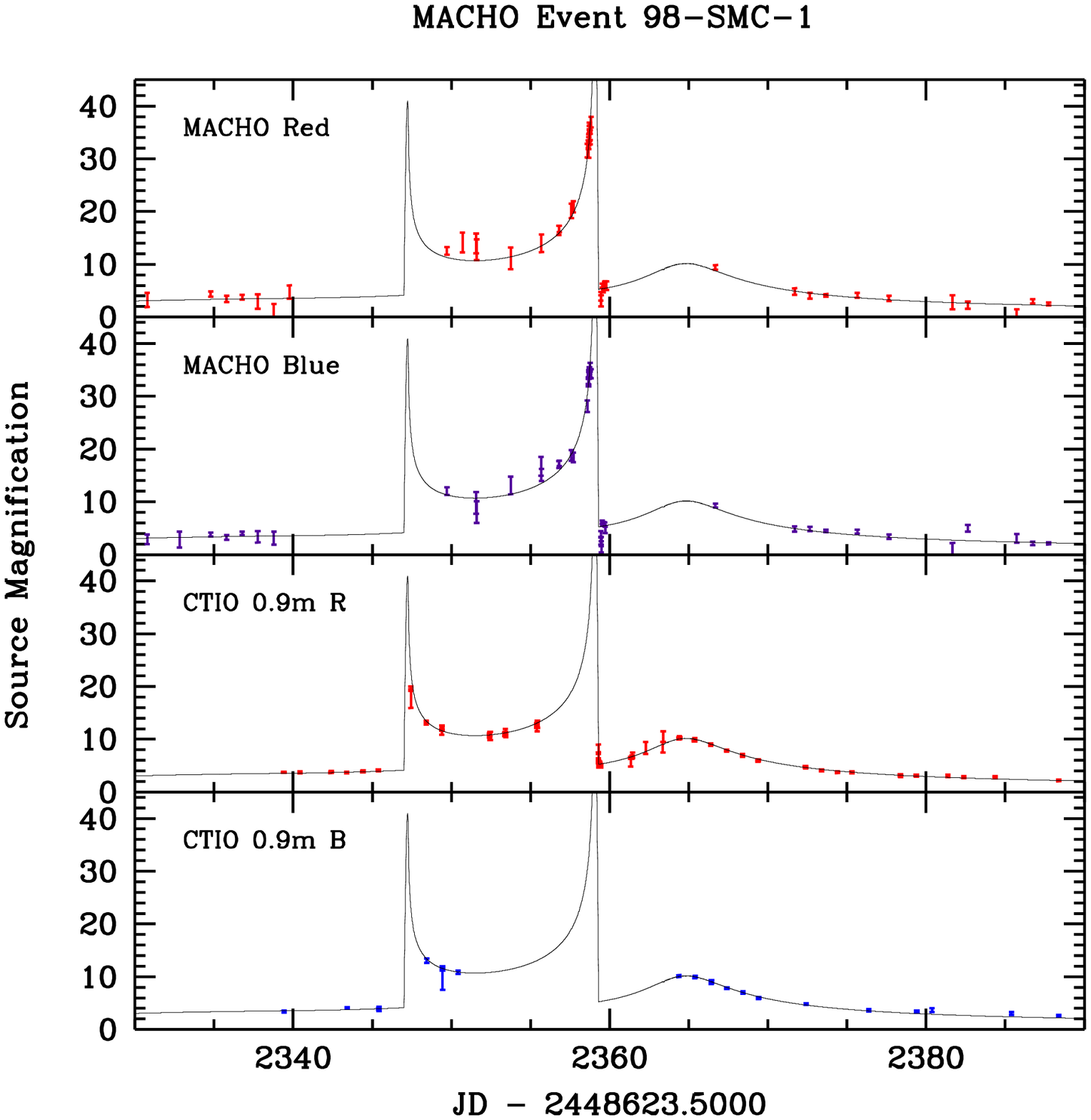}
\figcaption[f8.ps]{\label{fig-98smc1}
  Lightcurve of MACHO event 98-SMC-1, including our fit to binary microlensing.
}
\end{figure}

\begin{figure}
\plotone{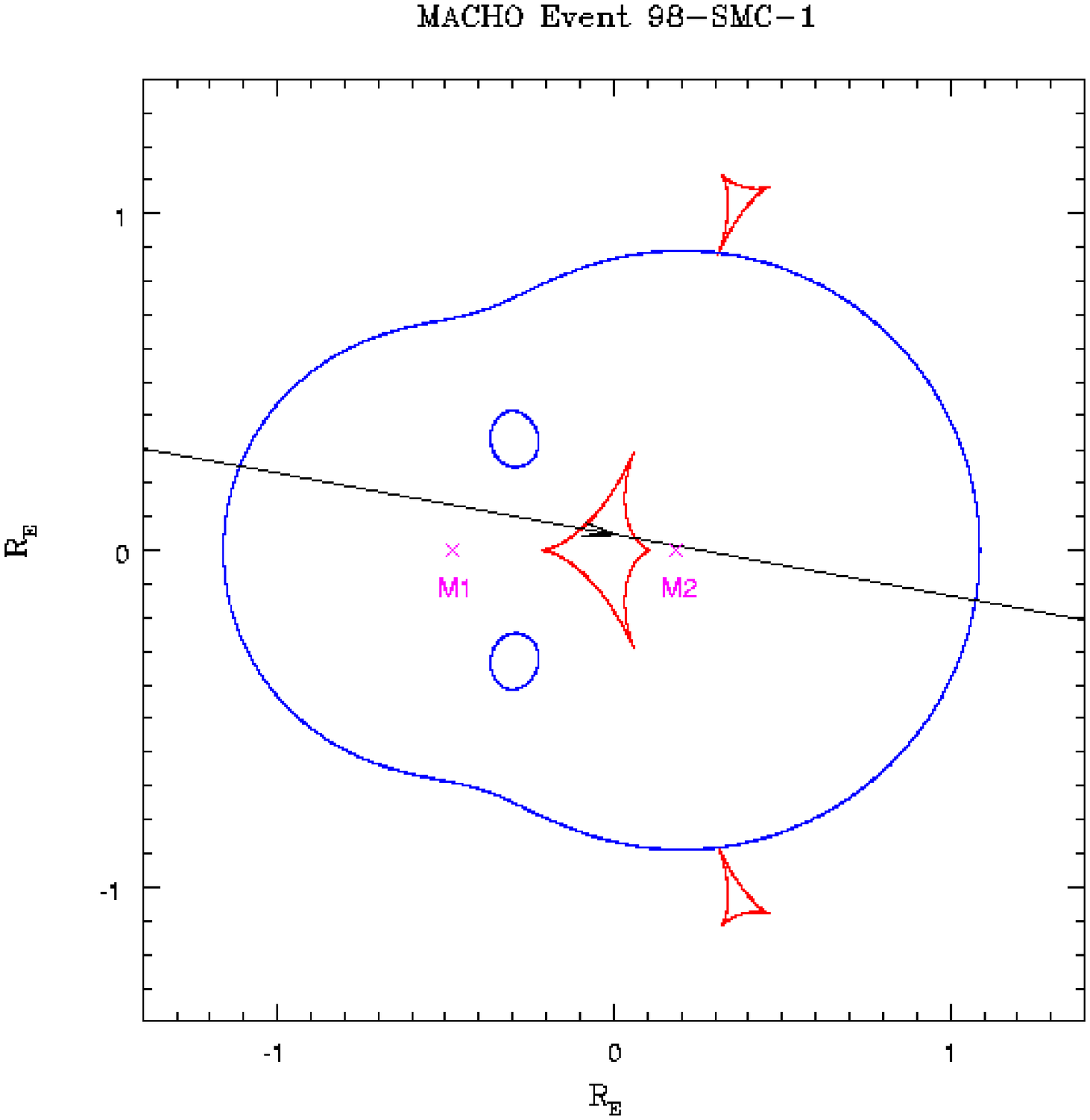}
\figcaption[f9.ps]{\label{fig-98smc1cc} Location of the
  (red) caustic and (blue) critical curves for the 98-SMC-1 binary lens fit
  presented in Fig.~\ref{fig-98smc1}.  The coordinate system, whose origin
  is at the center of mass, indicates distance in units of the system's
  Einstein Ring radius $\Re$.  Also shown are the locations of the
  lensing objects, and the trajectory of the source through the caustic
  structure.  }
\end{figure}

\begin{figure}
\plotone{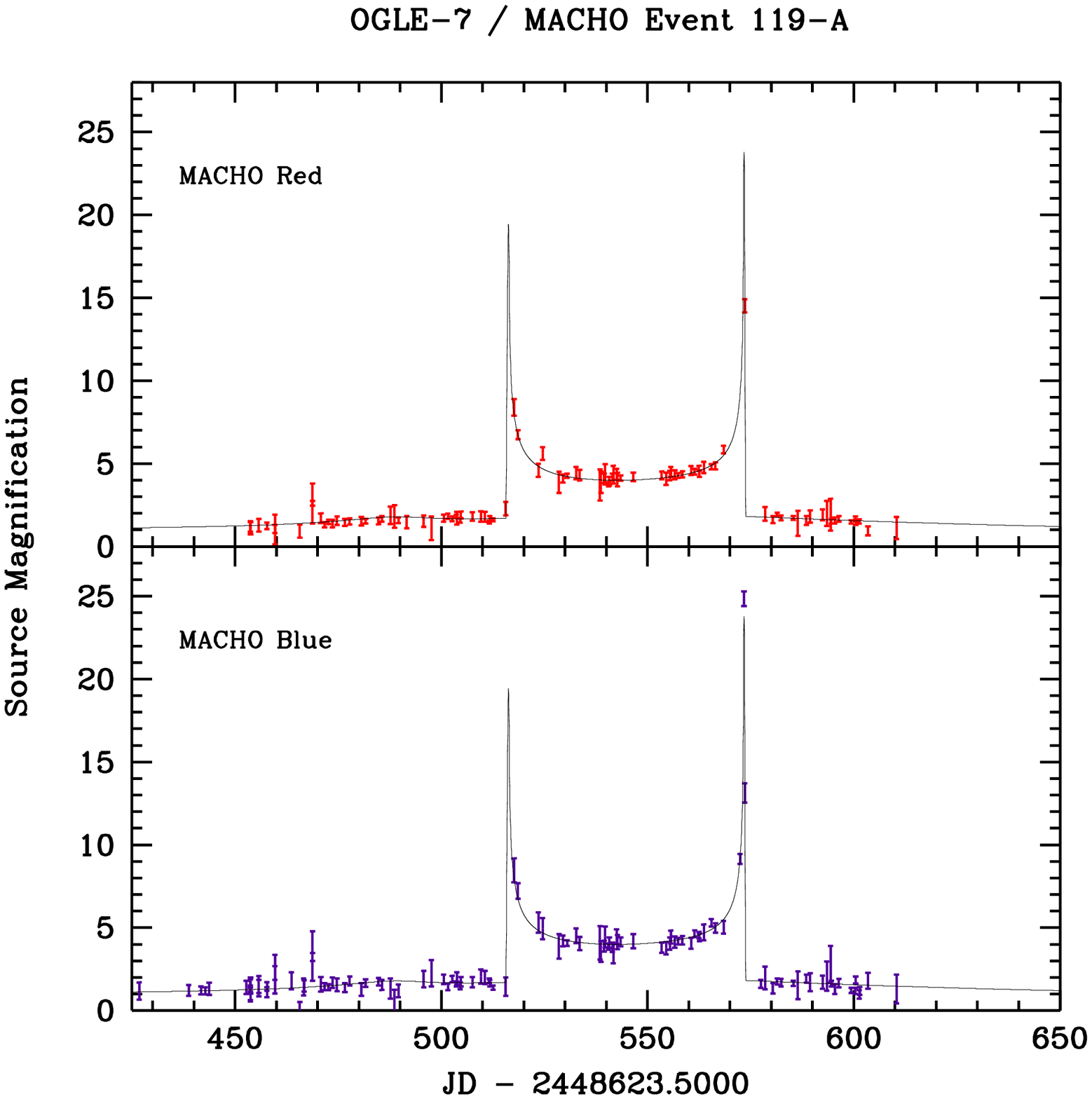}
\figcaption[f10.ps]{\label{fig-119a}
  Lightcurve of MACHO event 119-A, including our fit to binary microlensing.
}
\end{figure}

\begin{figure}
\plotone{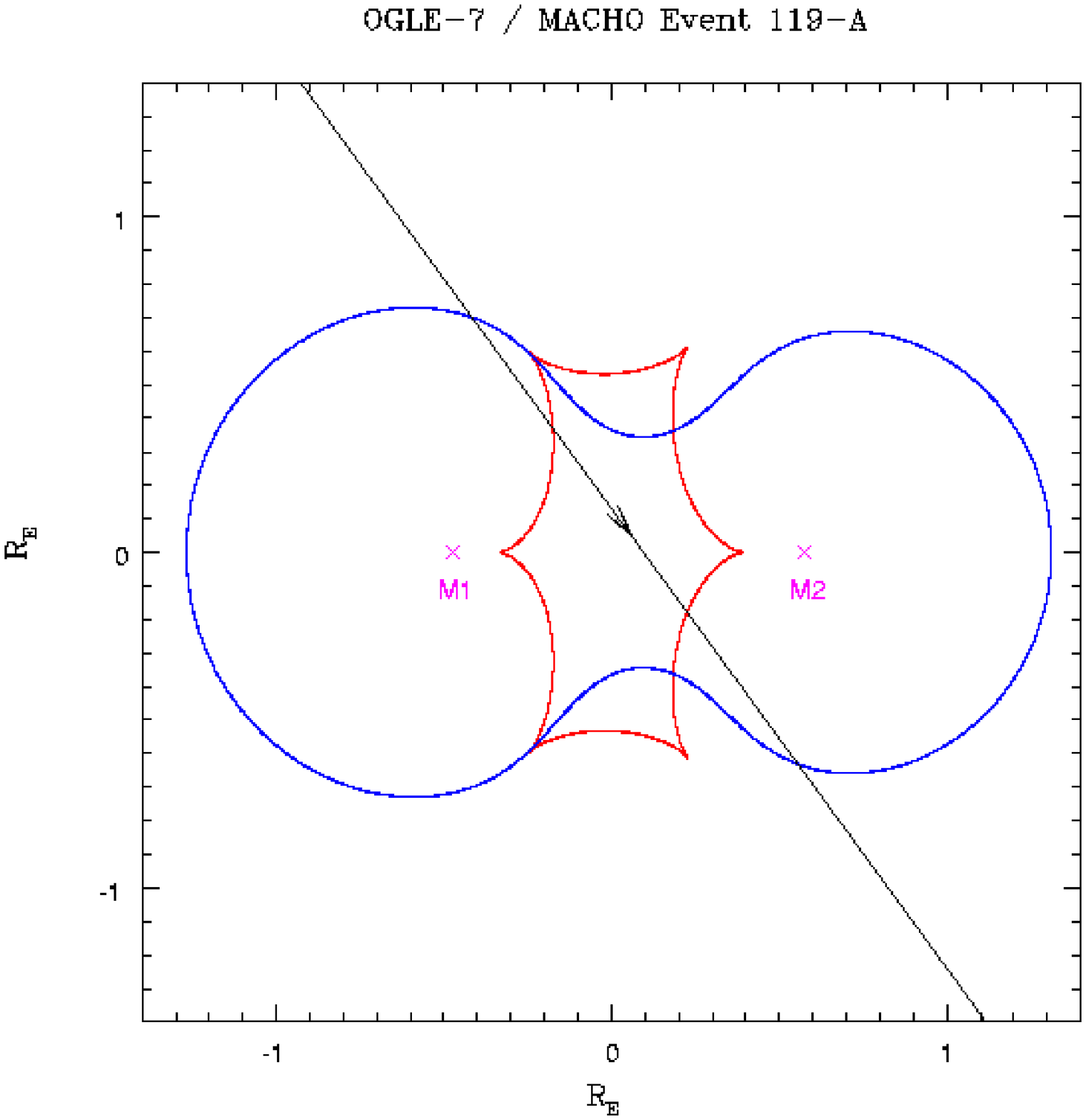}
\figcaption[f11.ps]{\label{fig-119acc} Location of the
  (red) caustic and (blue) critical curves for the 119-A binary lens fit
  presented in Fig.~\ref{fig-119a}.  The coordinate system, whose origin
  is at the center of mass, indicates distance in units of the system's
  Einstein Ring radius $\Re$.  Also shown are the locations of the
  lensing objects, and the trajectory of the source through the caustic
  structure.  }
\end{figure}

\begin{figure}
\plotone{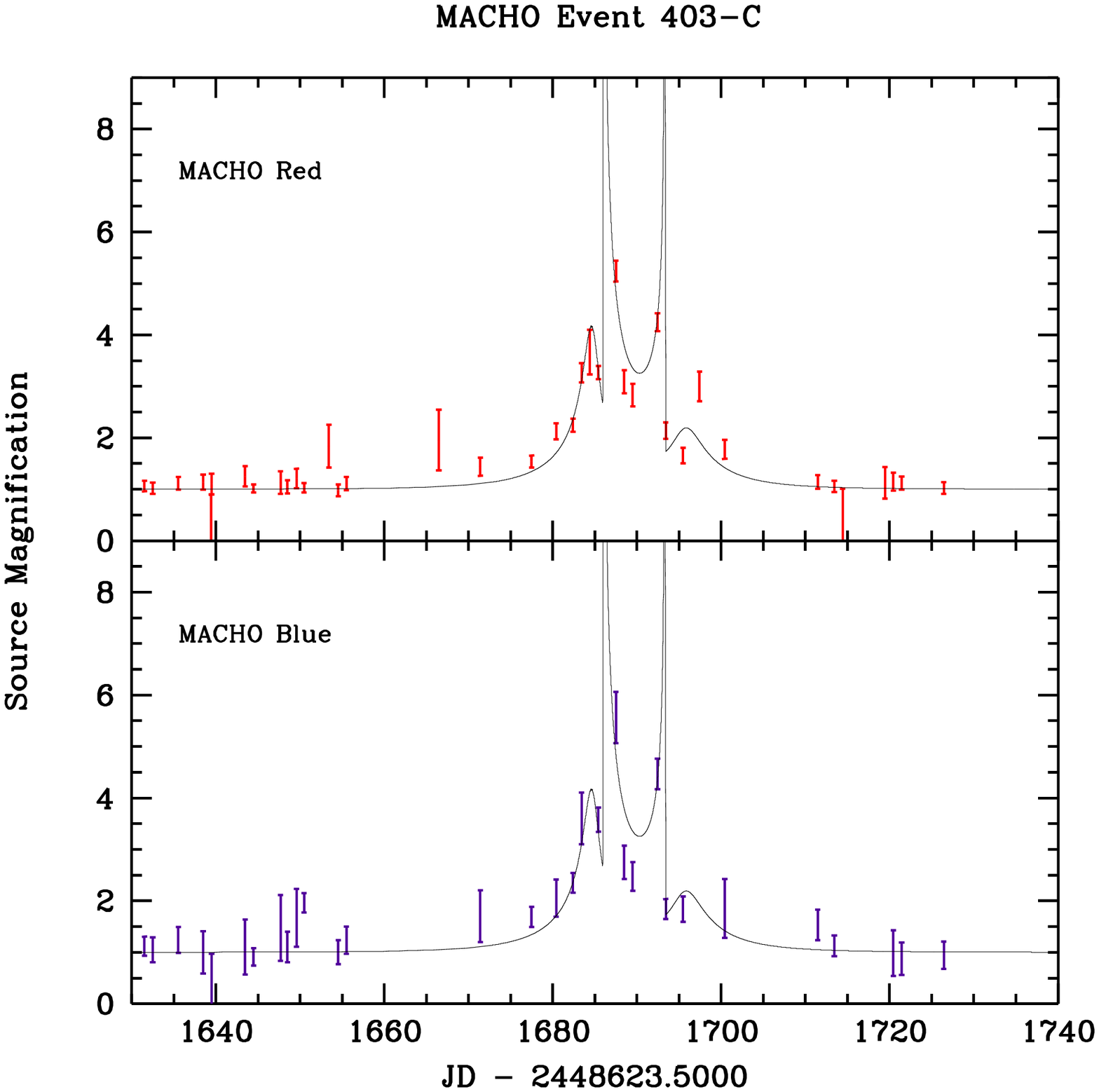}
\figcaption[f12.ps]{\label{fig-403c}
  Lightcurve of MACHO event 403-C, including our fit to binary microlensing.
}
\end{figure}

\begin{figure}
\plotone{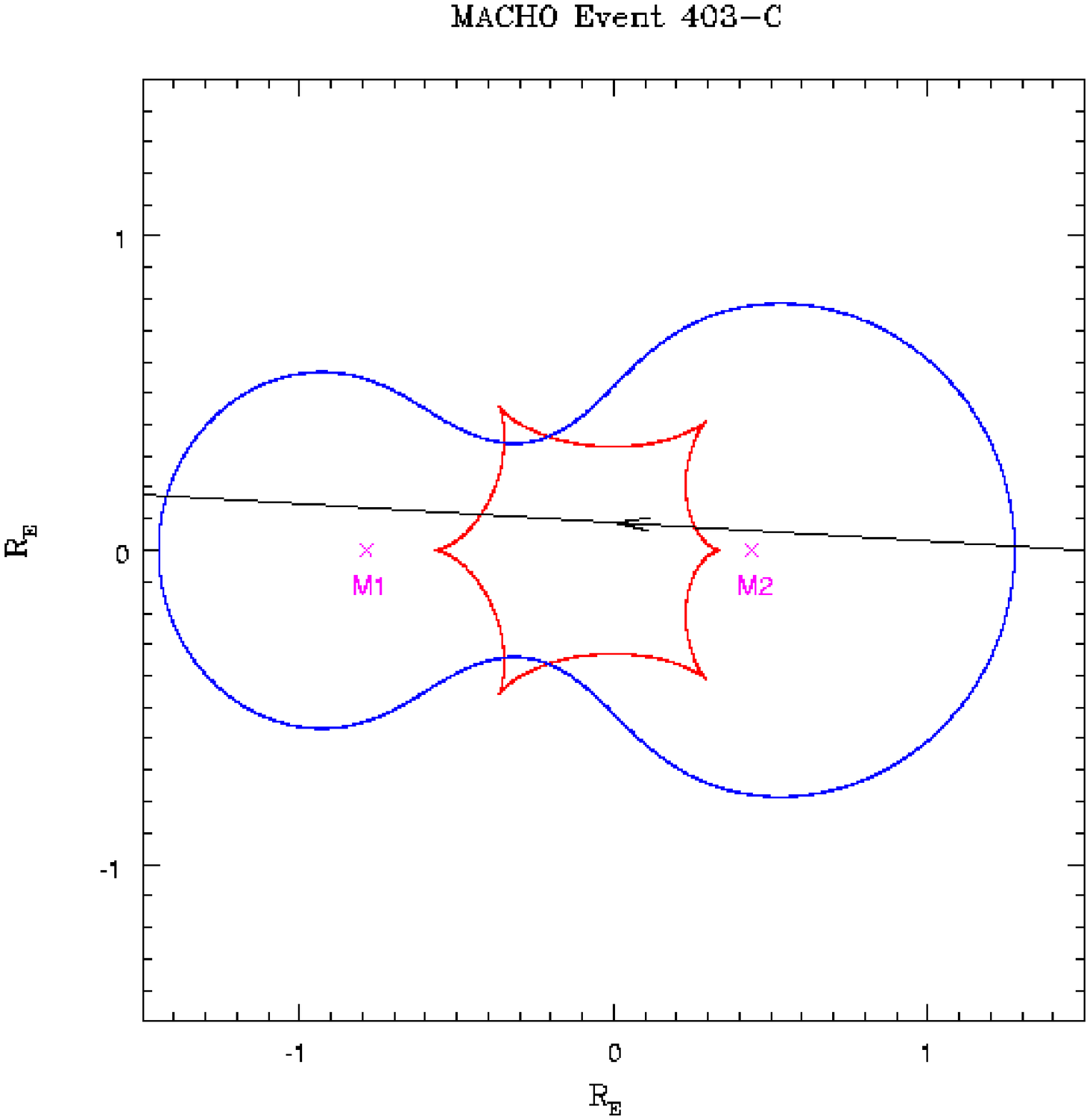}
\figcaption[f13.ps]{\label{fig-403ccc} Location of the
  (red) caustic and (blue) critical curves for the 403-C binary lens fit
  presented in Fig.~\ref{fig-403c}.  The coordinate system, whose origin
  is at the center of mass, indicates distance in units of the system's
  Einstein Ring radius $\Re$.  Also shown are the locations of the
  lensing objects, and the trajectory of the source through the caustic
  structure.  }
\end{figure}

\begin{figure}
\plotone{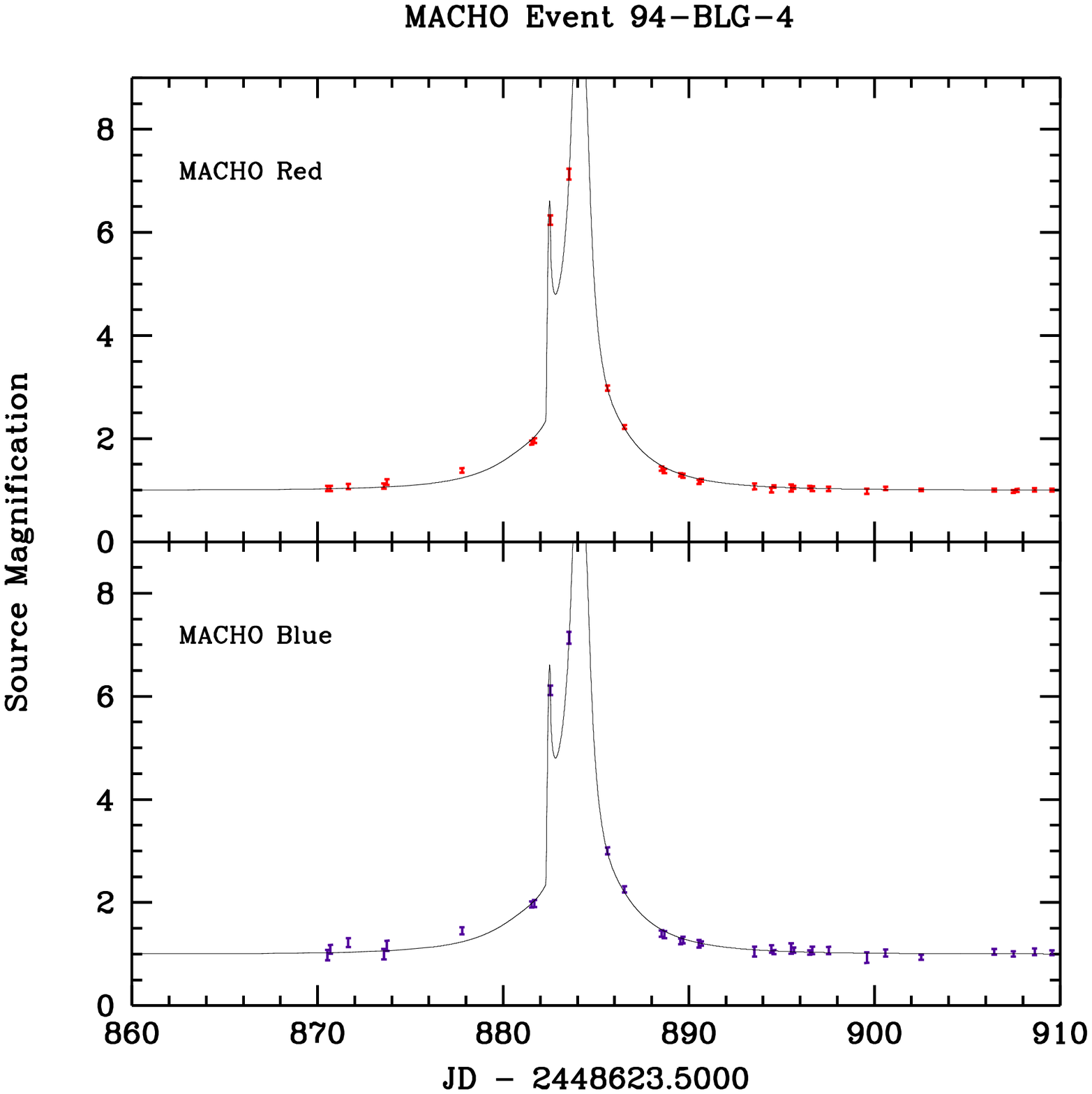}
\figcaption[f14.ps]{\label{fig-94blg4}
  Lightcurve of MACHO event 94-BLG-4, including our fit to binary microlensing.
}
\end{figure}

\begin{figure}
\plotone{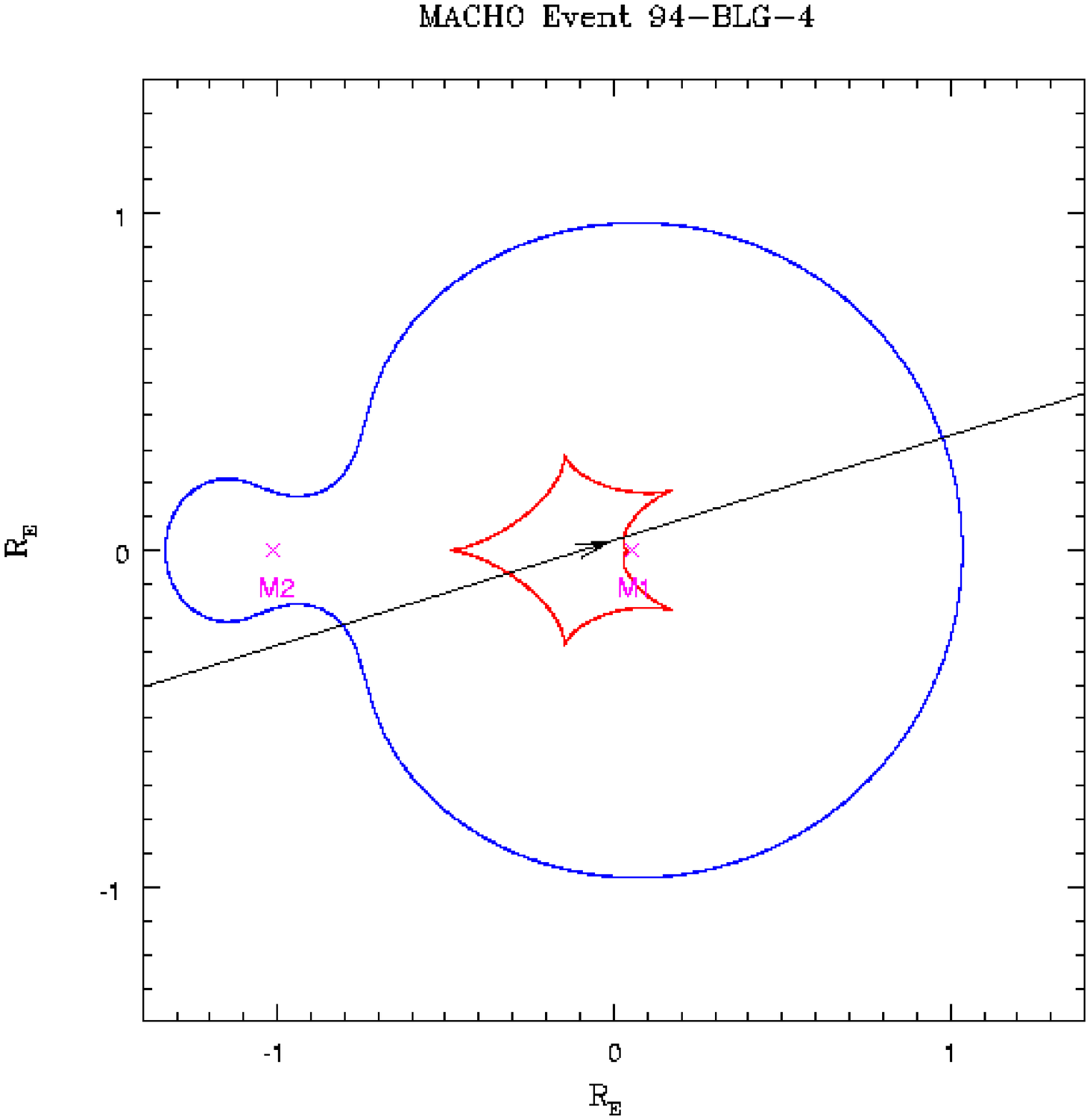}
\figcaption[f15.ps]{\label{fig-94blg4cc} Location of the
  (red) caustic and (blue) critical curves for the 94-BLG-4 binary lens fit
  presented in Fig.~\ref{fig-94blg4}.  The coordinate system, whose origin
  is at the center of mass, indicates distance in units of the system's
  Einstein Ring radius $\Re$.  Also shown are the locations of the
  lensing objects, and the trajectory of the source through the caustic
  structure.  }
\end{figure}

\begin{figure}
\plotone{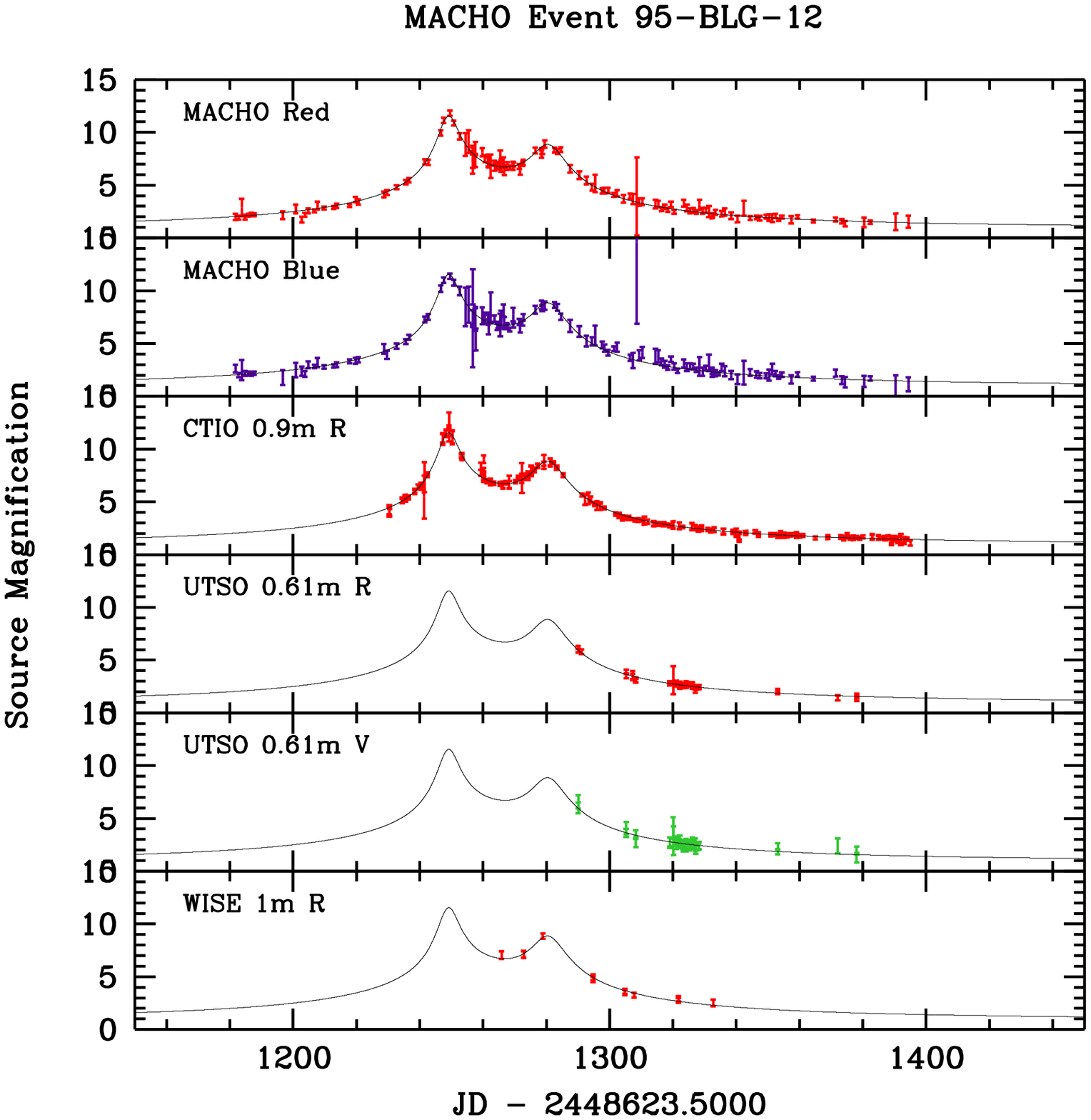}
\figcaption[f16.ps]{\label{fig-95blg12}
  Lightcurve of MACHO event 95-BLG-12, including our fit to binary microlensing.
}
\end{figure}

\begin{figure}
\plotone{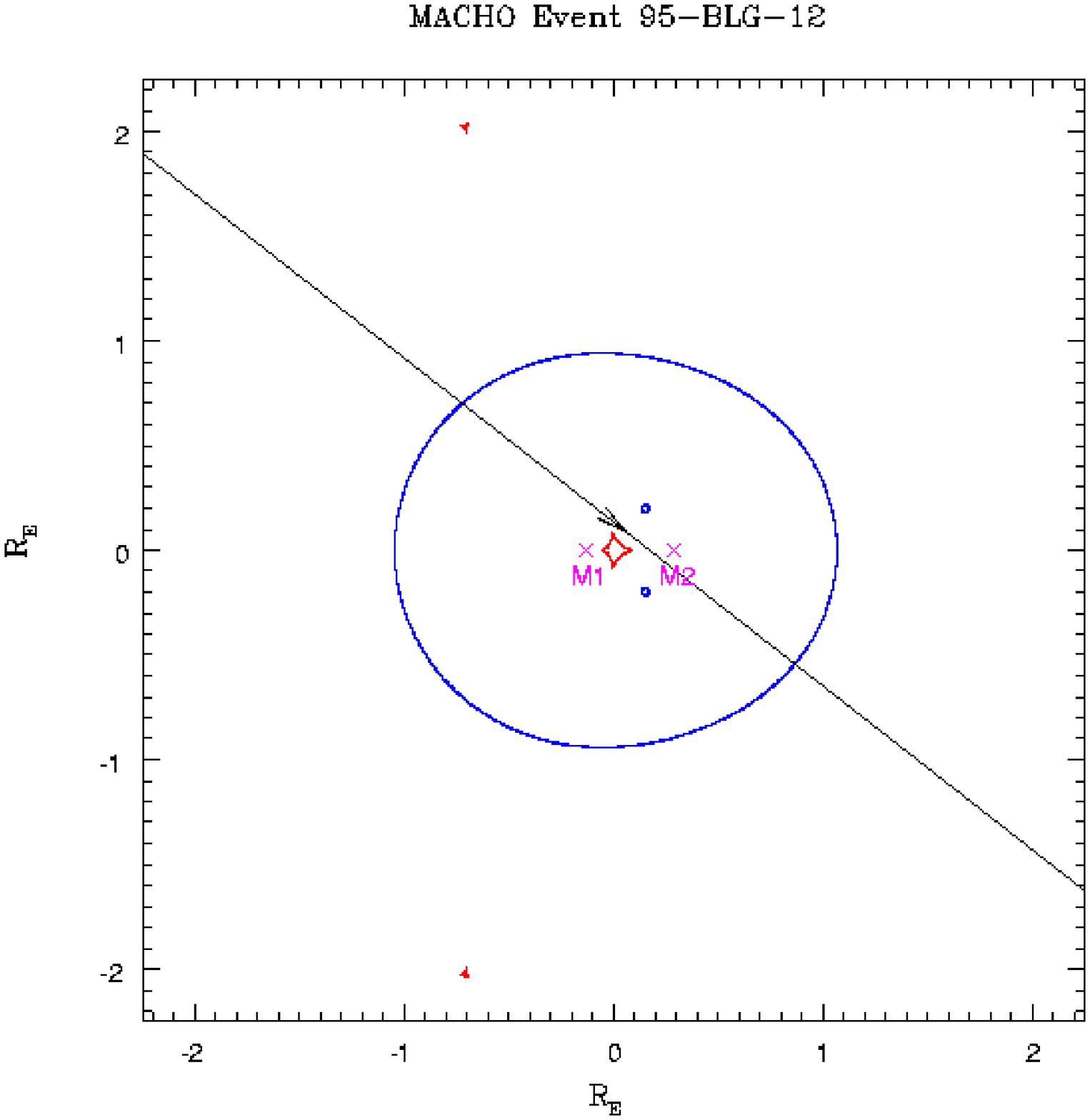}
\figcaption[f17.ps]{\label{fig-95blg12cc} Location of the
  (red) caustic and (blue) critical curves for the 95-BLG-12 binary lens fit
  presented in Fig.~\ref{fig-95blg12}.  The coordinate system, whose origin
  is at the center of mass, indicates distance in units of the system's
  Einstein Ring radius $\Re$.  Also shown are the locations of the
  lensing objects, and the trajectory of the source through the caustic
  structure.  }
\end{figure}

The lightcurve is consistent with single lens microlensing, except for
a shoulder shortly after the lightcurve peak.  This suggests a short
timescale structure which was not resolved at the sampling of the
survey system.  Binary fits for this event have been done by
\citeN{dominik96}, and the two fits displayed in Fig.~\ref{fig-lmc1}
were previously presented by \citeN{rhie96a}.  The similarity to
standard microlensing implies the fit resides in a class of fits with
small lens separations (fit~----- in Fig.~\ref{fig-lmc1}, and
Fig.~\ref{fig-lmc1ccn}) or large mass ratios (fit~-~-~- in
Fig.~\ref{fig-lmc1}, and Fig.~\ref{fig-lmc1ccp}).

The measured $\that \sim 35$ days for both of these fits yields an
expectation value for the mass of the lensing system of $\sim 0.1
\msun$ \cite{macho-nat93,dominik98b}.  For the large mass ratio fit,
this implies a secondary companion of $\sim 1$ Jupiter mass.  However,
since there were no follow-up efforts in place at the time of this
event, this degeneracy between fits cannot be broken.


\subsection{LMC-9 (80.6468.2746)}
\label{subsec-lmc9}

LMC-9 is the most anomalous lensing event discovered toward the LMC,
and it displays the lightcurve structure of a typical caustic crossing
binary lens event with similar mass lenses (see Fig.~\ref{fig-lmc9}).
Prior analysis of this event can be found in \citeN{macho-bennett96b}.
The lensed object appears to be on the red side of the LMC main
sequence, at the level of the clump.  However, the fits suggest only
$\sim 1/5$ of this light is lensed, and so the source likely resides
further down the main sequence as a late A or early F star.  The
lensed source does appear $0.4 \pm 0.1$'' from the MACHO object, based
upon DIP analysis.

The first caustic crossing appears to be resolved with two
observations on its rise.  If we assume the source is not itself a
binary, we can estimate the source radius as $1.5 \pm 0.2 \rsun$ from
its color and magnitude.  The measured $\tstar = 0.65 \pm 0.10$ days
yields a lens velocity projected to the LMC (assumed to be at $50
\kpc$) of $19 \pm 4 \kms$, considerably lower than what one would
expect from a halo lens.  In fact, this velocity is low enough that it
is only marginally consistent with a lens in the disk of the LMC, and
would not be consistent with any LMC models including velocity
dispersions in excess of the measured values.

\citeN{aubourg99} have suggested an LMC model with a self-lensing
optical depth large enough to explain the excess of lensing events
observed towards the LMC.  \citeN{aubourg99} achieve a high
microlensing optical depth without violating the virial relation
between velocity dispersion and microlensing optical depth
\cite{gould-disktau} by arguing that the velocity dispersion of the
lens population is much higher than the velocity dispersion of the
stars with measured radial velocities \cite{cowhart,zaritsky99}.
However, even in such a model our projected velocity of $19 \pm 4
\kms$ for LMC-9 would not be consistent with a lens in the
LMC\rlap.\footnote[3]{An early version of \citeN{aubourg99}
incorrectly suggested that their model was consistent with the LMC-9
projected velocity (Aubourg 1999, private communication).}

Thus, there are two possible interpretations of the LMC-9 caustic crossing
observations: 
\begin{enumerate}
   \item They imply that the lens resides in an LMC disk which
contains enough mass to generate only a fraction of the observed LMC
lensing events, or
   \item The source star is itself a binary, and the two caustic
crossing observations do not constrain the lens proper motion or its
location.
\end{enumerate}
Both of these possibilities have an {\it a priori} probability of
about 10\%. Clearly, better photometric coverage of the caustic
crossings could have resolved this issue.


\subsection{LMC-10 (18.3324.1765)}
\label{subsec-lmc10}

We include LMC-10 as a binary candidate as an example of variability
which passes our microlensing cuts \cite{macho-lmc2}, and yet is
almost certainly not standard single lens microlensing due to its
asymmetry (Fig.~\ref{fig-lmc10}).  Explanations for this event include
intrinsic stellar variability (we detect no recurrence of this
behavior in 5 further years of observations), a background supernova
(we find no obvious host galaxy down to R $\sim 21$), or possibly a
weakly perturbed binary microlensing event.  Such lightcurves are
thought to compose a significant fraction of all binary lensing events
\cite{distefano99}.  The best fit lightcurve does have a pair of
caustic crossing, but they are separated by about 4 hours and occurred
during daylight hours in Australia. For $\tstar < 0.2$ days, these
caustic crossings would not have been observable from Australia, and
the fit is a plausible explanation of the lightcurve.  The
microlensing interpretation of this event would be much more secure if
such a lightcurve were seen in a time reversed order, since the
observed rapid rise followed by a slow decay is common in known types
of stellar variability.  The one observational test that can still be
performed is to obtain high resolution HST images of this object.  The
centroid of the variable source is well constrained by the DIP method
of \citeN{austin-dip96}, and appears $1.0 \pm 0.2$'' from the MACHO
object.  If the binary lens interpretation is correct, then this
centroid should be centered on a $R = 21.4$ star.  Alternatively, the
HST frame might also reveal a background galaxy which will make the
supernova explanation most plausible.


\subsection{98-SMC-1 (208.15683.4237)}
\label{subsec-98smc1}

Alert 98-SMC-1, detected on May 25 1998, was the second microlensing
event seen towards the SMC.  Nightly GMAN followup observations of the
early lightcurve detected evidence of a caustic crossing near June 6.5
UT 1998 (day 2347.5 in Fig.~\ref{fig-98smc1}) \cite{macho-iauc98a}.
The prospect of resolving the second caustic crossing, allowing an
estimate of the lens location, prompted the microlensing community to
commit significant observational resources to this event.  Following
predictions of the second caustic crossing date by
\citeN{macho-iauc98b} and the PLANET collaboration, the event was
followed by the EROS \cite{eros-98smc1}, PLANET \cite{planet-98smc1},
MACHO/GMAN \cite{macho-98smc1}, OGLE \cite{ogle-98smc1}, and MPS
collaborations \cite{mps-98smc1}.  In each case, the analysis of event
data leads to a relatively long caustic crossing timescale, with the
robust conclusion that 98-SMC-1 is due to SMC-SMC self lensing.  A
joint paper incorporating all event data is in preparation
\cite{joint-98smc1}.

The fit we present here, originally from \citeN{macho-98smc1},
exhibits a $\tstar = 0.116\pm 0.010$ days.  We estimate $\rstar = 1.1
\pm 0.1 \rsun$, and assuming the source resides in the SMC at $60\kpc$
we find a lens velocity projected to the SMC of $\vhat = 76 \pm
10\kms$.  The probability for a Galactic halo lens to produce a $\vhat$
value this small is of order $0.1\,$\%.


\subsection{OGLE-7/MACHO-119-A (119.20226.2119)}
\label{subsec-ogle7}

This caustic crossing event was initially detected by the OGLE
collaboration \cite{udalski94c}, and later verified with data from the
MACHO database \cite{macho-mao94}.  As OGLE-7, it was the first
reported case of binary microlensing.  119-A is a moderately blended
event, with $f \sim 0.4$, and the source appears to be located near
the top of the bulge main sequence.  

The fit presented in Fig.~\ref{fig-119a}, and the geometry of the
lensing encounter in Fig.~\ref{fig-119acc}, are qualitatively similar
to the analysis of \citeN{udalski94c}.  A comparison of event
parameters (MACHO:OGLE) yields $\that = (169:160), \umin =
(0.08:0.05), a = (1.05:1.14), \epsilon_1 = (0.55:0.50), \theta =
(-0.94:0.84)$, where the difference in $\theta$ is due to a different
orientation of coordinate systems.  Fortunately, the MACHO dataset
provides better sampling of the lightcurve, and thus more tightly
constrains the global fit.  The data presented here include 2
observations on the falling portion of the second caustic crossing,
yielding an additional constraint of $\tstar = 0.21 \pm 0.03$ days.


\subsection{MACHO-403-C (403.47793.2961)}
\label{subsec-403c}

Event 403-C exhibits a series of photometric deviations near Aug 18,
1996 (day 1690 in Fig.~\ref{fig-403c}) which are, for the most part,
achromatic.  It is plausible these deviations are due to gravitational
microlensing.  However, this event does not appear to be consistent
with single lens microlensing, and is only marginally consistent with
binary microlensing.  The sparse sampling and relatively large error
bars prevent tight constraints on the binary microlensing event
parameters.


\subsection{94-BLG-4 (118.18141.731)}
\label{subsec-94blg4}

94-BLG-4 exhibits features similar to LMC-1, with a high-$\sigma$
deviation near the peak of the event which is unresolved by the survey
system.  This event was initially reported in
\citeN{macho-bennett97a}, after discovery during testing of the MACHO
Alert System.  The 2 observations apparently taken between caustic
crossings are not enough to completely constrain the lensing system,
thus Fig.~\ref{fig-94blg4} and Fig.~\ref{fig-94blg4cc} are probably
not unique interpretations of this sparsely sampled lightcurve.

The lensed object is a clump giant, with $V = 17.9$, $V-R = 1.1$,
based upon its position in Fig.~\ref{fig-cmdblg}.  The binary fit
blend fraction of $f \sim 1$ indicates the clump giant itself was
lensed, avoiding fit ambiguities introduced by unknown or
unconstrained blend fractions, which directly influence our
measurement of $\that$.  The event duration is relatively short,
$\that = 10.7$ days.  If we assume ``typical'' parameters for the
lensing object ($\vperp = 150 \kms$ and a distance 80\% of the way to
the Galactic bulge) we arrive at the relationship $\that \sim 70
\sqrt{m/\msun}$ days \cite{paczynski91}.  This implies an overall lens
mass of $0.02 \msun$ -- however, we expect the shortest of our events
to be drawn from lenses residing closer to the source, so the actual
mass is likely to be much larger than this.  If we take an upper limit
of $0.2 \msun$ as the total lens mass, then the mass ratio of 1/18
implies a secondary lens of $\lesssim 10$ Jupiter masses, for this
particular fit.


\subsection{95-BLG-12 (120.21263.1213) }
\label{subsec-95blg12} 

This 12th Alert of the 1995 bulge season was detected on May 15, at a
magnification of $A \sim 2$.  Real-time follow-up observations by both
GMAN \cite{macho-pratt96a} and PLANET \cite{albrow96} detected
deviations from the standard fit near Jun 5, making 95-BLG-12 the
first binary event detected in real-time.  Data on this event from the
PLANET collaboration are presented in \citeN{planet95}.

The lensed object is located on the subgiant branch of the bulge,
below the clump.  However, because subgiants are rare, we expect that
the majority of objects in this location of the color-magnitude
diagram are actually blends of multiple fainter stars. Thus, it is no
surprise that the binary lens fit indicates blend fractions of $\sim
0.2-0.3$.  95-BLG-12 is a good example of a significantly perturbed,
non-caustic crossing binary lensing event (Fig.~\ref{fig-95blg12}).
The multiple peaks are caused by the source approaching cusps in the
caustic curve, as displayed in Fig.~\ref{fig-95blg12cc}.  The
extensive follow-up data are able to constrain the event at nearly the
$1 \%$ level for much of its duration.


\subsection{96-BLG-3 (119.19444.2055)}
\label{subsec-96blg3}

\begin{figure}
\plotone{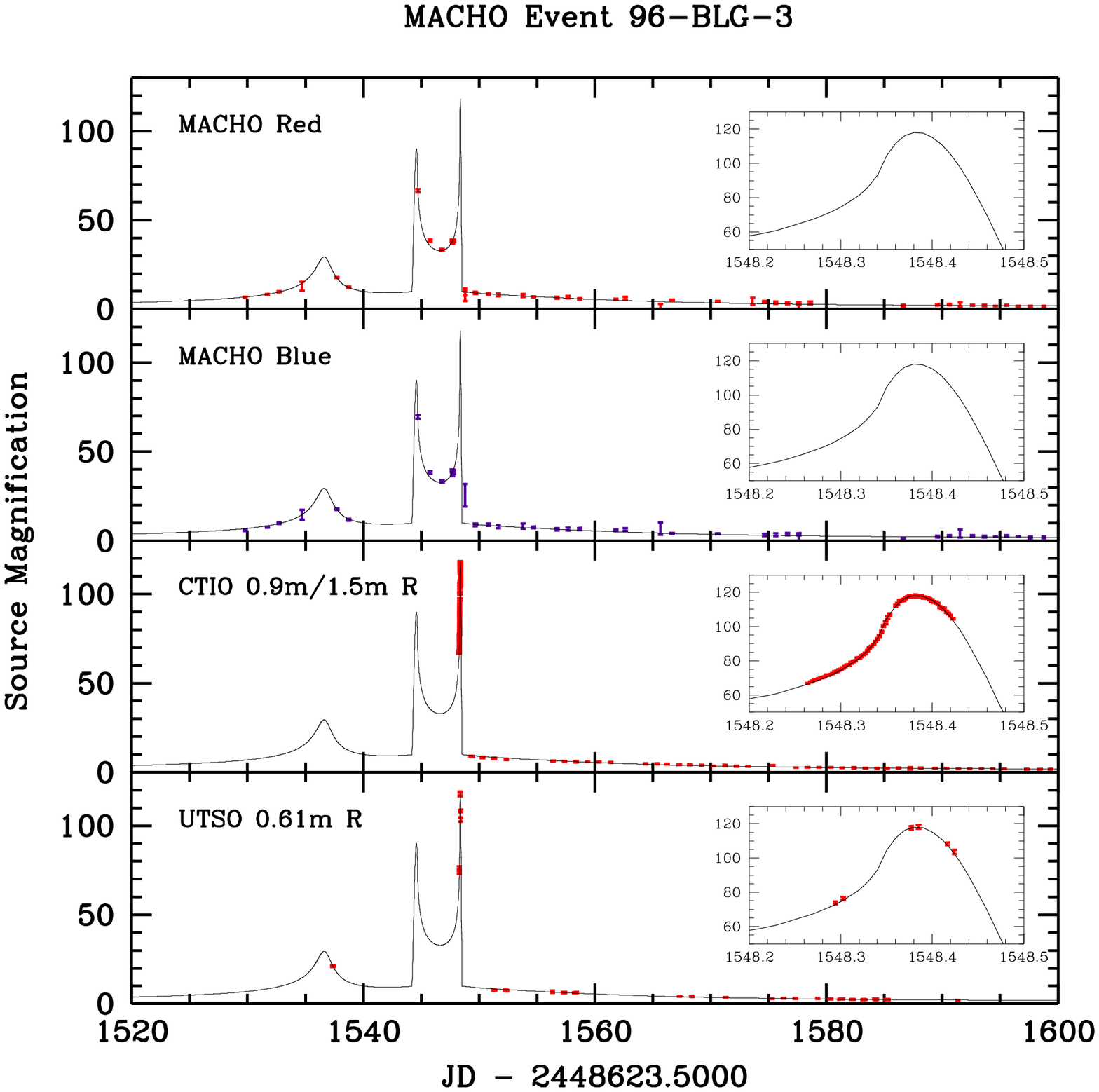}
\figcaption[f18.ps]{\label{fig-96blg3}
  Lightcurve of MACHO event 96-BLG-3, including our fit to binary microlensing.
}
\end{figure}

\begin{figure}
\plotone{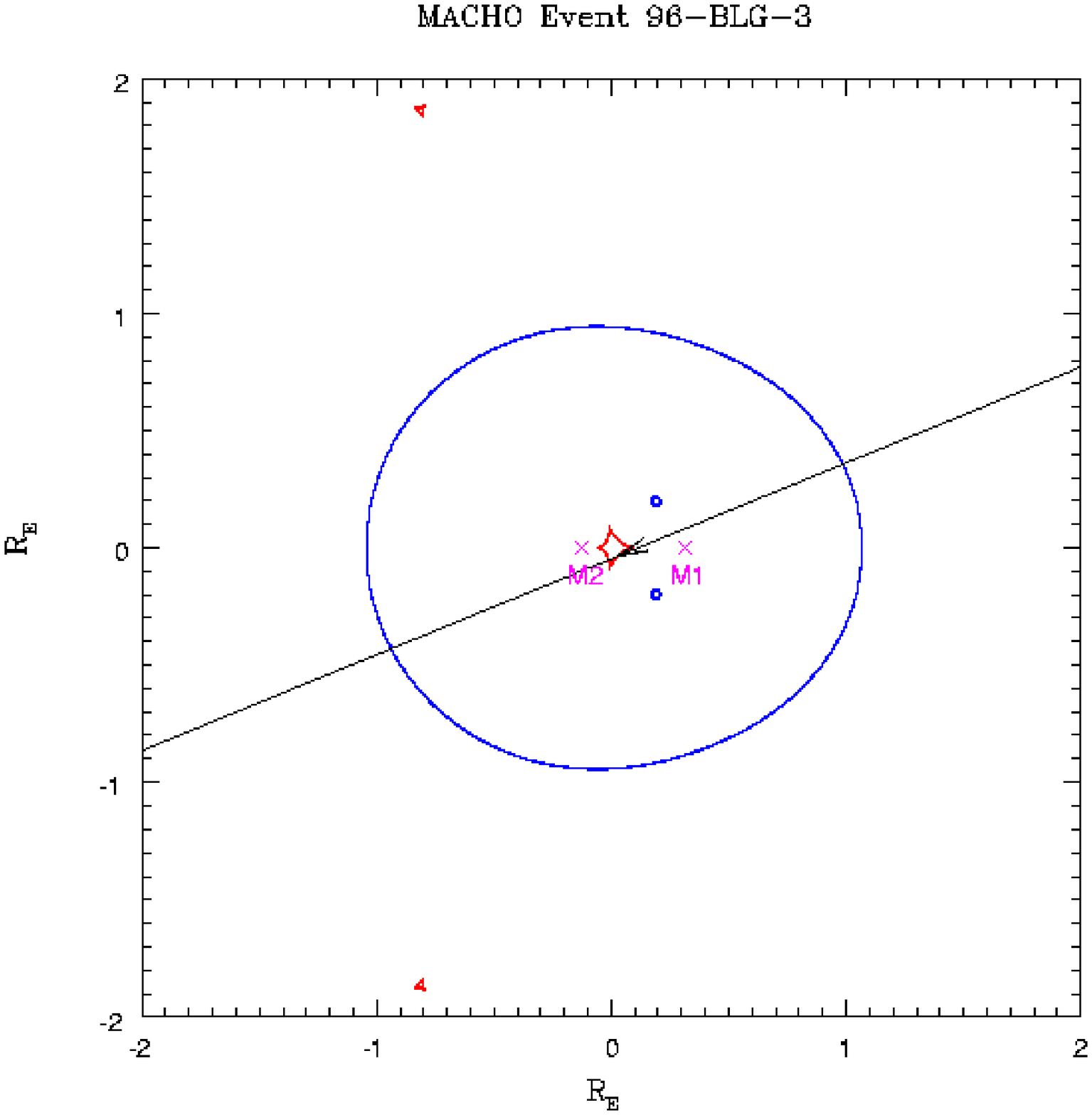}
\figcaption[f19.ps]{\label{fig-96blg3cc} Location of the
  (red) caustic and (blue) critical curves for the 96-BLG-3 binary lens fit
  presented in Fig.~\ref{fig-96blg3}.  The coordinate system, whose origin
  is at the center of mass, indicates distance in units of the system's
  Einstein Ring radius $\Re$.  Also shown are the locations of the
  lensing objects, and the trajectory of the source through the caustic
  structure.  }
\end{figure}

\begin{figure}
\plotone{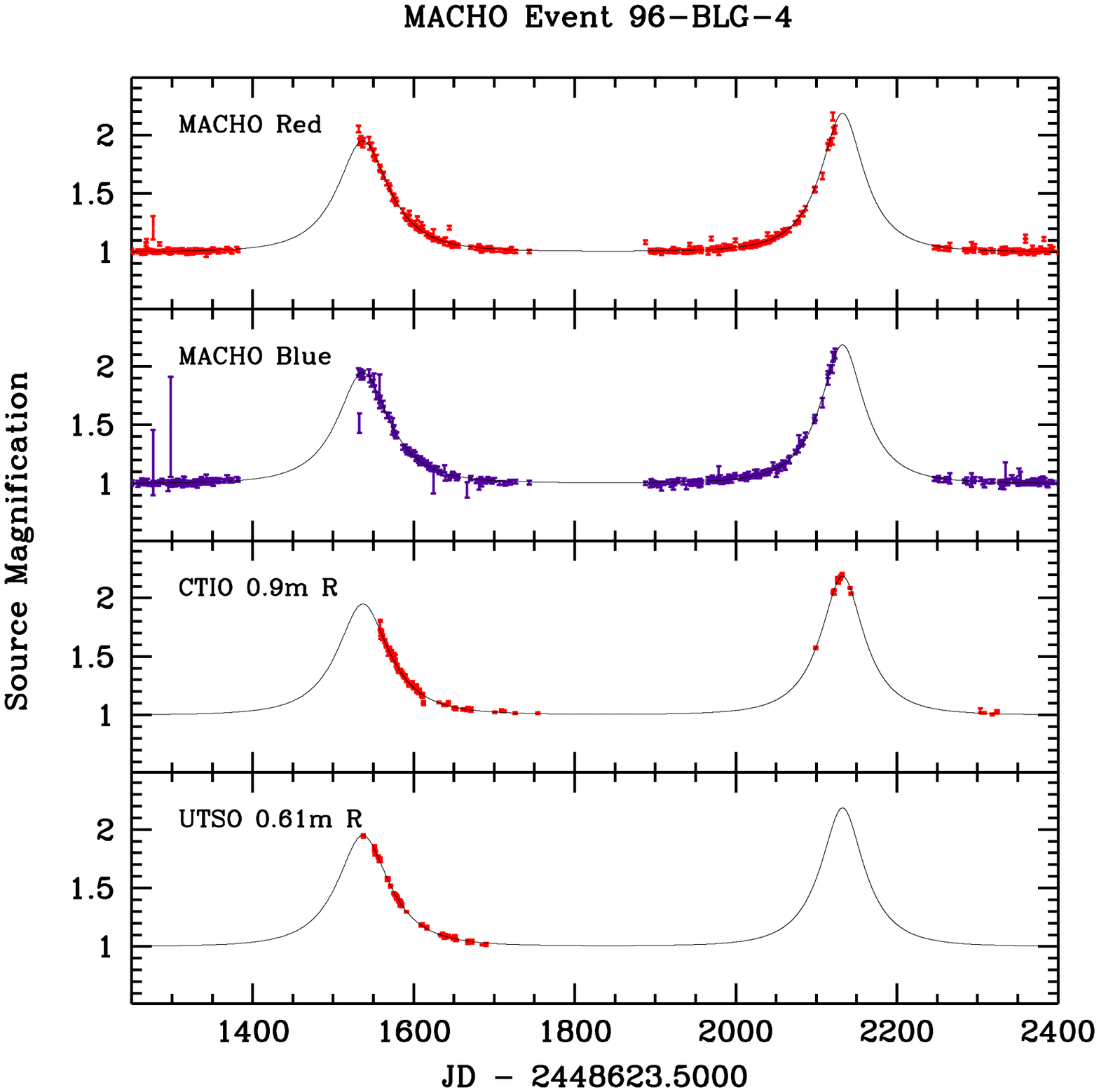}
\figcaption[f20.ps]{\label{fig-96blg4}
  Lightcurve of MACHO event 96-BLG-4, including our fit to binary microlensing.
}
\end{figure}

\begin{figure}
\plotone{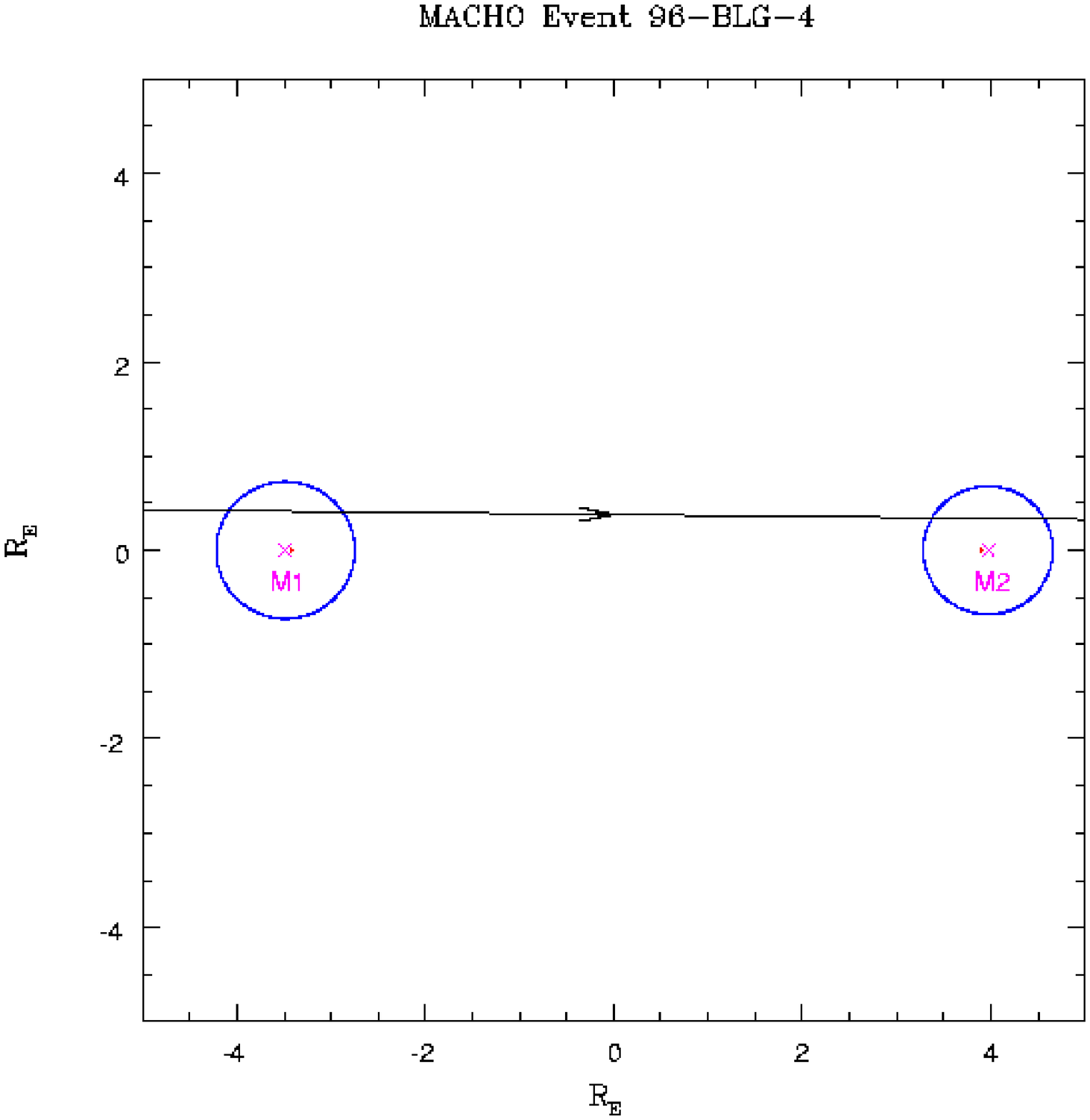}
\figcaption[f21.ps]{\label{fig-96blg4cc} Location of the
  (red) caustic and (blue) critical curves for the 96-BLG-4 binary lens fit
  presented in Fig.~\ref{fig-96blg4}.  The coordinate system, whose origin
  is at the center of mass, indicates distance in units of the system's
  Einstein Ring radius $\Re$.  Also shown are the locations of the
  lensing objects, and the trajectory of the source through the caustic
  structure.  }
\end{figure}

\begin{figure}
\plotone{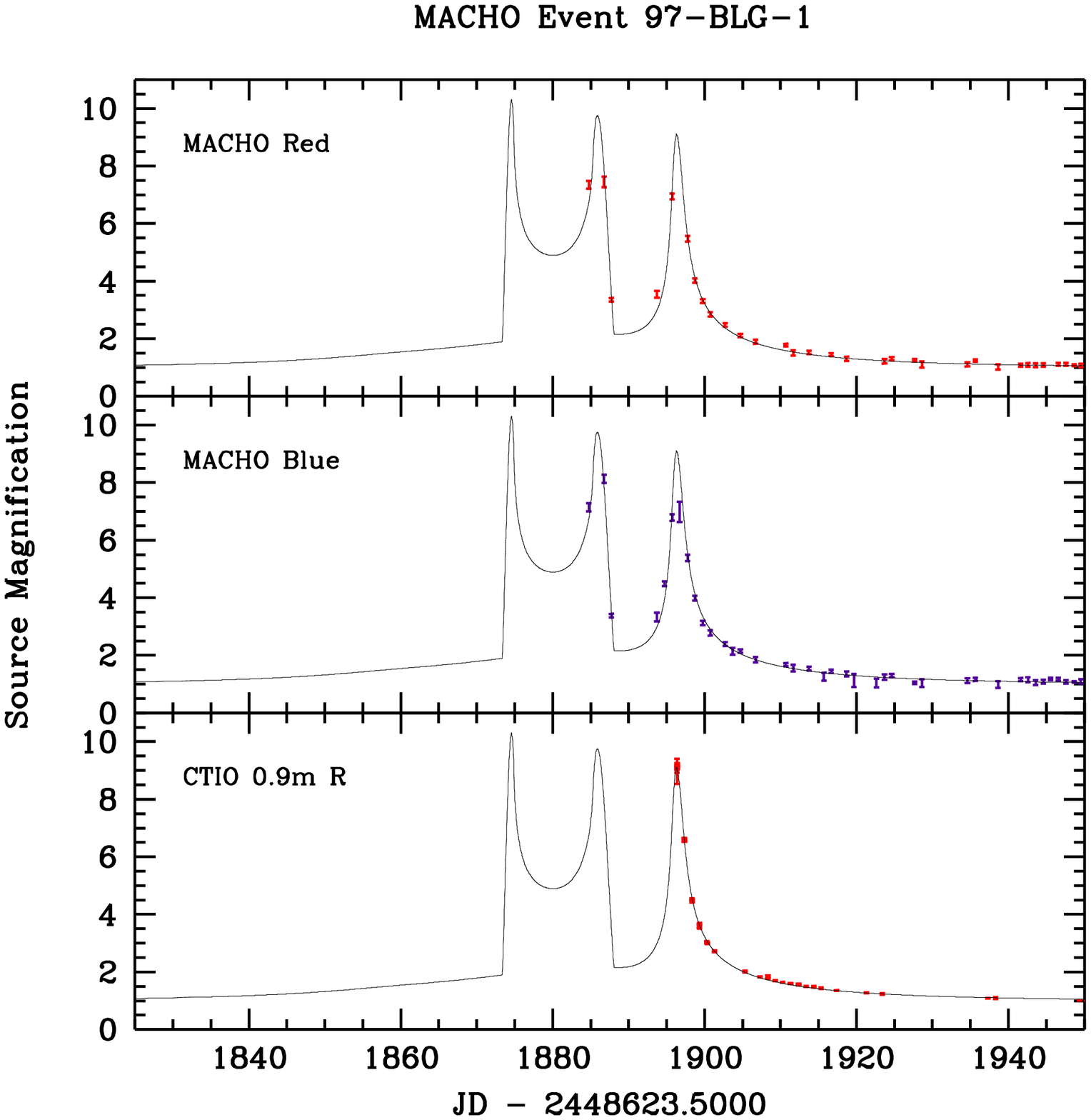}
\figcaption[f22.ps]{\label{fig-97blg1}
  Lightcurve of MACHO event 97-BLG-1, including our fit to binary microlensing.
}
\end{figure}

\begin{figure}
\plotone{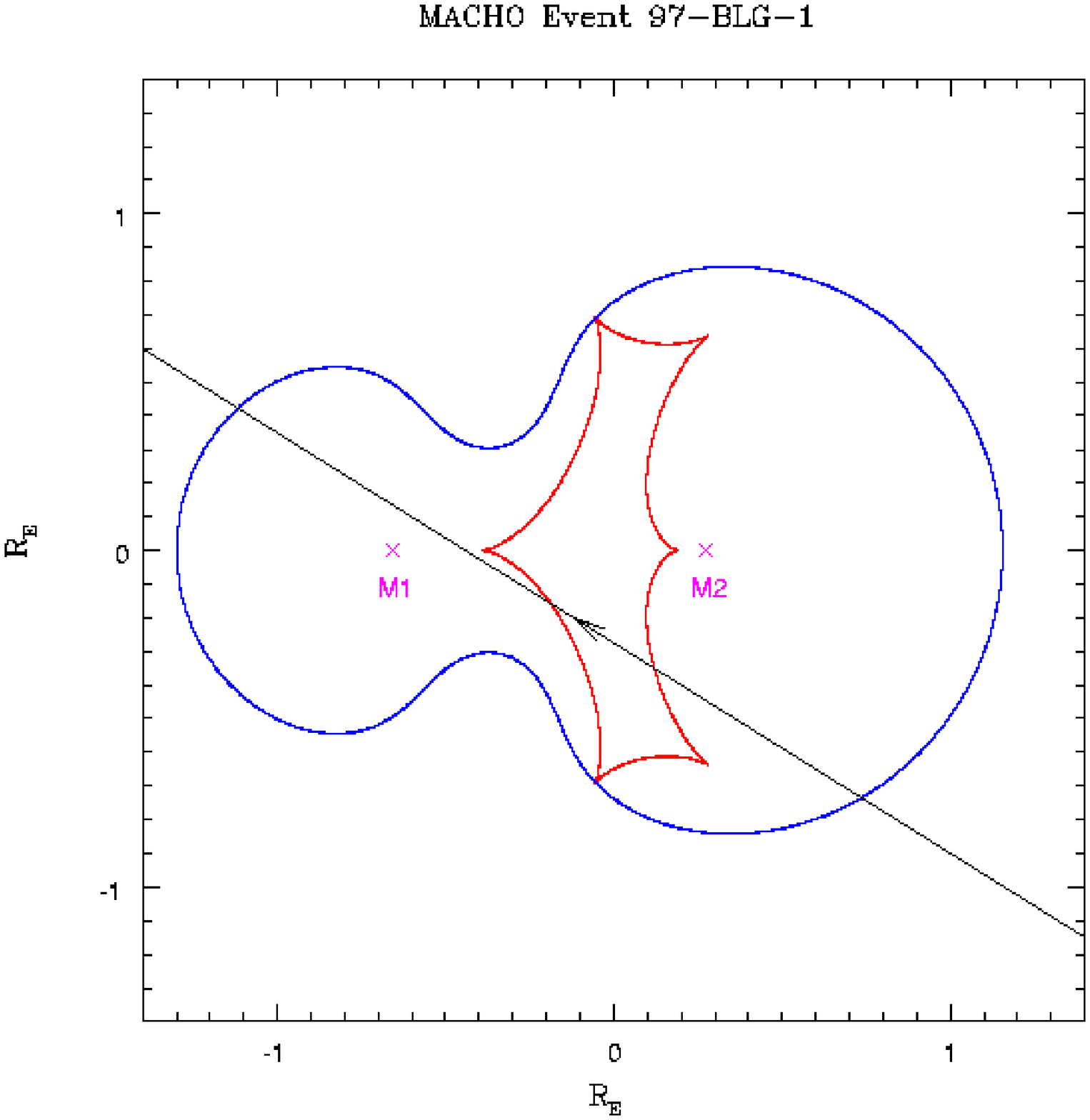}
\figcaption[f23.ps]{\label{fig-97blg1cc} Location of the
  (red) caustic and (blue) critical curves for the 97-BLG-1 binary lens fit
  presented in Fig.~\ref{fig-97blg1}.  The coordinate system, whose origin
  is at the center of mass, indicates distance in units of the system's
  Einstein Ring radius $\Re$.  Also shown are the locations of the
  lensing objects, and the trajectory of the source through the caustic
  structure.  }
\end{figure}

This event was discovered at the beginning of the 1996 bulge season,
and announced by the MACHO Alert System on Mar 12, 1996.  After an
initial peak (due to a cusp approach) which resembled a normal lensing
event, the star jumped to a large magnification on Mar 25, 1996,
implying a caustic crossing had likely occurred.  Based upon the
available MACHO data, \citeN{macho-iauc96} were able to successfully
predict the second caustic crossing to within 0.15 days.  The
important features at the time of this prediction were the initial
peak in Fig.~\ref{fig-96blg3}, due to a cusp approach, and the
sparsely sampled $\cup$-shape between caustics.  These global
constraints on the lightcurve provided enough leverage for an accurate
caustic crossing prediction, although this is not necessarily possible
with more local constraints, even a well sampled caustic crossing
\cite{planet-binsolns}.

Follow-up data taken at the CTIO 1.5m provided the first ever
resolution of a binary caustic crossing, as shown in
Fig.~\ref{fig-96blg3}.  In addition, spectroscopic observations of the
object were taken during the crossing \cite{lennon96}.  As the source
crossed out of the caustic region, its brightness peaked at extremely
high magnification, $\Amax \sim 120$.  However, the observed
magnification of the MACHO object was considerably less, as it is a
significant blend ($f \sim 0.2$ for MACHO).  Thus, while
\citeN{lennon96} report their spectrum of that of a G0 subgiant, the
blend fractions indicate that the source is a G0 dwarf.

\citeN{lennon96} showed that the source is not a spectroscopic binary,
and estimated from their spectra an effective temperature of $\teff =
6100$ K.  Their comparison to evolutionary tracks leads to an angular
source radius of $\theta_* = 0.94 \mu$as.  Since these spectra were
taken while the source was highly magnified, their $\teff$ should
represent the temperature of the lensed source.  However, their
estimate of the distance to the source, and hence its angular size,
depended upon the baseline brightness of the object ($V = 19.2$),
instead of the recovered brightness of the lensed source ($V = 20.8$).
Our analysis in Sec.~\ref{subsec-red}, Table~\ref{tab-size}, indicates
our estimate of $\teff = 6200$K is in excellent agreement with
\citeN{lennon96}, and de-blending the event reasonably leads to a
smaller $\theta_* = 0.53 \mu$as.


\subsection{96-BLG-4 (105.21417.101)}
\label{subsec-96blg4}

96-BLG-4 displays a repeating variability (Fig.~\ref{fig-96blg4}),
which would generally exclude it as a microlensing candidate.  The
first peak is well fit by standard microlensing ($\chi^2/dof = 1.2$),
and was not a binary candidate until the MACHO Alert System
re-triggered on this event $\sim 550$ days after the first peak.  The
color of this object ($V = 16.2, V-R = 1.1$) indicates it is unlikely
to be a long period variable (LPV), and any periodic variability is
ruled out by prior observations.  The object does appear to be a
bright giant, located close to the tip of the red giant branch.

As is the case with many of our giant sources, we do not detect a
significant color shift during the event, and the binary lens fit is
consistent with zero blending ($f = 1$).  The multiple achromatic
peaks suggest lensing of a single giant source by a binary lens, or
lensing of a binary system by a single lens.  Fig.~\ref{fig-cmdblg}
indicates that the source appears $\sim 1.3$ mag brighter than the red
clump in this region, so a single lens, binary source event seems
plausible (the probability of a chance superposition of giant stars is
extremely small).  Stars generally spend only a few percent of their
lives on the giant branch \cite{iben-67}, so it is unlikely for two
members of a binary system to reside on the giant branch at the same
time.  But, given the sample of $\sim 350$ candidate events that our
binary lens sample has been selected from, one binary giant source
event is plausible.

Another interpretation of this event is a very widely separated ($a =
7.5 \Re$) binary lens acting upon a single background giant source.
It has been recognized by \citeN{dismao-repeat} that lensing by widely
separated binary systems ($a > 2.5 \Re$) should occur with $\sim 10
\%$ the frequency of close binary events.  As shown in
Fig.~\ref{fig-96blg4cc}, the caustics for this event are extremely
small, which indicates that this event, unlike any of the others
presented in this paper, is very much like the superposition of two
single lens lightcurves.  This makes it difficult to unambiguously
discriminate between binary lens and binary source models for this
event.


\subsection{97-BLG-1 (113.18674.756)}
\label{subsec-97blg1}

97-BLG-1 was initially announced as microlensing on Mar 3, 1997.  A
substantial deviation from standard microlensing was noticed Mar 11,
1997.  This sharp decline, seen in Fig.~\ref{fig-97blg1}, signified
the source exiting the caustic region, and follow-up efforts were only
able to sample the final cusp approach (see Fig.~\ref{fig-97blg1cc}).

This MACHO object appears to be a clump giant, relatively unreddened
compared with the other events in our sample.  There is little
blending in this event, such that the giant is likely the lensed
source.  The second caustic crossing in this model is resolved with 2
MACHO observations, which leads to an estimate of $\tstar = 0.53 \pm
0.03$ days.  However, the lack of information prior to the start of
the bulge season severely limits our ability to parameterize this
event, so we expect that our fit may not be unique.  The only strong
constraints on this lensing encounter are the magnitude of the final
cusp approach and the likelihood of minimal blending for a clump giant
source.


\subsection{97-BLG-24 (101.20650.1216)}
\label{subsec-97blg24}

\begin{figure}
\plotone{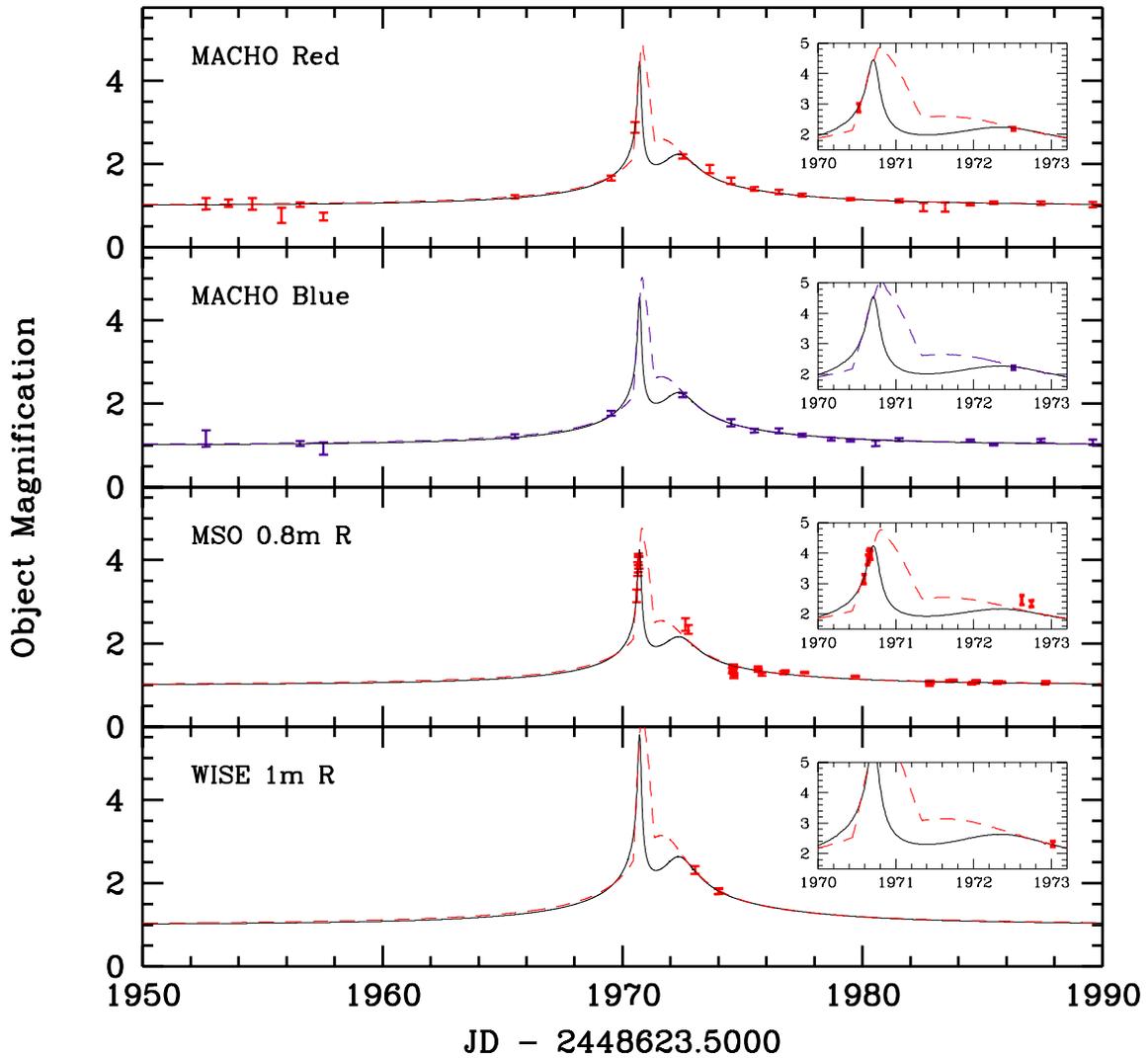}
\figcaption[f24.ps]{\label{fig-97blg24}
  Lightcurve of MACHO event 97-BLG-24, including our fits to binary microlensing.
  Due to the different blending parameters between fits, we plot the observed
  magnification of the MACHO object, as opposed to the actual lensed source.
}
\end{figure}

\begin{figure}
\plotone{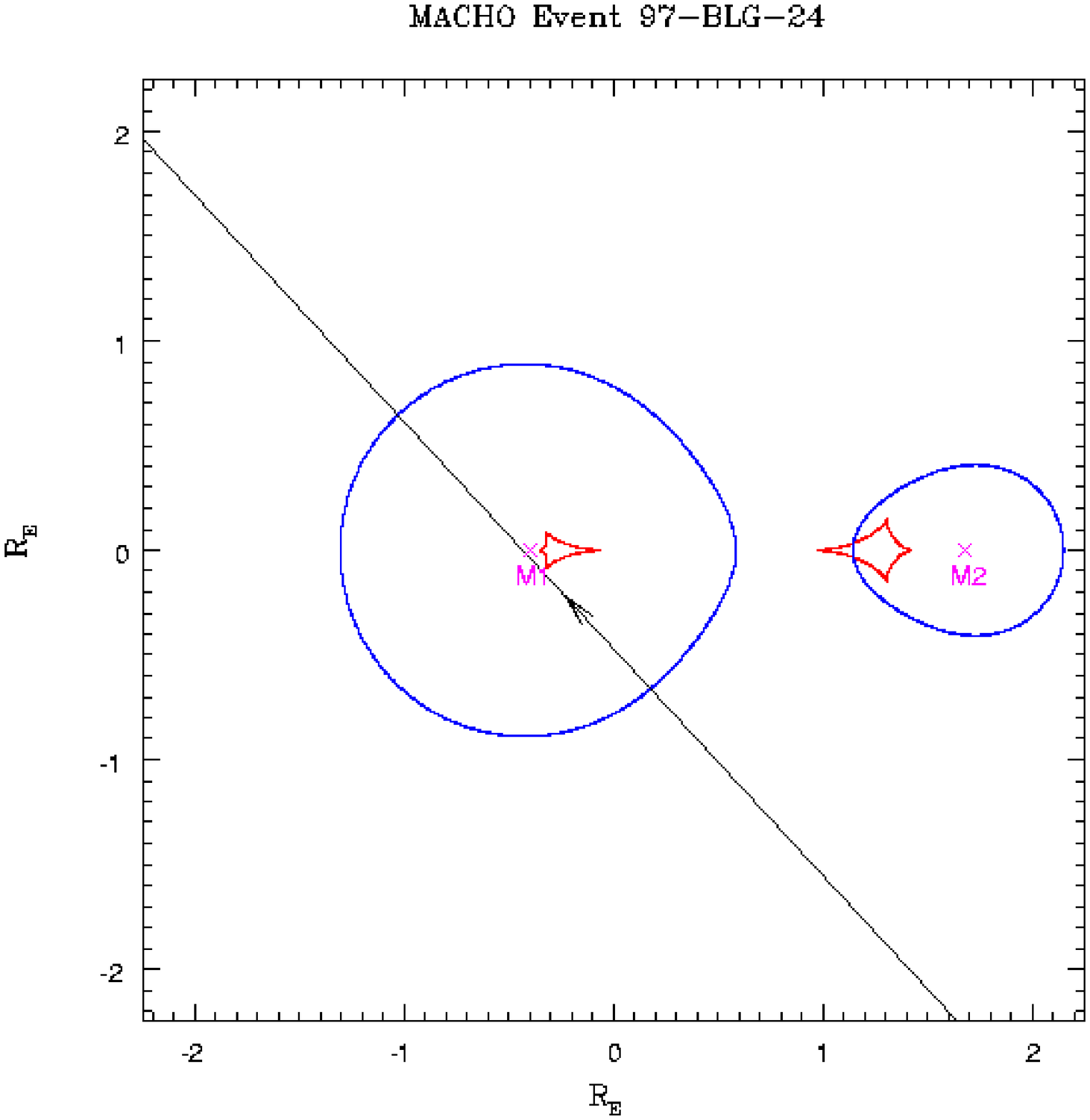}
\figcaption[f25.ps]{\label{fig-97blg24ccn} Location of the
  (red) caustic and (blue) critical curves for the 97-BLG-24 standard binary
  lens fit (-----) presented in Fig.~\ref{fig-97blg24}.  The coordinate
  system, whose origin is at the center of mass, indicates distance in
  units of the system's Einstein ring radius $\Re$.  Also shown are the
  locations of the lensing objects, and the trajectory of the source
  through the caustic structure.  }
\end{figure}

\begin{figure}
\plotone{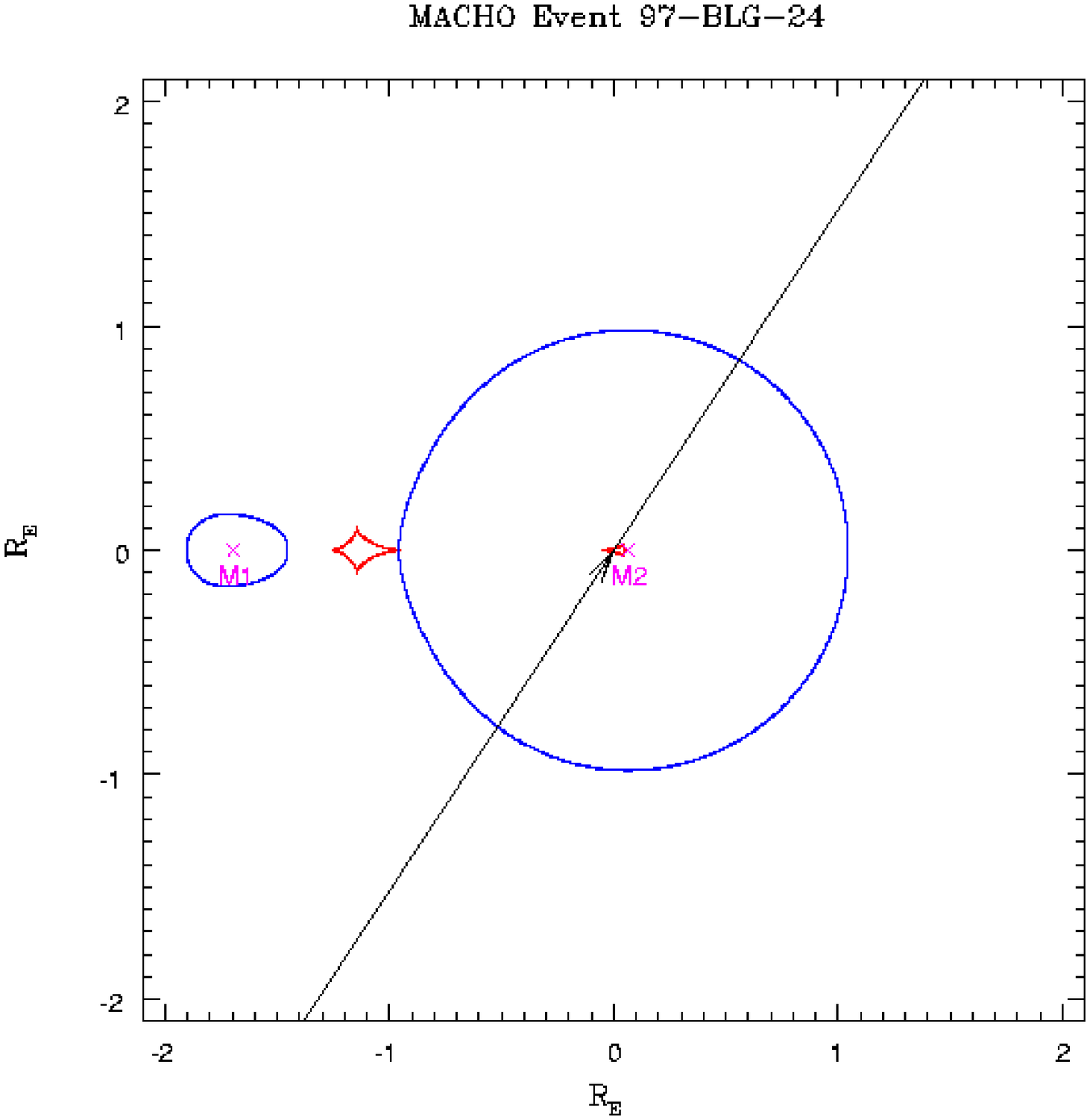}
\figcaption[f26.ps]{\label{fig-97blg24ccp} Location of the
  (red) caustic and (blue) critical curves for the 97-BLG-24 'planetary'
  binary lens fit (-~-~-) presented in Fig.~\ref{fig-97blg24}.  The
  coordinate system, whose origin is at the center of mass, indicates
  distance in units of the system's Einstein ring radius $\Re$.  Also
  shown are the locations of the lensing objects, and the trajectory of
  the source through the caustic structure.  }
\end{figure}

\begin{figure}
\plotone{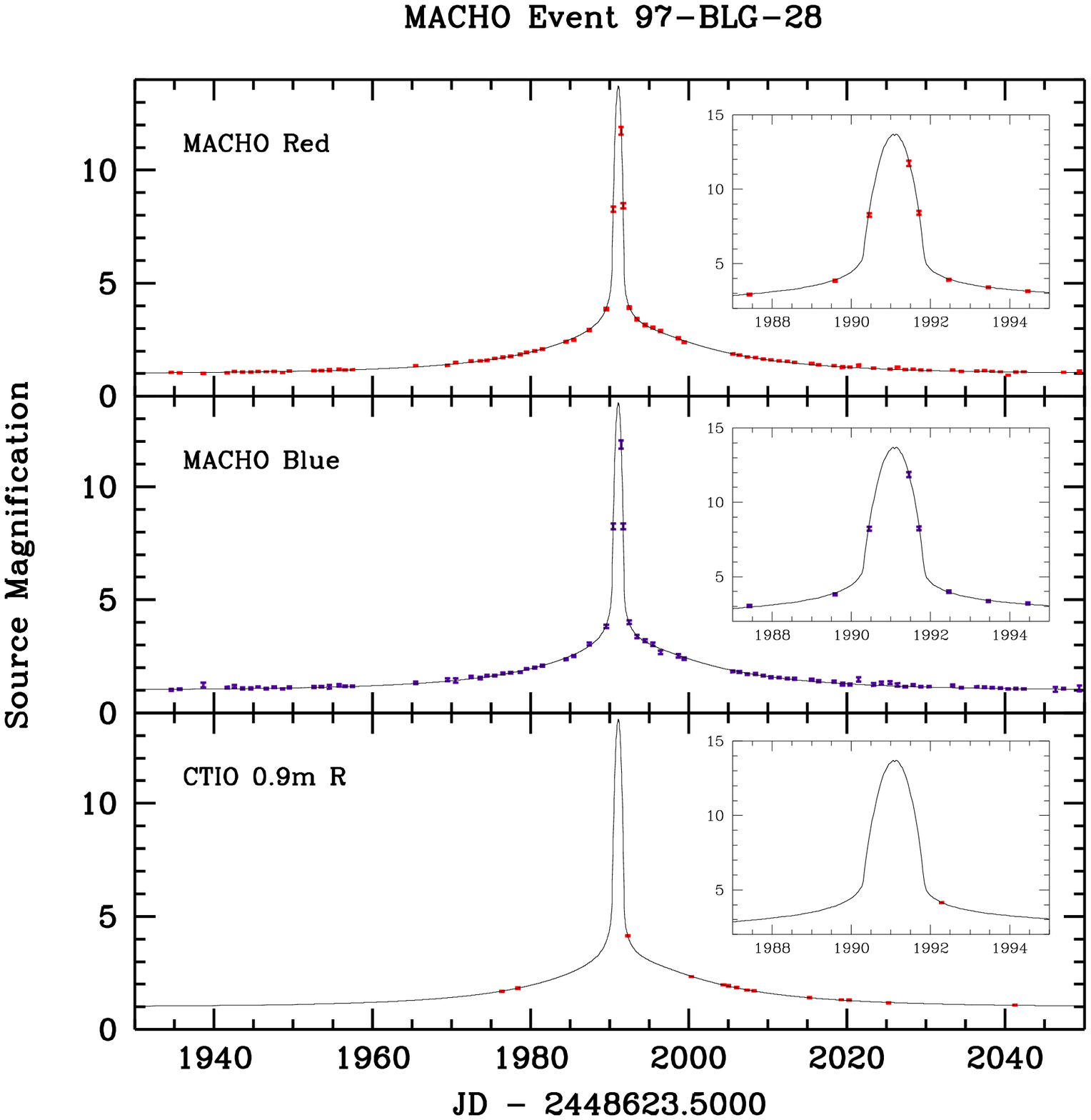}
\figcaption[f27.ps]{\label{fig-97blg28}
  Lightcurve of MACHO event 97-BLG-28, including our fit to binary microlensing.
}
\end{figure}

\begin{figure}
\plotone{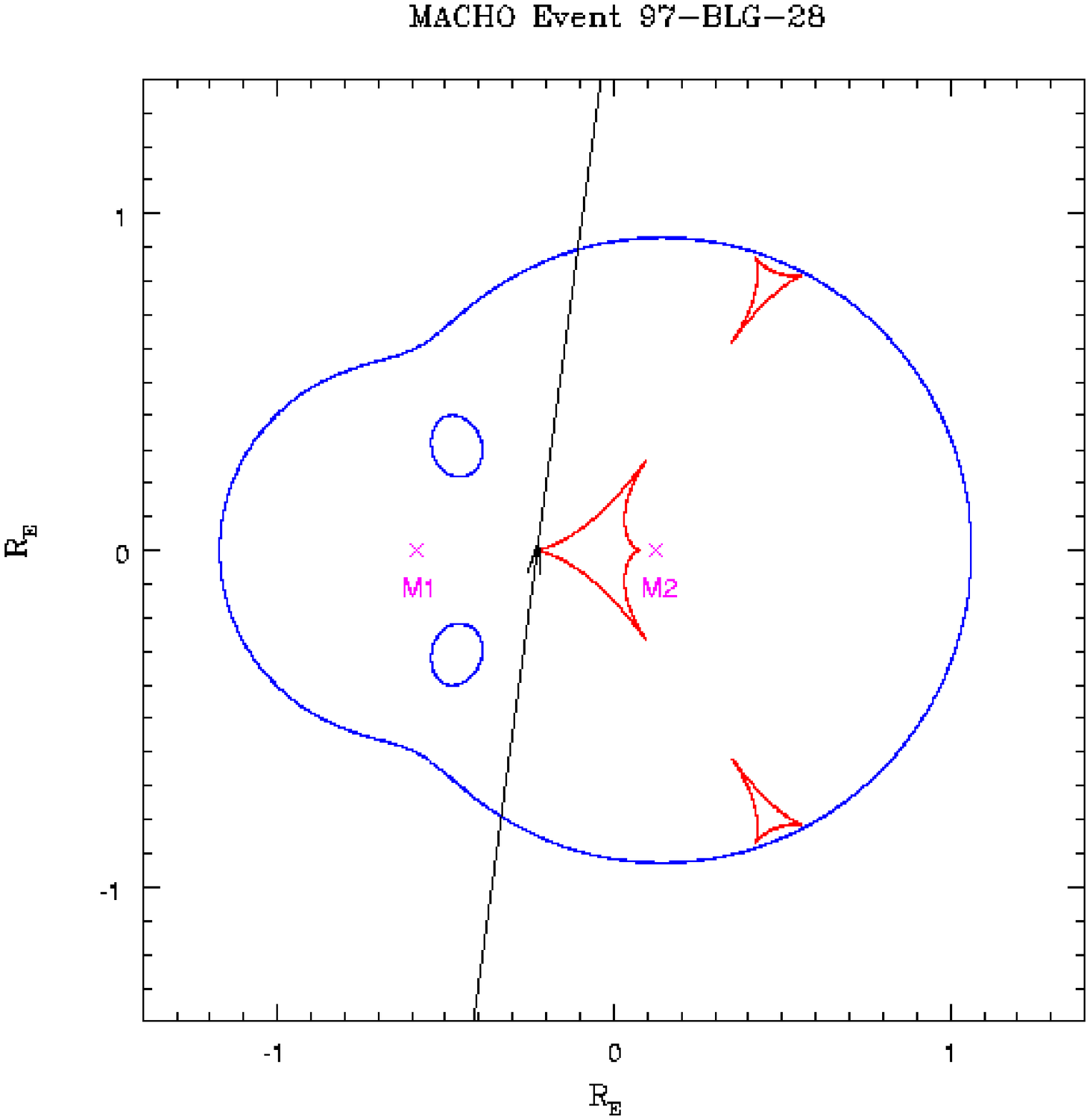}
\figcaption[f28.ps]{\label{fig-97blg28cc} Location of the
  (red) caustic and (blue) critical curves for the 97-BLG-28 binary lens fit
  presented in Fig.~\ref{fig-97blg28}.  The coordinate system, whose origin
  is at the center of mass, indicates distance in units of the system's
  Einstein Ring radius $\Re$.  Also shown are the locations of the
  lensing objects, and the trajectory of the source through the caustic
  structure.  }
\end{figure}

\begin{figure}
\plotone{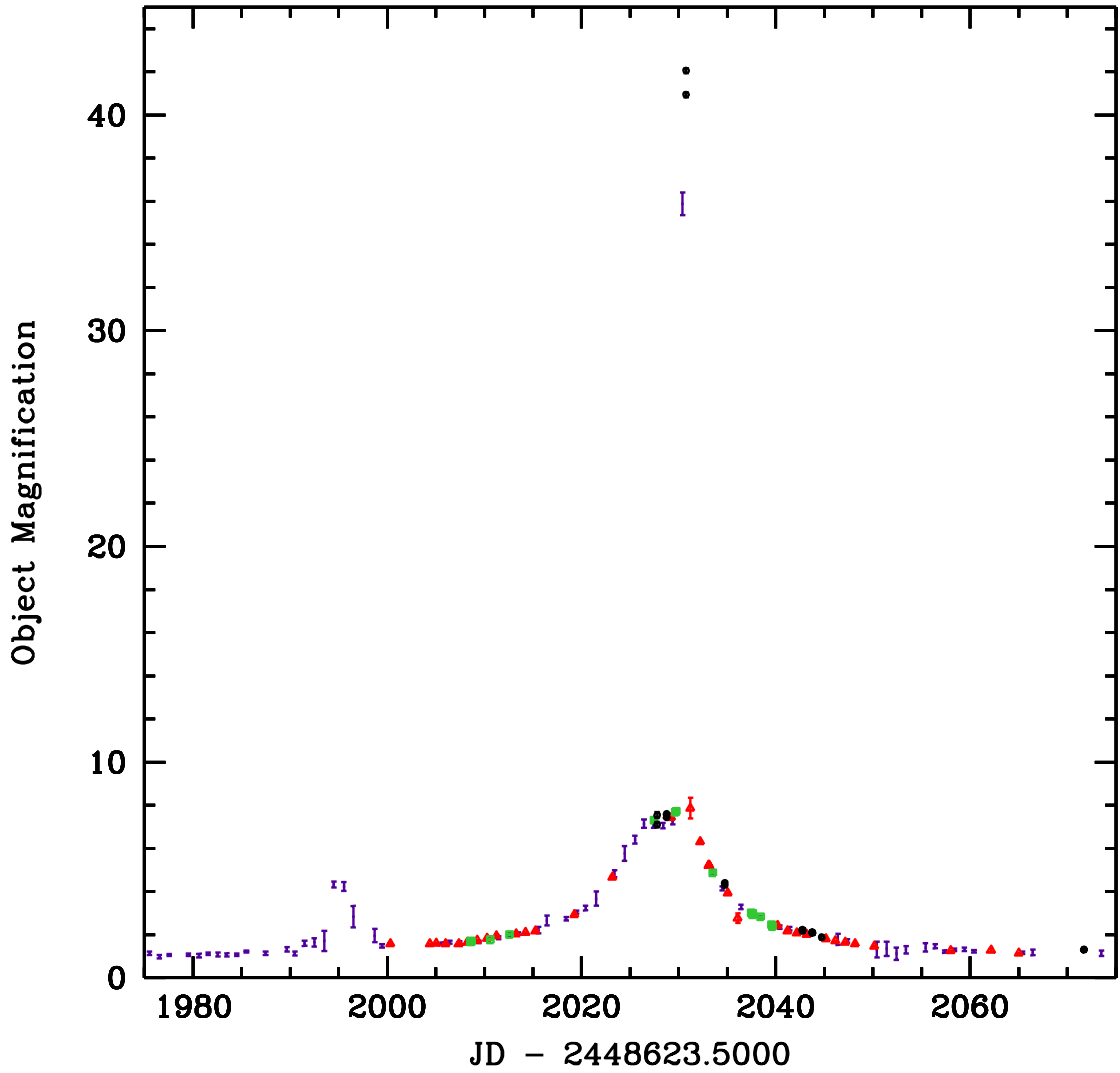}
\figcaption[f29.ps]{\label{fig-97blg41}
  Lightcurve of MACHO event 97-BLG-41, with an approximate baseline
  determined from a fit to the second peak, disregarding the caustic
  features.  Plotted are the MACHO-B (blue points), CTIO-r (red
  triangles), MSO30-r (green squares) and WISE-r (black circles) data.
}
\end{figure}

\begin{figure}
\plotone{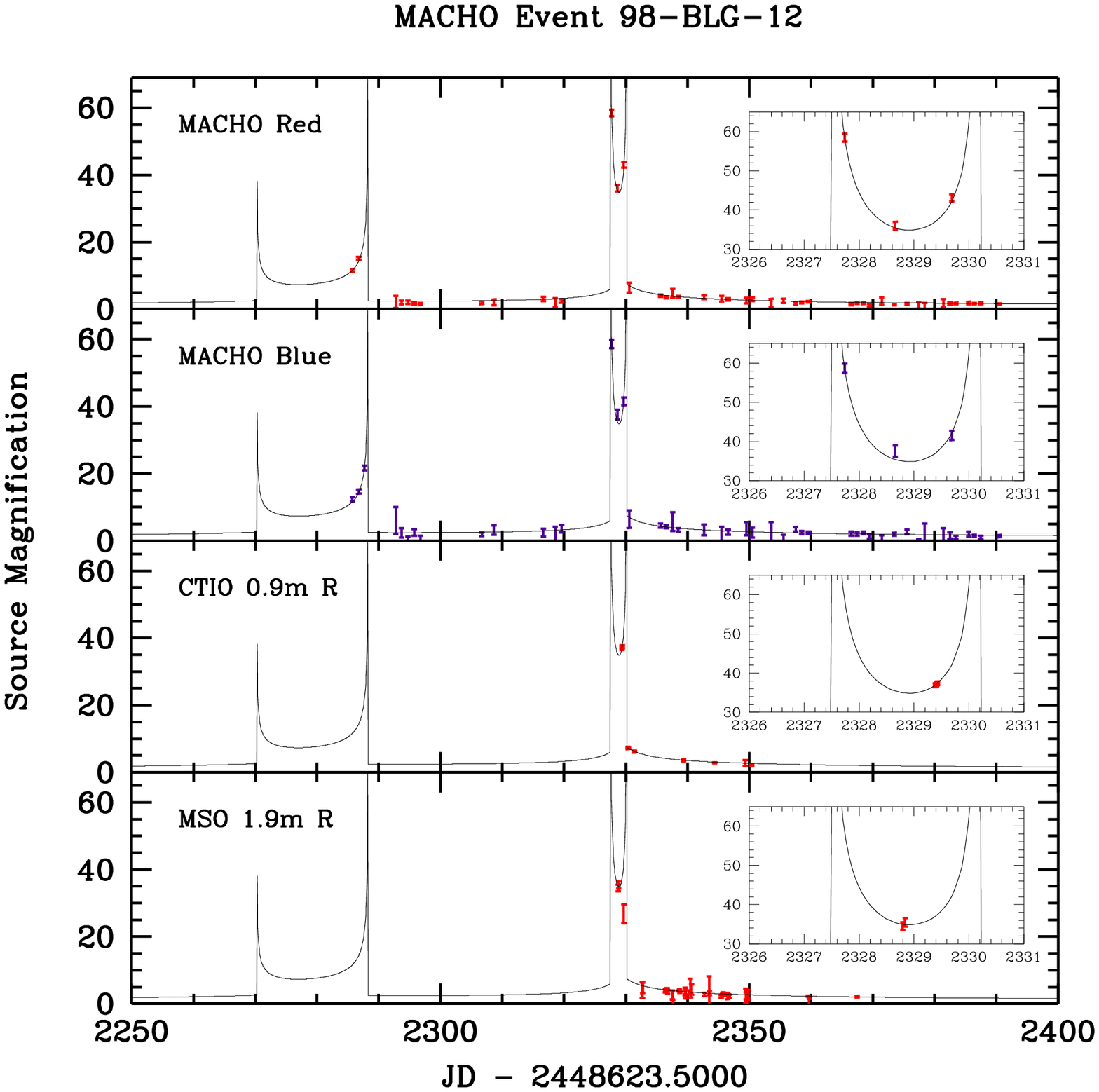}
\figcaption[f30.ps]{\label{fig-98blg12}
  Lightcurve of MACHO event 98-BLG-12, including our fit to binary microlensing.
}
\end{figure}

\begin{figure}
\plotone{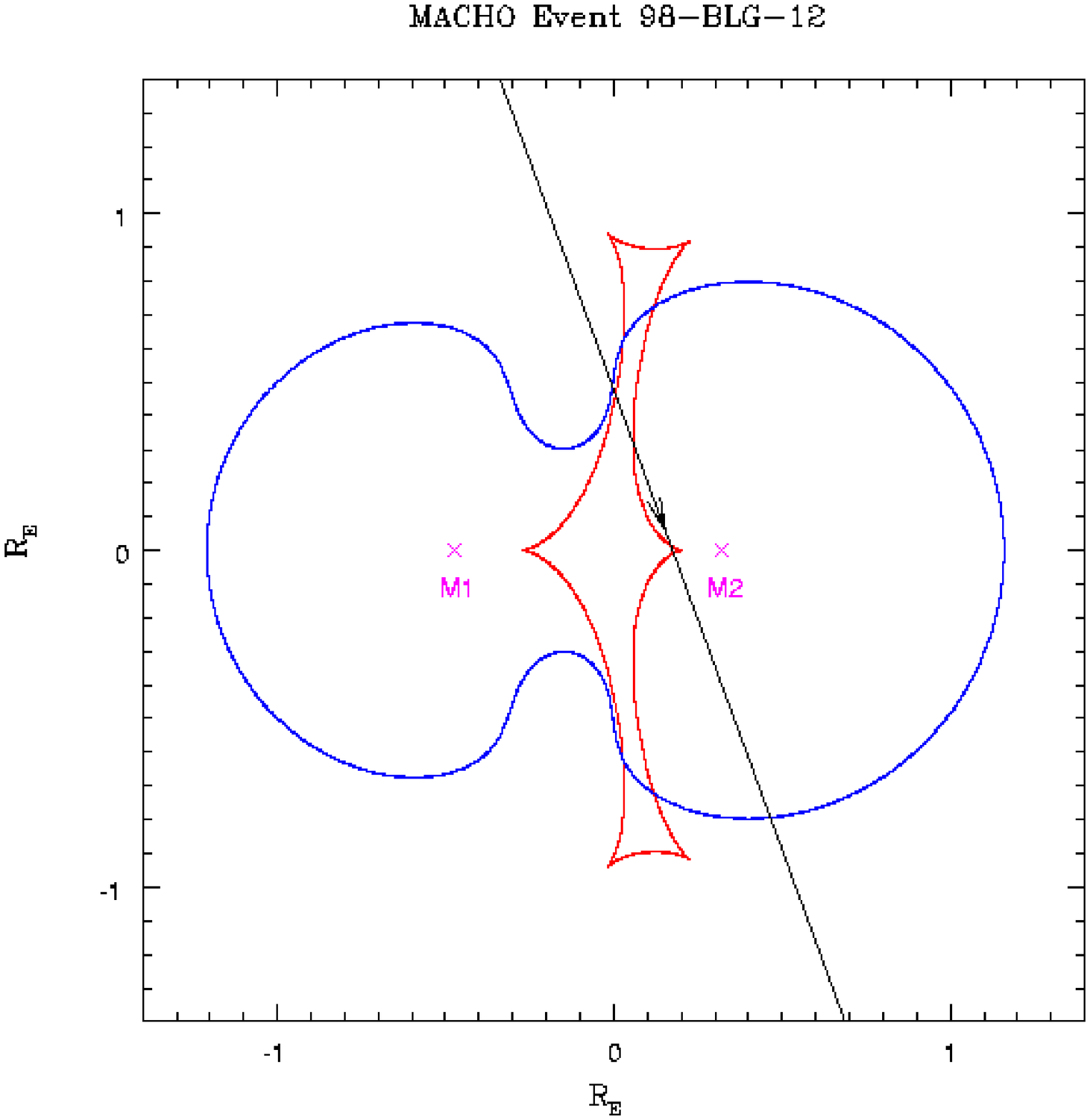}
\figcaption[f31.ps]{\label{fig-98blg12cc} Location of the
  (red) caustic and (blue) critical curves for the 98-BLG-12 binary lens fit
  presented in Fig.~\ref{fig-98blg12}.  The coordinate system, whose origin
  is at the center of mass, indicates distance in units of the system's
  Einstein Ring radius $\Re$.  Also shown are the locations of the
  lensing objects, and the trajectory of the source through the caustic
  structure.  }
\end{figure}

The MACHO lightcurve for 97-BLG-24 (Fig.~\ref{fig-97blg24}) exhibits a
significant deviation from point source microlensing, similar to LMC-1
(Fig.~\ref{fig-lmc1}).  However, this deviation was noticed in
real-time, allowing immediate follow-up observations to be undertaken
with the MSO 30'' telescope.  These data reveal an unusual increase in
the object's brightness before observations ended for the evening.

Unfortunately, the lensed source appears $\sim 1.2''$ from the MACHO
object, in a region of high crowding.  It is difficult to
independently photometer the lensed source even in the follow-up
photometry, and the data presented here represent the change in
brightness of the brightest neighbor to the lensed source -- i.e. the
event is strongly blended.  Nevertheless, we are able to resolve
evidence of lens binarity in the deviation near the peak of the
lightcurve.  This deviation is more heavily sampled than LMC-1, but
still suffers from the same model degeneracy.  (We note that
Fig.~\ref{fig-97blg24} displays the apparent brightening of the MACHO
object, not the lensed source, due to the different blend fractions
between the two fits presented in Tables~\ref{tab-params}
and~\ref{tab-bparams}.)

The solid and dashed line fits in Fig.~\ref{fig-97blg24} display
$\that$ of 30.7 and 45.5 days, implying overall lens masses of $0.19$
and $0.42 \msun$, respectively.  The former fit includes a binary
system with components of $0.16$ and $0.03 \msun$, likely a stellar -
brown dwarf system.  The latter dashed fit, with a mass ratio of
$29:1$, implies lens masses of $0.41$ and $0.014 \msun$, consistent
with a stellar lens with a companion of $\sim 14$ Jupiter masses.


\subsection{97-BLG-28 (108.18951.593)}
\label{subsec-97blg28}

After being detected and alerted upon May 29, 1997, this event began
to increase in brightness at an unexpected rate on Jun 14, 1997 (day
1990 in Fig.~\ref{fig-97blg28}), and both MACHO/GMAN and PLANET issued
secondary alerts for a binary lensing event in progress.  The PLANET
collaboration was able to obtain nearly constant coverage of this
event, resulting in parameterization of the limb-darkening
coefficients for the source, and an estimate of the lens proper motion
of $\mu = 19.4 \pm 2.6 \kmskpc$ \cite{planet-97blg28}.

The results presented here are similar to those of
\citeN{planet-97blg28}.  The fit to microlensing suggests a moderate
amount of un-lensed blue light in the photometered object, and the
object is likely a lensed clump giant source blended with objects of
bluer color.  The trajectory of the source plotted in
Fig.~\ref{fig-97blg28cc} indicates the lightcurve deviation was due to
a cusp crossing.  Resolution of the source face during this crossing
allows a measurement of $\tstar = 0.760 \pm 0.014$ days.  A comparison
of event parameters (MACHO:PLANET) with model LD1 of
\citeN{planet-97blg28} yields $\that = (52.8:54.4), \umin =
(0.225:0.215), a = (0.71:0.69), \epsilon_1 = (0.17:0.19), \theta =
(1.44:1.42), \tstar = (0.76:0.78)$.

Following the procedure outlined in Section~\ref{subsec-red} and
Table~\ref{tab-size}, we estimate the reddening to be $E(V-R) = 0.67
\pm 0.04$, which yields de-reddened source magnitudes of $V=15.52 \pm
0.19$ and $R=14.94 \pm 0.15$.  From this we determine $\teff = 4500
\pm 200$K for the source star.  Using a bolometric correction of $BC_V
= -0.48$, we find an angular source radius of $\theta_* = 6.58 \pm
0.90 \mu$as.  For comparison, the PLANET group find a de-reddened
$V=15.27$, $\teff = 4350$K, and $\theta_* = 8.74\pm 1.17\mu$as, so our
results are in reasonably good agreement.


\subsection{97-BLG-41 (402.47862.1576)}
\label{subsec-97blg41}

This event was detected and alerted upon on 18 Jun, 1997.  This event
exhibited what appeared to be a fairly normal rise and fall for a
microlensing event.  However, after peak the event did not decline to
baseline magnification, but leveled off at A $\sim 1.5$, and began a
slow rise, which itself was fit well by a longer duration microlensing
event.  The deviation from a normal single lens lightcurve was noted
and announced by both the MACHO/GMAN and PLANET collaborations.  Near
the peak of the event is an apparent caustic or cusp crossing.  The
MACHO and GMAN data have been plotted in Fig.~\ref{fig-97blg41}.

Considerable effort has been made to fit this lightcurve to a binary
lens model, but no satisfactory model has been found, even when the
possible orbital motion of the lens was included.  However, a
satisfactory multiple lens fit has been found by \citeN{mps-97blg41}.


\subsection{98-BLG-12 (179.21577.1740)}
\label{subsec-98blg12}

98-BLG-12 was detected on Apr 8, 1998, and initially thought to be a
rapidly rising, high magnification event.  This behavior, evident in
Fig.~\ref{fig-98blg12}, was a result of the source exiting its first
passage through the caustic structure (Fig.~\ref{fig-98blg12cc}).  It
was not recognized as a binary lensing event until the source
re-entered the caustic structure near May 17.5 UT, 1998 (day 2327.5),
and was subsequently observed at high magnification by the survey
telescope on May 17.74 UT, 1998 (day 2327.74).  The $< 3$ days spent
between caustics allowed little time for follow-up observations to
constrain event parameters.  Interestingly, in all passbands, this
event appears heavily blended ($f \sim 0.2$).


\subsection{98-BLG-14 (401.48408.649)}
\label{subsec-98blg14}

\begin{figure}
\plotone{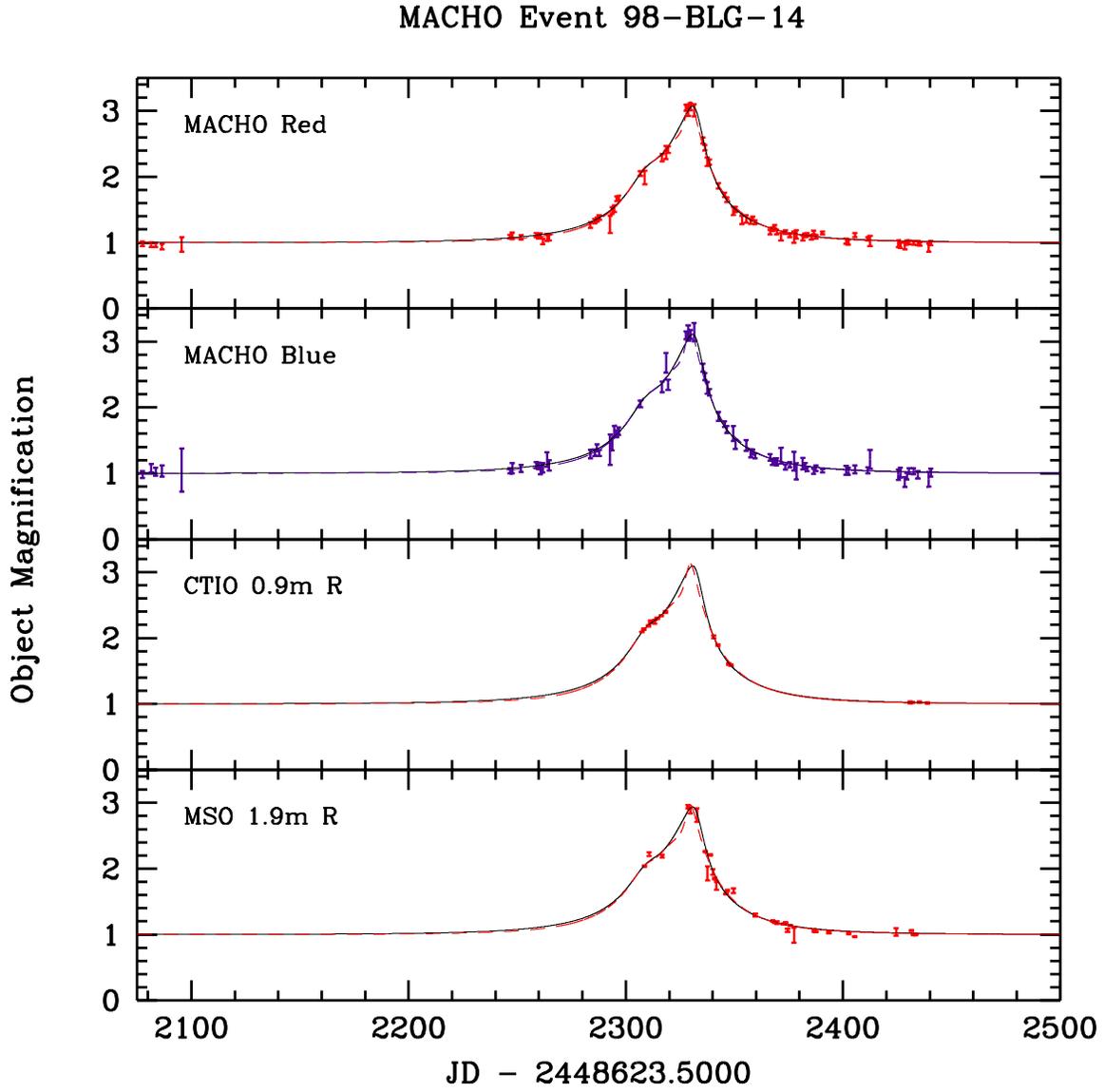}
\figcaption[f32.ps]{\label{fig-98blg14}
  Lightcurve of MACHO event 98-BLG-14, including our fits to binary
  microlensing.  Due to the different blending parameters between fits,
  we plot the observed magnification of the MACHO object, as opposed to
  the actual lensed source.  The MSO74 data have been averaged into 1 day
  bins.
}
\end{figure}

\begin{figure}
\plotone{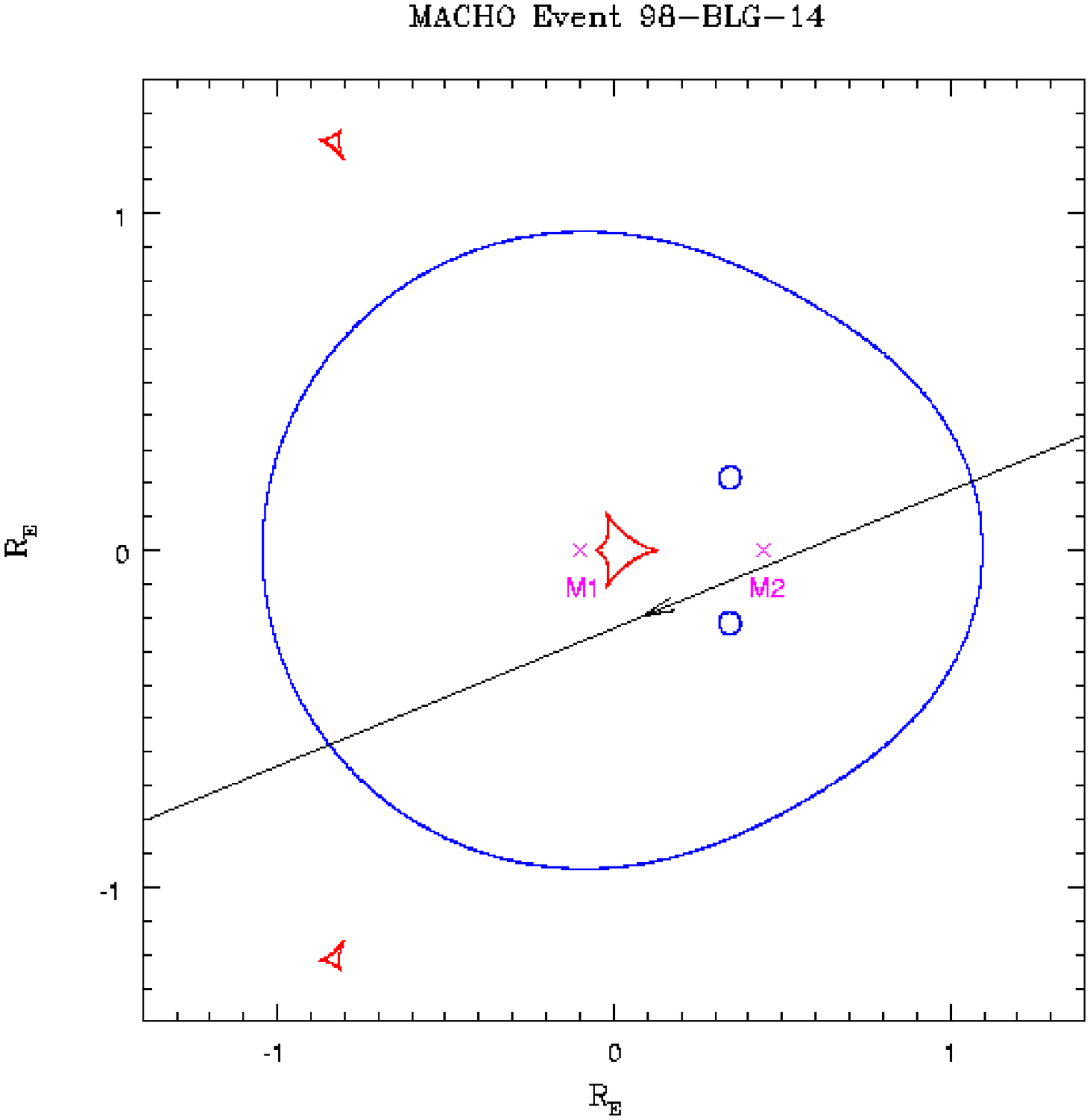}
\figcaption[f33.ps]{\label{fig-98blg14ccn} Location of
  the (red) caustic and (blue) critical curves for the solid 98-BLG-14
  binary lens fit (-----) presented in Fig.~\ref{fig-98blg14}.  The
  coordinate system, whose origin is at the center of mass, indicates
  distance in units of the system's Einstein Ring radius $\Re$.  Also
  shown are the locations of the lensing objects, and the trajectory of
  the source through the caustic structure.  }
\end{figure}

\begin{figure}
\plotone{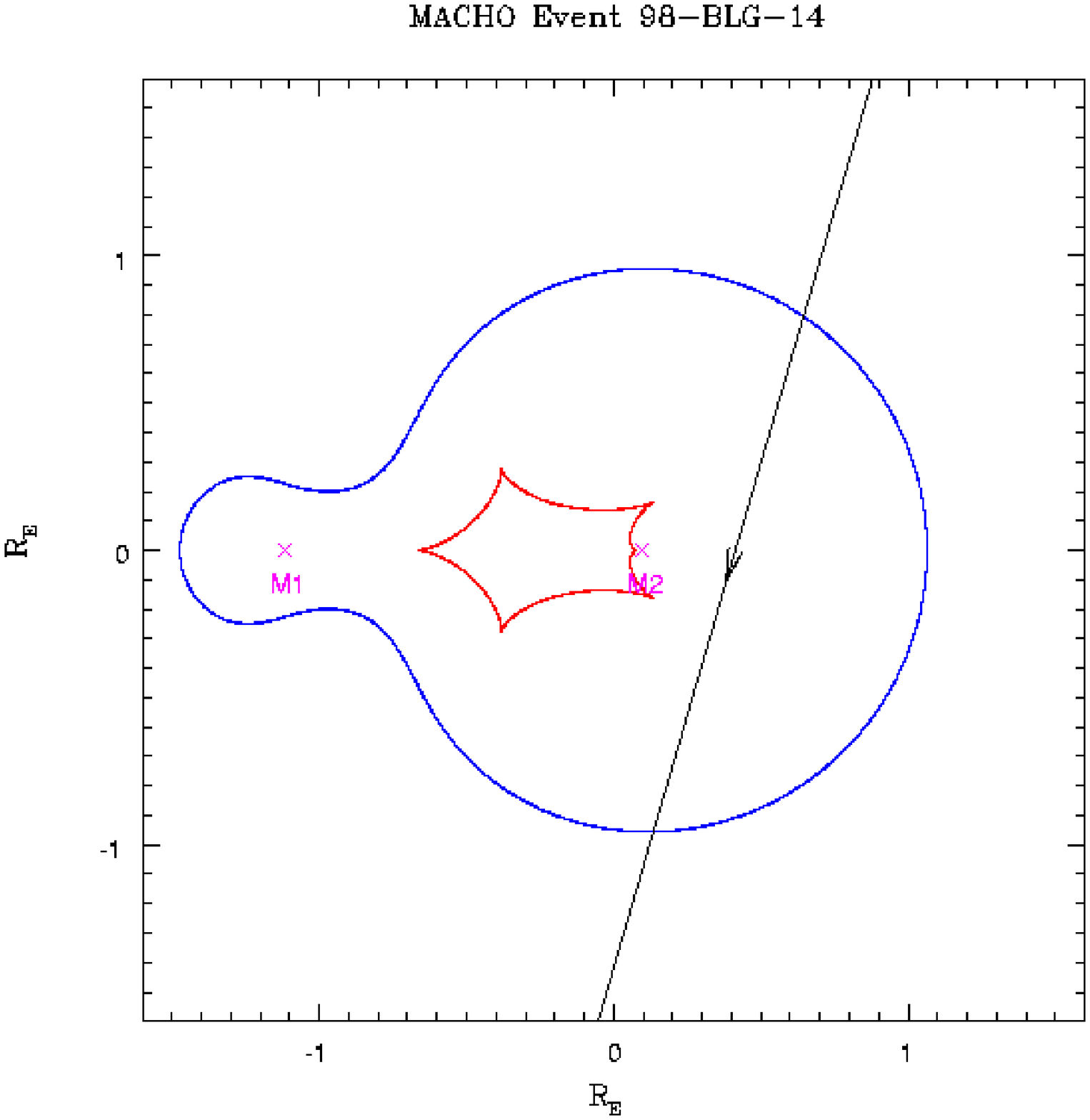}
\figcaption[f34.ps]{\label{fig-98blg14ccp} Location of
  the (red) caustic and (blue) critical curves for the dashed (large
  mass ratio) 98-BLG-14 binary lens fit (-~-~-) presented in
  Fig.~\ref{fig-98blg14}.  The coordinate system, whose origin is at the
  center of mass, indicates distance in units of the system's Einstein
  Ring radius $\Re$.  Also shown are the locations of the lensing
  objects, and the trajectory of the source through the caustic
  structure.  }
\end{figure}

\begin{figure}
\plotone{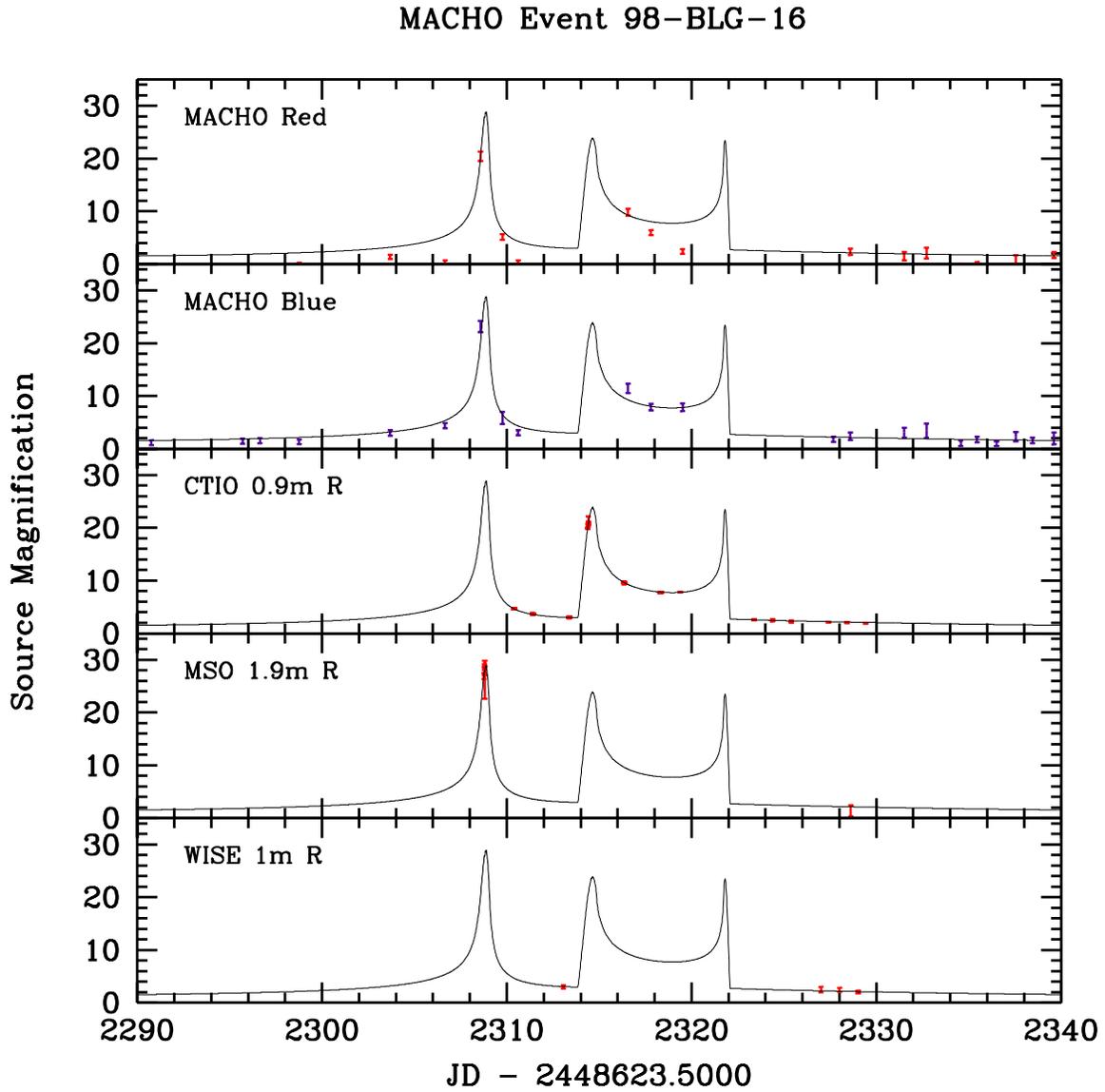}
\figcaption[f35.ps]{\label{fig-98blg16}
  Lightcurve of MACHO event 98-BLG-16, including our fit to binary
  microlensing.  The MACHO Red data exhibit excessive scatter due to a
  defective amplifier.
}
\end{figure}

\begin{figure}
\plotone{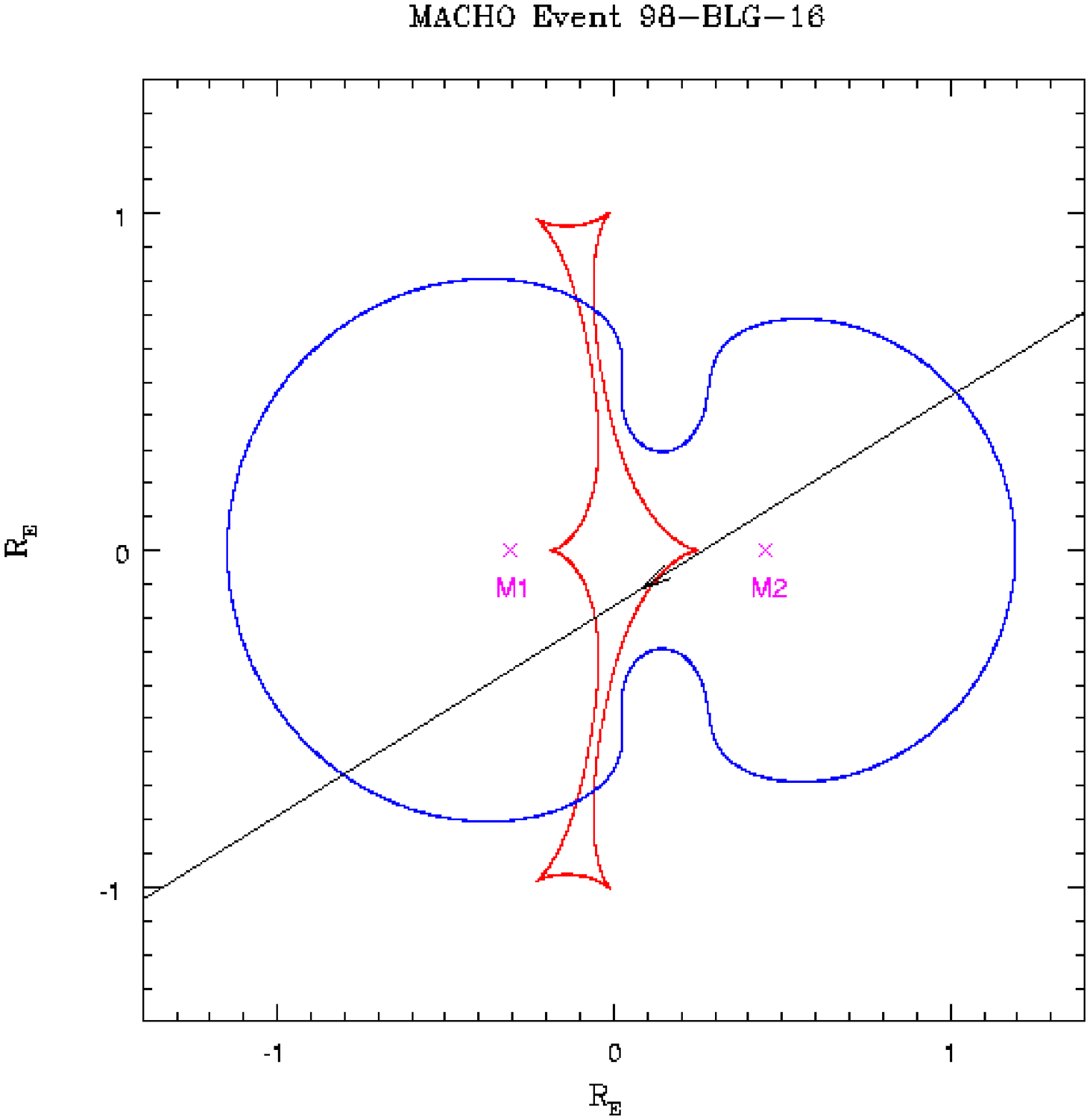}
\figcaption[f36.ps]{\label{fig-98blg16cc} Location of the
  (red) caustic and (blue) critical curves for the 98-BLG-16 binary lens fit
  presented in Fig.~\ref{fig-98blg16}.  The coordinate system, whose origin
  is at the center of mass, indicates distance in units of the system's
  Einstein Ring radius $\Re$.  Also shown are the locations of the
  lensing objects, and the trajectory of the source through the caustic
  structure.  }
\end{figure}

\begin{figure}
\plotone{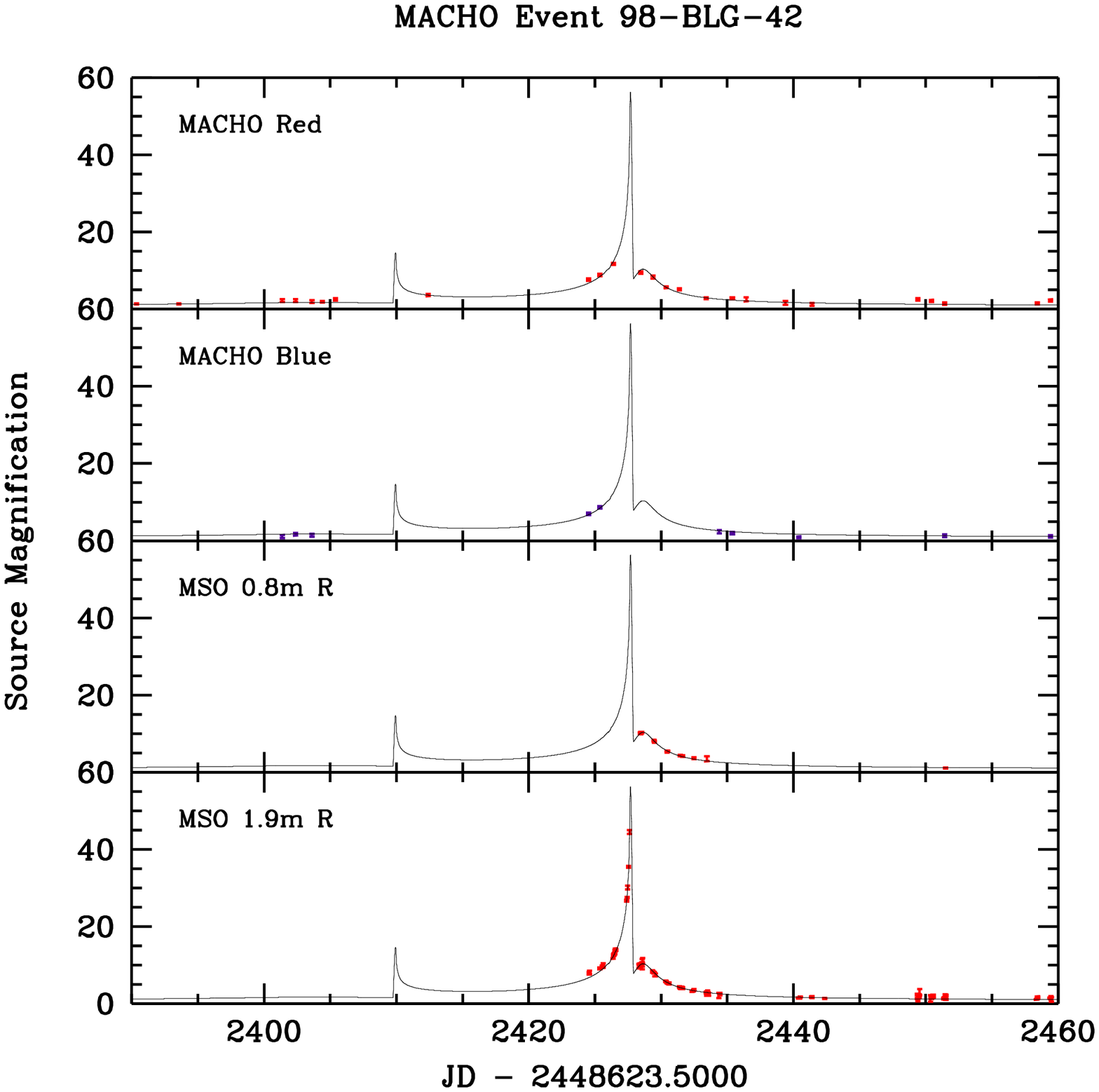}
\figcaption[f37.ps]{\label{fig-98blg42}
  Lightcurve of MACHO event 98-BLG-42, including our fit to binary microlensing.
}
\end{figure}

\begin{figure}
\plotone{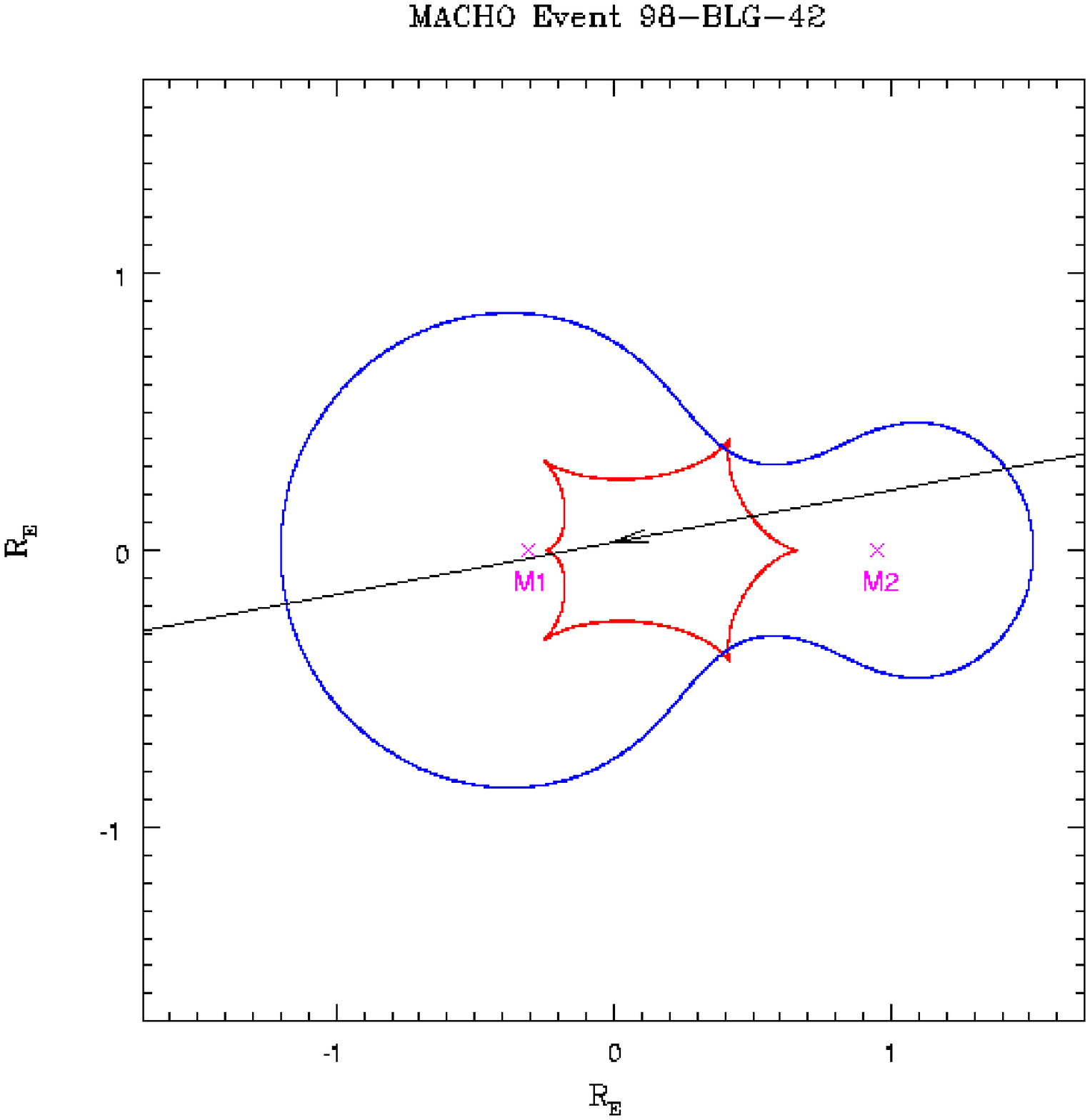}
\figcaption[f38.ps]{\label{fig-98blg42cc} Location of the
  (red) caustic and (blue) critical curves for the 98-BLG-42 binary lens fit
  presented in Fig.~\ref{fig-98blg42}.  The coordinate system, whose origin
  is at the center of mass, indicates distance in units of the system's
  Einstein Ring radius $\Re$.  Also shown are the locations of the
  lensing objects, and the trajectory of the source through the caustic
  structure.  }
\end{figure}

\begin{figure}
\plotone{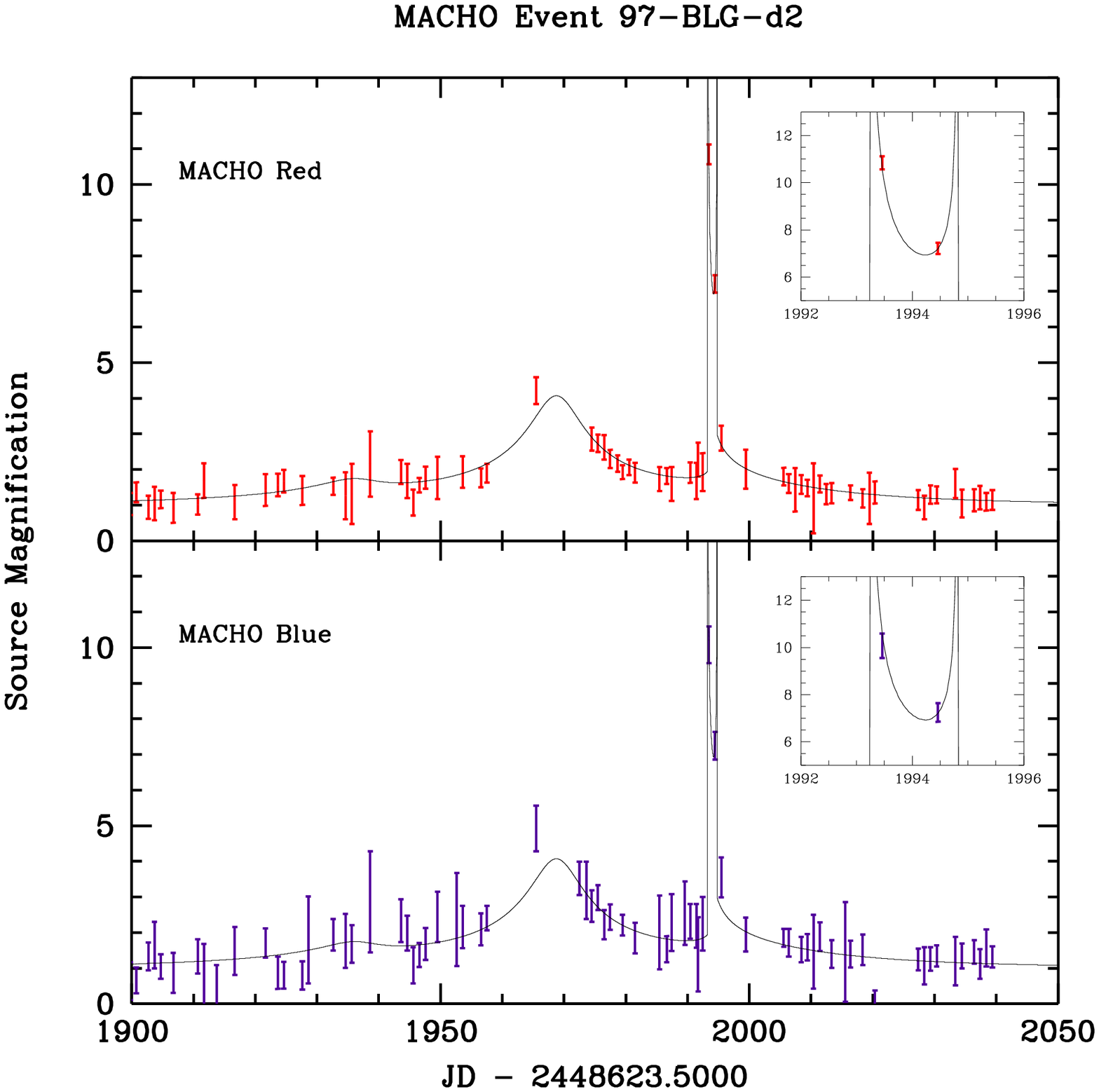}
\figcaption[f39.ps]{\label{fig-97blgd2}
  Lightcurve of MACHO event 97-BLG-d2, including our fit to binary microlensing.
}
\end{figure}

\begin{figure}
\plotone{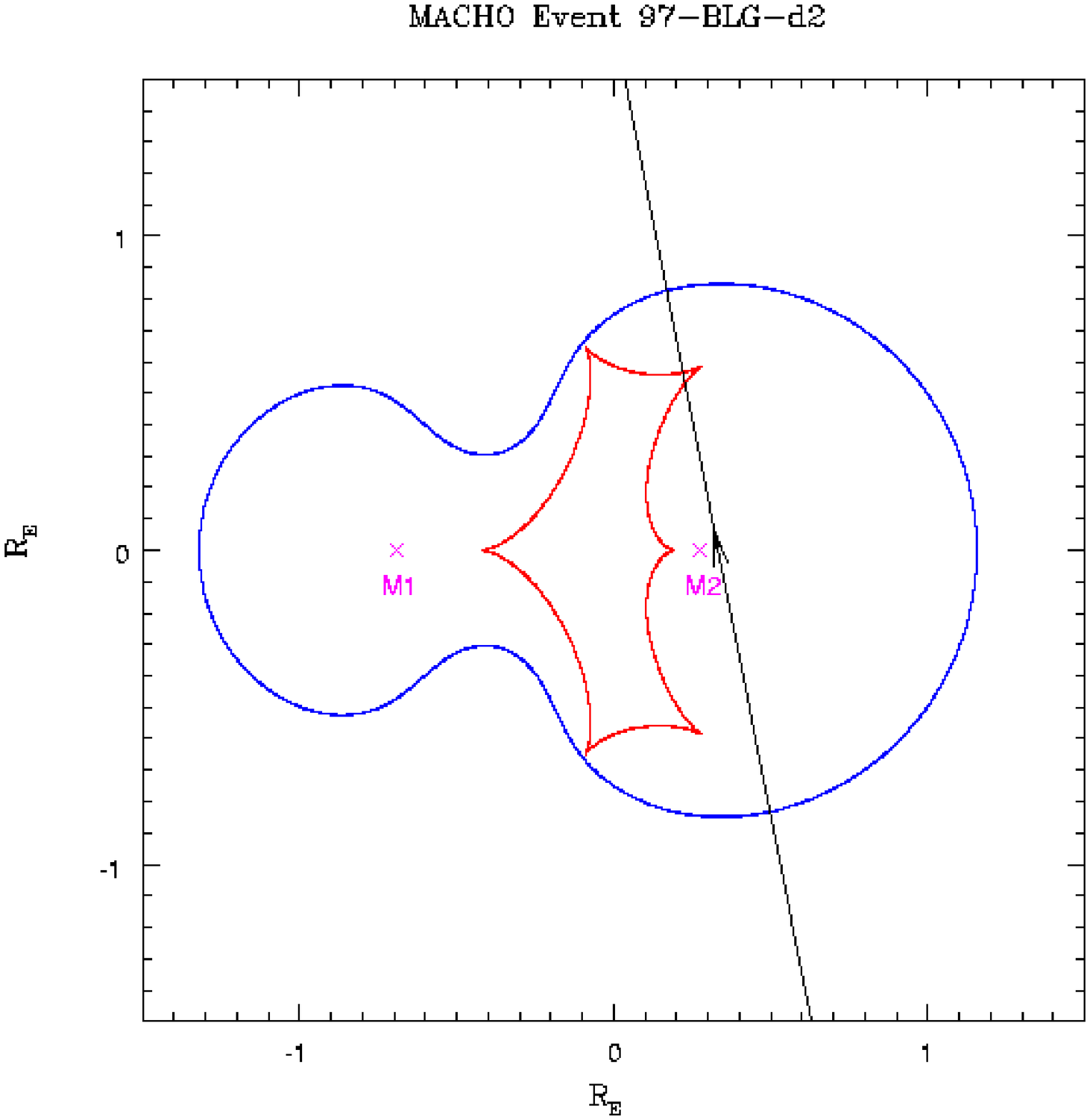}
\figcaption[f40.ps]{\label{fig-97blgd2cc} Location of the
  (red) caustic and (blue) critical curves for the 97-BLG-d2 binary lens fit
  presented in Fig.~\ref{fig-97blgd2}.  The coordinate system, whose origin
  is at the center of mass, indicates distance in units of the system's
  Einstein Ring radius $\Re$.  Also shown are the locations of the
  lensing objects, and the trajectory of the source through the caustic
  structure.  }
\end{figure}

\begin{figure}
\plotone{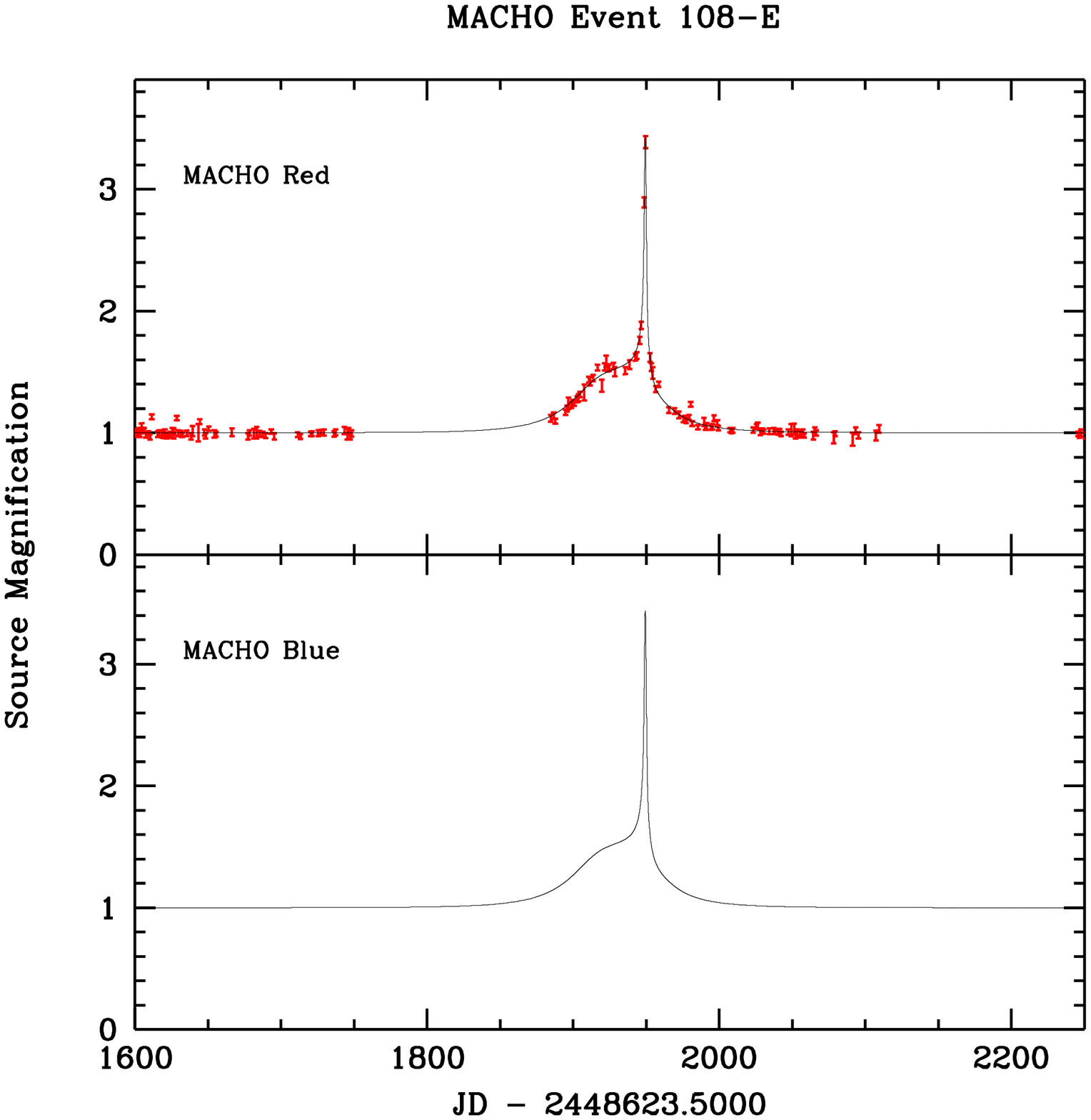}
\figcaption[f41.ps]{\label{fig-108193331878}
  Lightcurve of MACHO event 108-E, including our fit to binary microlensing.
}
\end{figure}

\begin{figure}
\plotone{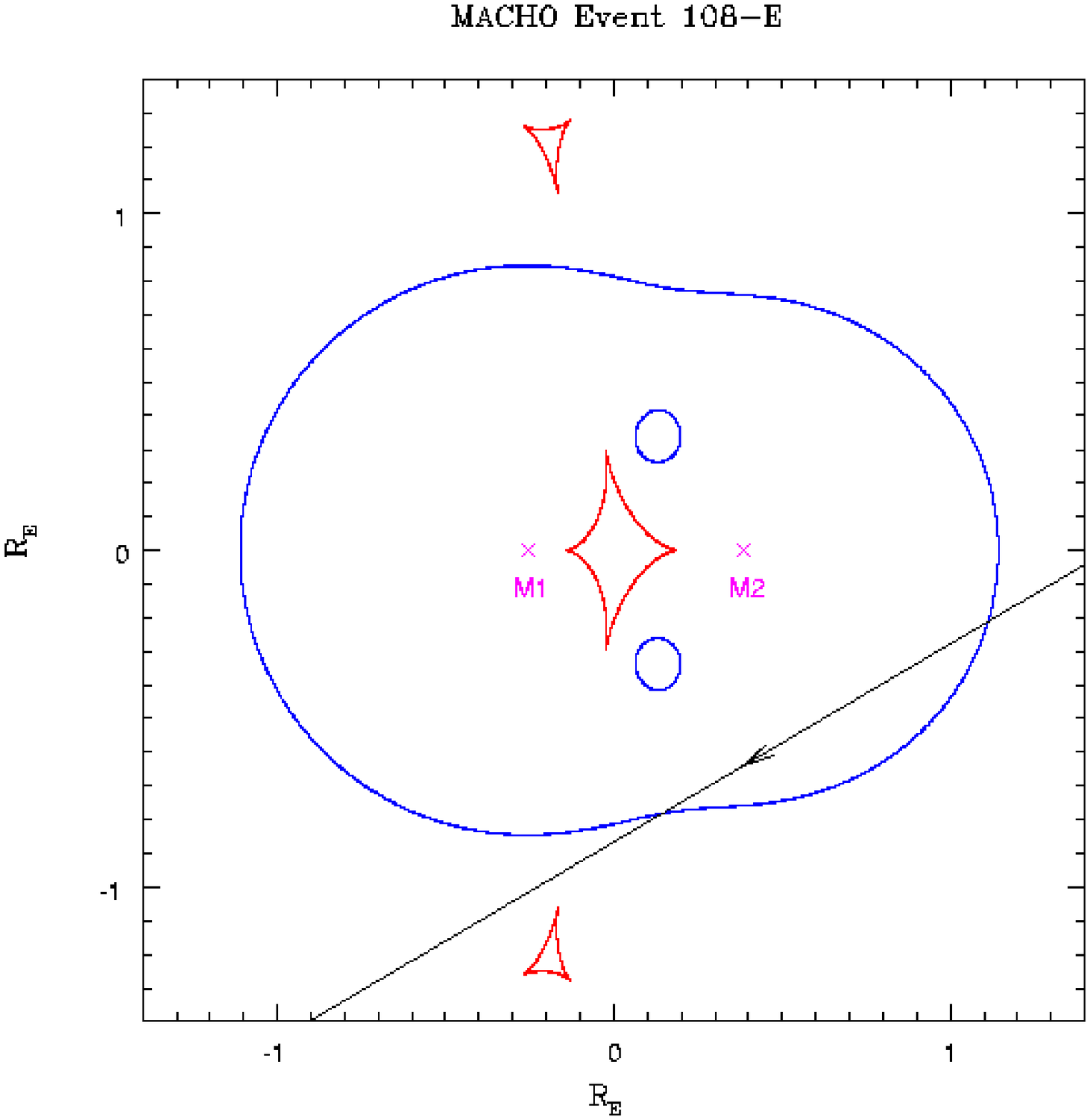}
\figcaption[f42.ps]{\label{fig-108193331878cc} Location of the
  (red) caustic and (blue) critical curves for the 108-E binary lens fit
  presented in Fig.~\ref{fig-108193331878}.  The coordinate system, whose origin
  is at the center of mass, indicates distance in units of the system's
  Einstein Ring radius $\Re$.  Also shown are the locations of the
  lensing objects, and the trajectory of the source through the caustic
  structure.  }
\end{figure}

\begin{figure}
\plotone{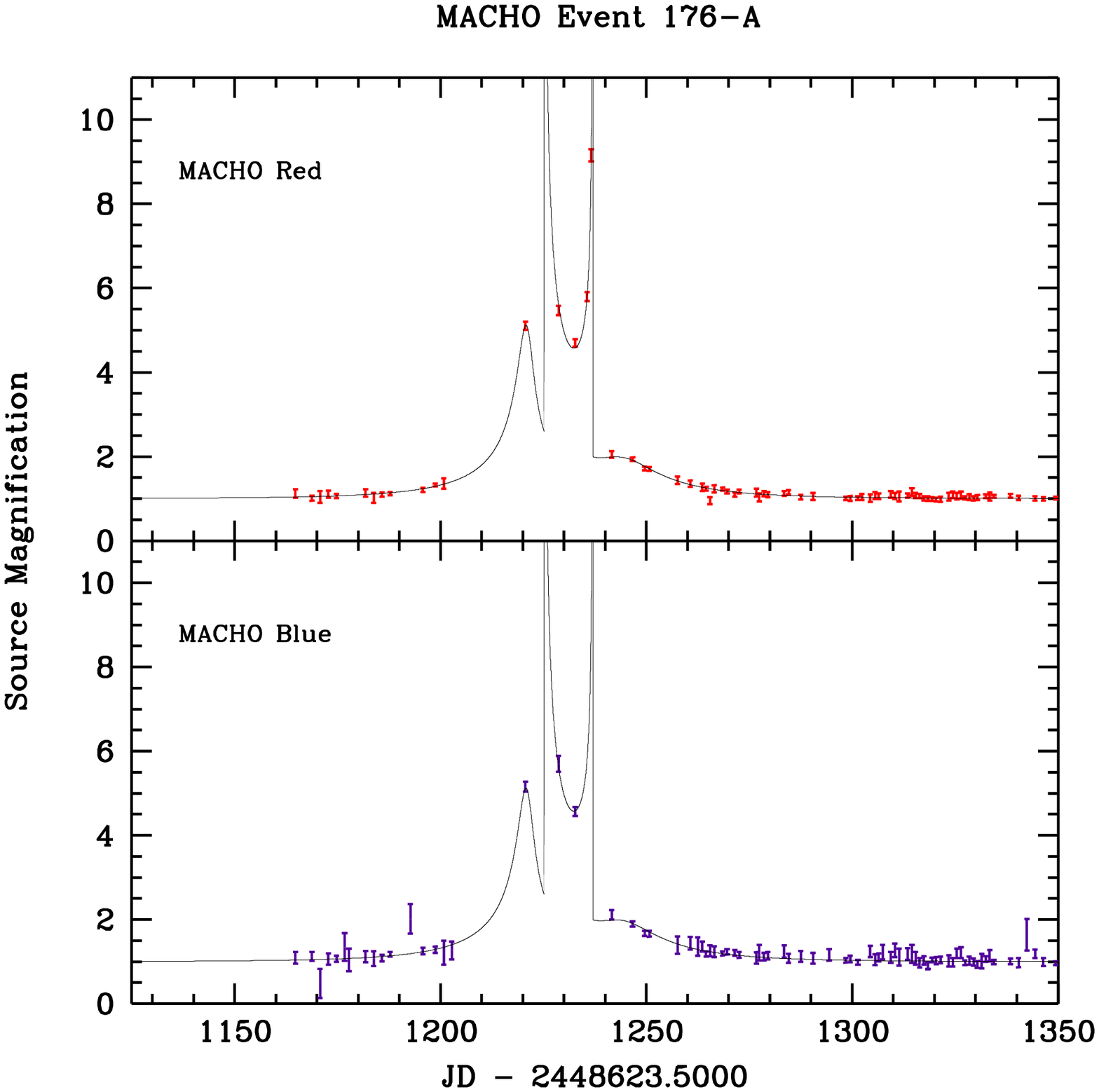}
\figcaption[f43.ps]{\label{fig-17619219978}
  Lightcurve of MACHO event 176-A, including our fit to binary microlensing.
}
\end{figure}

\begin{figure}
\plotone{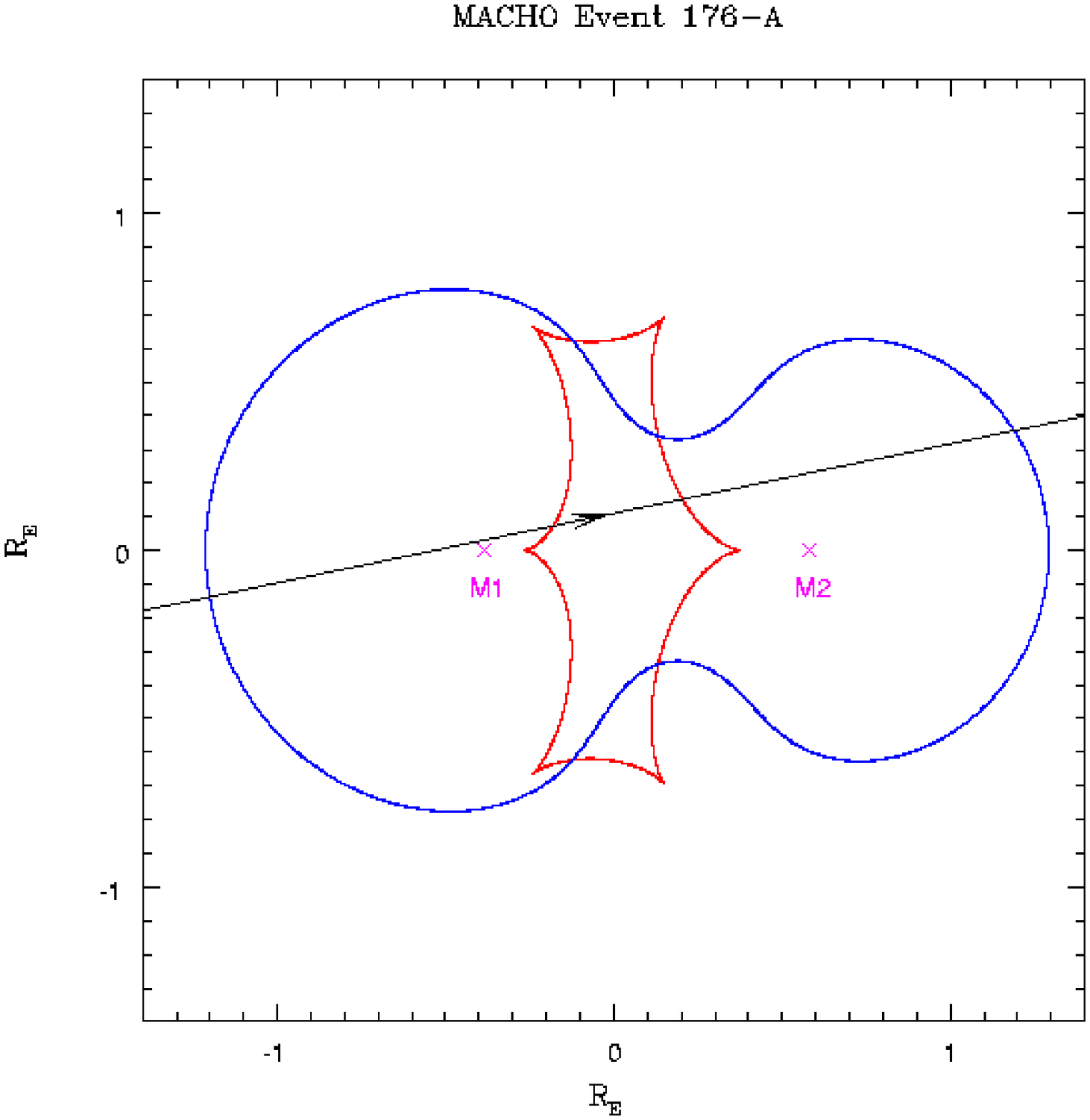}
\figcaption[f44.ps]{\label{fig-17619219978cc} Location of the
  (red) caustic and (blue) critical curves for the 176-A binary lens fit
  presented in Fig.~\ref{fig-17619219978}.  The coordinate system, whose origin
  is at the center of mass, indicates distance in units of the system's
  Einstein Ring radius $\Re$.  Also shown are the locations of the
  lensing objects, and the trajectory of the source through the caustic
  structure.  }
\end{figure}

\begin{figure}
\plotone{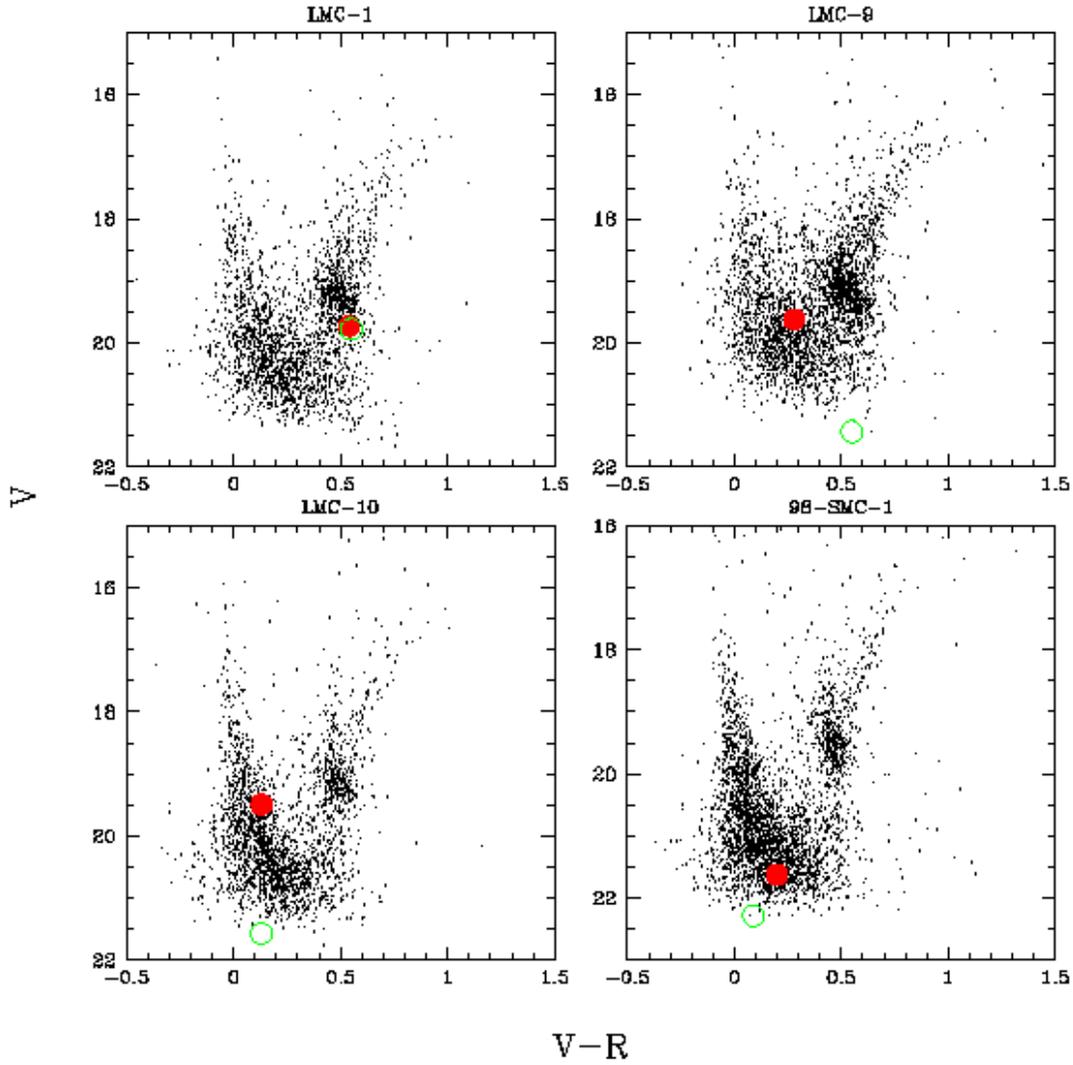}
\figcaption[f45.ps]{\label{fig-cmdmc}
  Distribution of $V, V-R$ for $\sim 3,000$ stars neighboring the
  lensed object, indicated with the (red) filled circle, for objects in
  the Magellanic Clouds.  The de-blended source location, determined
  from blending parameters in the microlensing fit, is indicated with
  the (green) open circle.
}
\end{figure}

\begin{figure}
\plotone{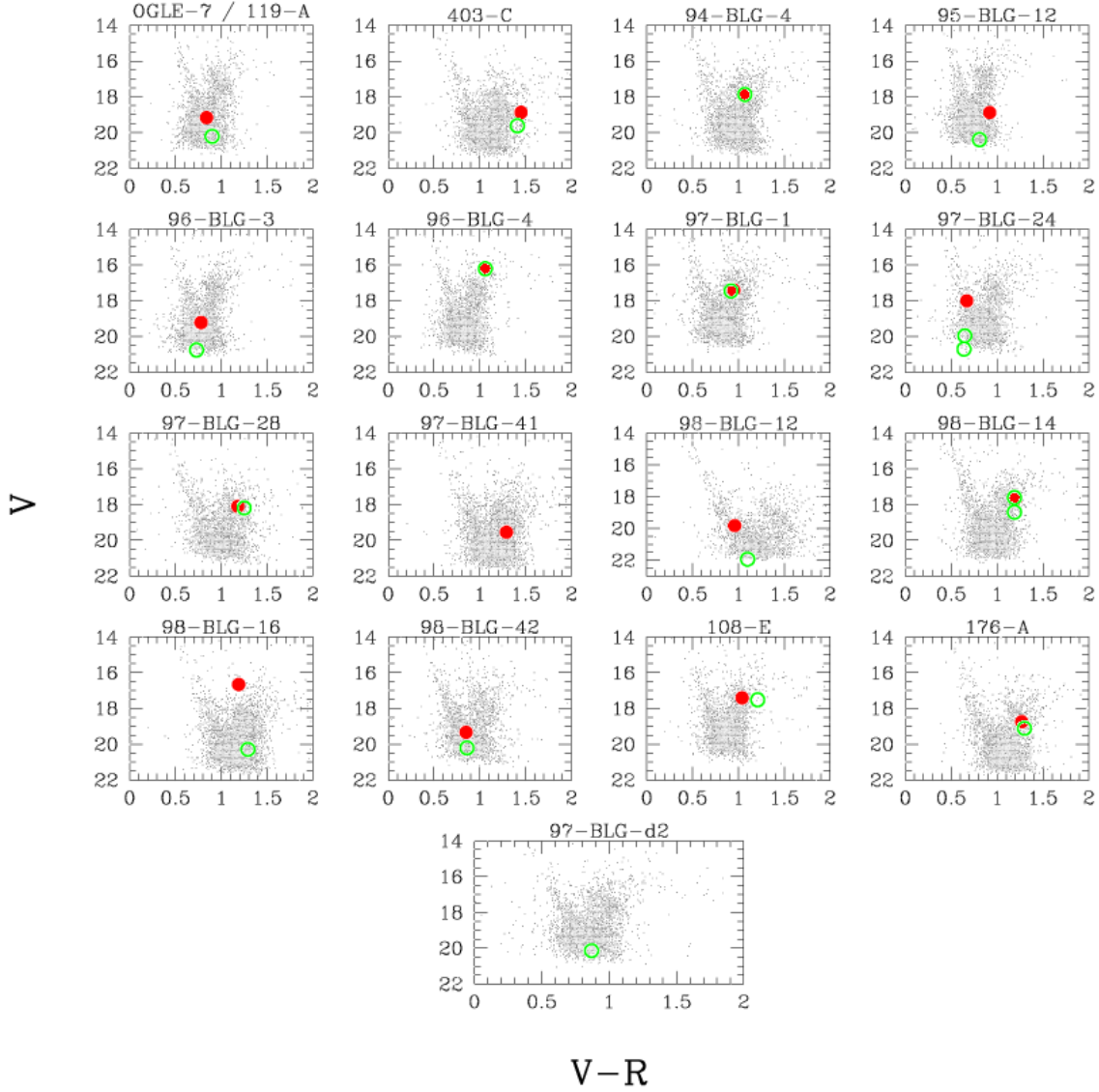}
\figcaption[f46.ps]{\label{fig-cmdblg}
  Distribution of $V$, $V-R$ for $\sim 3,000$ stars neighboring the
  lensed object, indicated with the (red) filled circle, for objects in
  the Galactic bulge.  The de-blended source location, determined from
  blending parameters in the microlensing fit, is indicated with the
  (green) open circle.  Notable cases: 97-BLG-24, where the brighter
  source is associated with the low mass ratio fit in
  Fig.~\ref{fig-97blg24}; 97-BLG-41, where we have no estimate of the
  source properties; 98-BLG-14, where the brighter source is associated
  with the large mass ratio fit in Fig.~\ref{fig-98blg14}; 97-BLG-d2,
  where we only have an estimate of the lensed source properties.
}
\end{figure}

This brightening of this apparent clump giant object was detected and
alerted on Apr 26, 1998.  Initially, it was not clear if the asymmetry
in Fig.~\ref{fig-98blg14} was due to the parallax effect
\cite{refsdal-parallax,macho-par}, and initial data allowed fits of
similar significance for both binary and parallax models.  However,
the higher precision photometry from CTIO and MSO74 observations
clearly favor the binary interpretation over the best fit parallax
lightcurve ($\delchi = 115.75$ with 1 less degree of freedom), while
the MACHO data also provide $\delchi = 30.32$.  Unfortunately, there
does remain a degeneracy between binary lens models.
Fig.~\ref{fig-98blg14ccn} and Fig.~\ref{fig-98blg14ccp} indicate this is
a non-caustic crossing binary event, similar to 95-BLG-12.

The blend fraction for the best fit (fit~----- in
Fig.~\ref{fig-98blg14}, and Fig.~\ref{fig-98blg14ccn}) is $f \sim 0.5$
for all 4 passbands of coverage, but there is another fit that is
almost as good with $f \simgt 1$ (fit~-~-~- in Fig.~\ref{fig-98blg14},
and Fig.~\ref{fig-98blg14ccp}).  The best fit suggests a blend of
clump giant stars, where they are constrained to lie within a seeing
disk, or $\sim 1''$.  98-BLG-14 is located closer to the Galactic
center than most of our events at $l=1.96$, $b = -2.29$, where the
surface density of giants is quite high.  The average separation of
clump giants is $5.3''$ in the vicinity of 98-BLG-14, so we expect
that $\sim 10$\% of clump giants will be blended with another clump
giant.  The de-reddened brightness of the MACHO object would place it
at the bright tip of the clump.  In the best fit, the lensed source
would be somewhat lower than the mean clump brightness.  It is thus
reasonable for the lensed clump giant source to be blended with an
unlensed clump giant.  The second-best fit differs from the first by
$\delchi = 5$, and from the event timescale of $\that = 74$ days and
lens mass ratio of $12:1$, we can estimate a lens system comprised of
a $1.03 \msun$ primary and $0.088 \msun$ companion.


\subsection{98-BLG-16 (402.47863.110)}
\label{subsec-98blg16}

98-BLG-16 was detected and alerted on Apr 28, 1998, and thought to be
a high magnification, short timescale event.  The initial sharp rise
in Fig.~\ref{fig-98blg16} was due to a cusp approach
(Fig.~\ref{fig-98blg16cc}).  Subsequent lightcurve interpretation was
hindered by a significant amount of scatter in the MACHO Red passband,
due to a defective amplifier, and a centroid offset of $\sim 1.3''$
between the MACHO object and the lensed source.  Follow-up data from
the CTIO 0.9m were able to resolve these objects, which allowed them
to be independently photometered.  Inspection of the CTIO dataset
indicates the lensed source has a baseline flux of $\sim 7\%$ of the
brightest (constant) neighboring star which serves as the target of
MACHO photometry.  The binary lens fit determines blend fractions
consistent with this, $f_{MACHO} \sim 0.04$ and $f_{CTIO} \sim 1$,
indicating there is no significant flux contribution from the lensing
objects.  The CTIO data from May 4.3 UT, 1998 (day 2314.3) provide a
constraint on the lens proper motion for this particular fit, $\tstar
= 0.163 \pm 0.003$ days.


\subsection{98-BLG-42 (101.21045.2528)}
\label{subsec-98blg42}

98-BLG-42 was detected and alerted on Aug 22, 1998 at a magnification
of $\sim 4.0$.  The source at this point in time was inside the
caustic region depicted in Fig.~\ref{fig-98blg42cc}.  The event was
immediately followed up by the MPS effort, on the MSO 74'' telescope.
Over the next 3 nights the source began to rapidly increase in
brightness.  The PLANET collaboration issued an Anomaly Alert on Aug
26.0 UT, 1998 (day 2428.0) indicating the source underwent a caustic
crossing between Aug 25.0 and 25.7 UT, 1998.  Our fit presented here
indicates a caustic crossing date of Aug 25.77 UT, with a source
radius crossing time of $\tstar = 0.109 \pm 0.016$ days.


\subsection{97-BLG-d2 (108.19073.2291)}
\label{subsec-108190732291}

This event was detected in the 3-year difference image analysis (DIA)
of MACHO field 108, originally presented in \citeN{macho-dia}.  The
event is most closely associated with MACHO object 108.19073.2291,
however the DIA technique uniquely identifies the time-varying source.
The lightcurve associated with this analysis is purely a lightcurve of
residuals around the source's baseline flux, which is not determined.
For consistency with the notation of the other fits in this paper, we
have added an arbitrary amount of flux to the residual lightcurve, and
fit for the fraction of the combined lightcurve that is lensed.  In
this way we are able to estimate the baseline brightness of the lensed
source.  The source trajectory in Fig.~\ref{fig-97blgd2cc} includes 2
cusp approaches, and passage through a caustic structure, which is
suggested by 2 MACHO data points at high magnification.


\subsection{MACHO-108-E (108.19333.1878)}
\label{subsec-108193331878}

This event was detected in the course of the MACHO bulge 5-year
analysis \cite{macho-bulge5}.  The deviation occurred with the Alert
System in place.  However, the Alert System was not triggered since
all but 3 of the MACHO Blue data points are removed from the
lightcurve due to the object's proximity to the edge of a detector.
For this reason we are unable to realistically estimate properties of
the lensed source, or set meaningful limits on the lens brightness.
The lightcurve (Fig.~\ref{fig-108193331878}) is characterized by
approaches to 2 of the 3 caustics in the source plane
(Fig.~\ref{fig-108193331878cc}).


\subsection{MACHO-176-A (176.19219.978)}
\label{subsec-17619219978}

Event 176-A was also detected in the bulge 5-year analysis, and is a
good example of a poorly sampled binary lens event.  The magnitude of
the initial caustic approach is unconstrained, and the fit presented
here places the one relevant data point at its peak
(Fig.~\ref{fig-17619219978}).  The $\cup$-shape of the subsequent data
suggests a caustic crossing and gradual decline to baseline.  The
critical and caustic curves are presented in
Fig.~\ref{fig-17619219978cc}.


\section{What Can We Learn?}
\label{sec-summary}

With our ensemble of 21 binary microlensing candidates, we can begin
to consider mapping the properties of the binary events to the lensing
population as a whole (\citeNP{distefano99,kerins-evans}).  This is
most difficult towards the Magellanic Clouds, where we have the
additional uncertainty of an unknown or unmodelled lensing population.

\subsection{Towards the Magellanic Clouds}
\label{subsec-sumLMC}

We have presented 3 candidate LMC binary events out of the 8 events
published in \citeN{macho-lmc2}, and detect no more unambiguous binary
lens candidates in the $\sim 20$ events \cite{macho-lmc5} in our
5-year analysis (we do however detect one binary source candidate,
96-LMC-2 \cite{becker-aas-gman}).  A color-magnitude diagram,
incorporating the de-blended magnitudes of the Magellanic sources from
Table~\ref{tab-sources}, can be found in Figure~\ref{fig-cmdmc}.

LMC-9 is a resolved caustic crossing event, where the measured
$\tstar$ (assuming a single lensed source) is consistent with the
lensing system residing in the LMC \cite{macho-bennett96b}, but only
if the velocity dispersion and the self-lensing optical depth
\cite{gould-disktau} of the LMC are both small.  If the LMC
self-lensing optical depth is large enough to explain most of the
microlensing events seen towards the LMC, as in the model of
\citeN{aubourg99}, then the proper motion of LMC-9 is not consistent
with an LMC lens, unless the source star is actually a pair of binary
stars of similar brightness.  However, in this case, we can no longer
constrain the proper motion of the lens.

As emphasized by \citeN{distefano99}, we should also expect events
without obvious caustic crossings, similar to LMC-10.  This event is
consistent with a binary lens event, but the asymmetry of the
lightcurve also resembles what might be expected for some types of
stellar variability.  A future HST frame of this object could confirm
the microlensing prediction for the amount of blending.

Event 98-SMC-1 was recognized to be a caustic crossing event in real
time with the GMAN follow-up observations presented here.  An
unprecedented response by the majority of the microlensing community
resulted in dense coverage of the event, including resolution of the
second caustic crossing by the PLANET and EROS collaborations.
Important constraints on the initial caustic crossing date were
provided by the OGLE and MPS collaborations.  The lens proper motion
derived from each of these datasets is most consistent with a SMC
lens.  This is the strongest constraint that has yet been placed on
the location of the lensing population towards the Magellanic Clouds.
However, as \citeN{eros-smc} point out, the SMC is expected to have a
large self-lensing optical depth, so that a large fraction of SMC
events are likely to be due to self-lensing, even if most of the LMC
events are due to halo lenses.

\citeN{kerins-evans} reach the conclusion that, if the caustic
crossing events LMC-9 and 98-SMC-1 are both due to Magellanic lenses,
than the bulk of lensing seen so far towards the Magellanic Clouds is
most likely due to self-lensing, where the lenses may reside in
Magellanic stellar or dark halos.  However, the suggestive evidence
that the LMC-9 lens may reside in the LMC only applies for LMC models
with a low self-lensing optical depth.  Furthermore, as
\citeN{honma99} has pointed out, there is probably a bias in favor of
detecting long timescale caustic crossing events.  Thus, we may be
more likely to detect and characterize caustic crossing features for
self-lensing events than for halo lensing events, if the latter tend
to have shorter timescales.  A potentially more serious bias may be
that there may be very few binary lenses in the halo.  If most of the
LMC events are due to lenses in the Galactic halo, then they comprise
a previously unknown population with an unknown binary fraction.  So,
it is possible that the sample of binary events themselves selects
against the halo lensing events.  If so, we might expect a smaller
fraction of binary lensing events towards the LMC than towards the
bulge when the event samples get larger.

\subsection{Towards the Galactic Bulge}
\label{subsec-sumbulge}

A more representative sample of binary lenses can be found amongst the
17 Galactic bulge candidates.  For the duration 1994-1998, the Alert
System triggered on 196 Galactic bulge microlensing events.  Twelve of
the Alert events are presented here as binary lens candidates.  Since
the Alert System is tuned to detect generic variability, it is
reasonable to make the assumption that the 12 binary events out of the
196 Alert events are representative of the detectable binary fraction
of the lensing population as a whole.  This is consistent with
theoretical estimates of a binary lensing rate of order $10\%$
\cite{mao91}.

A color-magnitude diagram, incorporating the de-blended magnitudes of
the lensed sources from Table~\ref{tab-sources}, can be found in
Figure~\ref{fig-cmdblg}.  It is interesting to note that clump giants
are over-represented in our sample, implying our binary lens detection
efficiency is highest with these bright sources.  It is also important
to note that for the 6 clump giants lensed, in most cases the blending
fraction is quite close to 1 (with the possible exception of
98-BLG-14), indicating there is insignificant contamination from
neighboring sources and, most importantly, from the lensing system.
This is contrary to the distributions of blending in the majority of
binary microlensing events that have been reported in the literature
thus far \cite{distefano99}.

Two of the binary events (97-BLG-24 and 98-BLG-16) are heavily blended
in the MACHO data, but the lensed source star is resolved in the
follow-up GMAN data, at separations of $1.2''$ and $1.3''$,
respectively.  However, the 97-BLG-24 field is crowded enough that it
is difficult to independently photometer the MACHO object and lensed
source.  For 98-BLG-16, the microlensing fit to CTIO data exhibits no
blending, indicating the lens is in fact relatively dark.  Candidates
95-BLG-12, 96-BLG-3, and 98-BLG-12 are strongly blended in all
passbands ($70-80 \%$ contamination), and it is possible in these
cases that the lens contributes significant flux to the source
brightness.  High resolution photometry of strongly blended events can
disentangle blending due to crowding, and blending due to lens
luminosity.

Excluding event 97-BLG-24 for which we have two fits, we find 12 (9)
caustic crossing events from our ensemble of 16 (only the 11 Alert)
bulge candidates.  While not all of the GMAN follow-up data for the
interval 1995-1998 have yet been reduced, we so far have found no
weakly perturbed systems, at the $\sim 1 \%$ level (see however
\citeNP{becker-thesis}).  \citeN{distefano99} indicates that several
of these weakly perturbed events should exist for each caustic
crossing event (but see Sec.~\ref{subsec-rate}).  That we have
detected no such events indicates our efficiency at detecting them, or
characterizing them as binary microlensing, is currently quite low.

\subsubsection{Mass Ratio Distribution}
\label{subsec-mrat}

\begin{figure}
\plotone{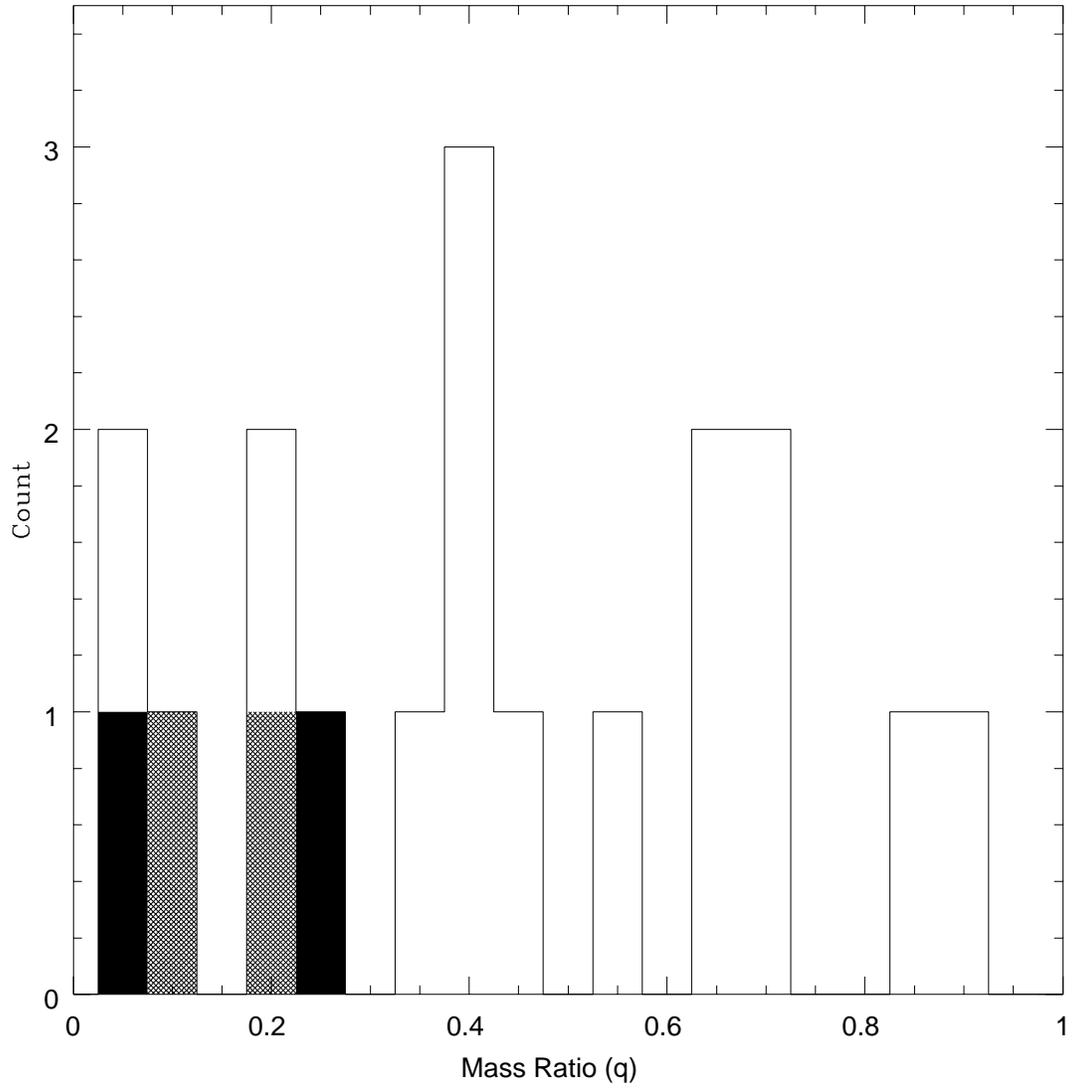}
\figcaption[f47.ps]{\label{fig-mass}
  The distribution of mass ratios ($q \leq 1$) for our Galactic bulge
  binary microlensing events.  The contribution of our 2 fits each to events
  97-BLG-24 and 98-BLG-14 are represented by the dark and light shaded
  areas, respectively.
}
\end{figure}

While some of the events here are poorly constrained, and thus can be
characterized by multiple combinations of event parameters, we can
begin to probe the distribution of mass ratios of the binary lensing
population towards the Galactic bulge, given the above fits.
Distributions of the binary parameters $\theta$ and $a$ are less
informative, with $\theta$ representing a random orientation between
the lens separation axis and motion with respect to the source, and
$a$ the projection of the lens separation at a random (unknown)
orbital phase.  Figure~\ref{fig-mass} contains a histogram of the
distribution of mass ratios ($q$, defined to be $\leq 1$) which we
find from our bulge events.  The 2 fits each for events 97-BLG-24 and
98-BLG-14 are represented by additional shaded areas.  The
distribution here is free of inclination uncertainties present in
studies of spectroscopic binaries (SB), but does likely suffer from
non-uniqueness of fits in several cases.

The binary mass ratio distribution function N($q$) has been studied
by, e.g. \citeN{trimble-bin90}, who examines SB systems with both
giant and relatively bright main sequence primaries.  Similar
morphologies are found between samples, generally characterized by
N($q$) $\propto q^{-1}$, with a possible peak near $q \sim 0.3$.
However, we expect many of our lenses to be drawn from the lower, more
populated portion of the main sequence.  Given the apparent dependence
of N($q$) on spectral type \cite{abt-bin83}, it would be more
appropriate to compare Fig.~\ref{fig-mass} with the study of
solar-type SB by \citeN{mazeh-bin92}.  They find, with a considerably
smaller sample than \citeN{trimble-bin90}, a relatively flat N($q$),
possibly rising towards larger $q$.  We therefore compare two models,
N($q$) $\propto q^{-1}$ and N($q$) = constant, against the events with
$0.1 < q < 1$ using the one-sided Kolmogorov-Smirnov (KS) test.  We
find probabilities of $0.07$ and $0.41$, respectively, of finding a KS
deviation between data and model as large as that observed.  A N($q$)
$\propto q^{-1}$ is clearly inconsistent with our measured
distribution, while a flat N($q$) is consistent with our data.

\subsubsection{Reddening Estimates}
\label{subsec-red}

For 10 of our bulge events, the microlensing fits provide a
measurement of $\tstar$.  However, it is our coverage of events 119-A
(Fig.~\ref{fig-119a}), 96-BLG-3 (Fig.~\ref{fig-96blg3}), 97-BLG-28
(Fig.~\ref{fig-97blg28}), and 98-BLG-42 (Fig.~\ref{fig-98blg42}) which
most reliably constrain $\tstar$.

We therefore have 4 events where we believe we have a reliable
measurement of the time it takes the lens to transit the source face,
$2 \tstar$.  To arrive at an estimate of the proper motion of the
lensing system, we must first determine the angular size of the
source.  This can be done by assuming the source is a blackbody and
estimating its effective temperature $\teff$ and apparent bolometric
magnitude $\mbol$.  We first need to determine the extinction, $\Av$,
and reddening, $\Evr$, to the source -- in the following, we assume
$\Av = 3.97 \Evr$ \cite{rieke85}.

\citeN{stanek-map}, in a study of red clump stars in Baade's Window,
finds a range of $\Av$ from 1.26 to 2.79, even though this is the
clearest and most uniform window through the bulge.  This suggests it
would be unwise to apply an ``average'' extinction correction to all
of our events.  Instead, we estimate the reddening for each source
using neighboring RR Lyrae stars, whose intrinsic colors are well
known.  An intrinsic color-period ($P$) relationship has been
established for field RR Lyrae by \citeN{caputo-92}, in the form of
\begin{equation}
  (B-V)_0 = 0.658 + 0.097 ~ {\rm [Fe/H]} + 0.710 ~ {\rm log} P.
\end{equation}
We assume [Fe/H] = $-1$, after \citeN{walker-terndrup}, and transform
to $(V-R)_0$ with the relation \cite{macho-apj97c}
\begin{equation}
  (V-R)_0 = 0.004 + 0.566 (B-V)_0.
\end{equation}

We then compute the reddening to all RR Lyrae within $10'$ of the
source, excluding significant outliers, and apply the reddening and
extinction corrections to our lensed source star.  The results of this
are listed in the first columns of Table~\ref{tab-size}.  We note that
two of the events, 119-A and 96-BLG-3, occurred within Baade's Window,
but both sources are just outside the extinction grid reported by
\citeN{stanek-map}.

We next interpolate the ATLAS9 and NMARCS model atmospheres presented
in \citeN{bessell-98} (assuming solar metallicity, ${\rm log}~g$ = 2.0
for giants, and ${\rm log}~g$ = 4.5 otherwise) to determine the
source's $\teff$ and bolometric correction in V ($BC_V$), given its
unreddened $(V-R)_0$.  It is then straightforward to calculate the
angular size of the source $\theta_*$ using the bolometric flux method
of \citeN{gray92}, and to determine the relative proper motion $\mu =
\theta_* / \tstar$ between the lens and the source, further described
in Sec.~\ref{subsec-ppm}.

\subsubsection{Lens Proper Motions}
\label{subsec-ppm}

\begin{figure}
\plotone{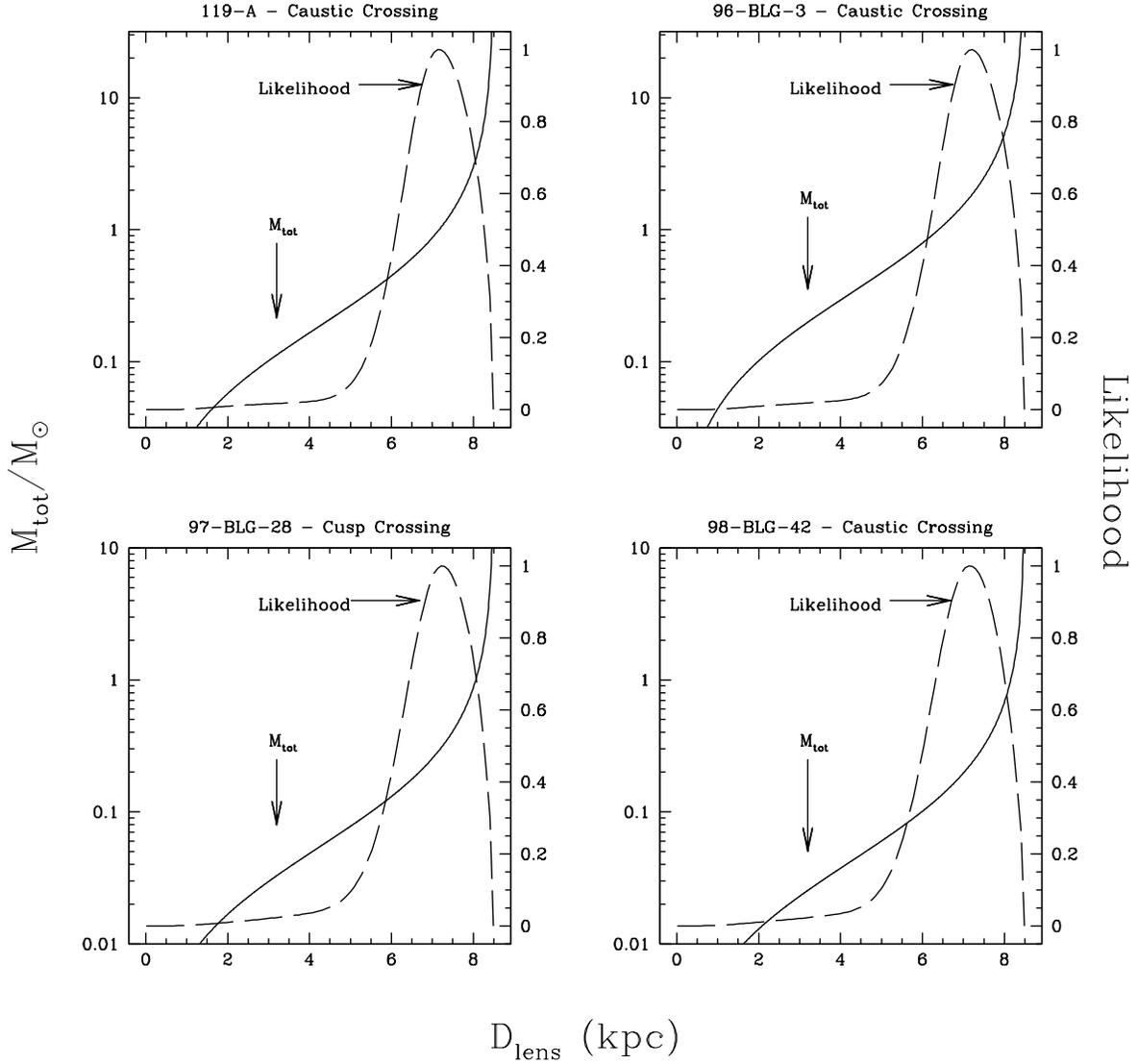}
\figcaption[f48.ps]{\label{fig-masslike}
  The relation between the lens distance, $\dlens$, and the total
  mass, $\mtot$, is shown for the four lensing events with reliable
  proper motion determinations.  Also shown are the results of a
  likelihood analysis to estimate the lens parameters.  The most
  likely lens distances and masses are given in Table~\ref{tab-ppm}.
}
\end{figure}

The determination of the lens proper motion with respect to the
source, $\mu$, yields an additional constraint on the parameters of
the lensing system.  For most lensing events, the only observable
parameter which constrains the distance, $D_{ol}$, total mass,
$\mtot$, and the transverse velocity, $\vperp$, of the lens system is
the event timescale, $\that$.  However, for caustic crossing events,
we have seen that it is also possible to measure the lens proper
motion, which is related to the distance and transverse velocity of
the lens by
\begin{equation}
\mu = {\vperp\over D_{ol}}  \ .
\end{equation}
Thus, for the four events with measured $\mu$ values, we have two
constraints on three parameters, so there is a one-parameter family of
solutions, namely
\begin{equation}
\mtot = {\mu^2 \that^2\over 16G}
          {D_{ol} \left(D_{ol}+D_{ls}\right)\over D_{ls}}  \ .
\end{equation}
These solutions are shown in Figure~\ref{fig-masslike}. We see that
$\mtot$ grows as function of the lens distance, and passes through the
expected lens masses of $0.1-2\msun$ at distances of $2-7.5\kpc$.

We can make a more accurate estimate of the lens parameters if we make
use of our knowledge of the velocity distributions of the source and
lens populations.  This requires that we specify a Galactic model, and
we select the simple bar model of \citeN{han-gould95}.  The kinematics
of the bulge using giants has been measured by
\citeN{dante-kinematics}.  We assume velocity dispersions of $30\kms$
in each transverse direction for disk lenses, and $80\kms$ in each
transverse direction for both lenses and source stars located in the
central bar.  We have assumed a distance of $R_0 = 8 \kpc$ to the
Galactic center and $8.5 \kpc$ to each source star.  We have done a
likelihood analysis to determine the most probable lens distance and
mass for each of the four events with reliable proper motion
measurements, and the results are summarized in
Figure~\ref{fig-masslike} and Table~\ref{tab-ppm}.  The measured
proper motions of $\mu = 1.8-3.2\,$mas/yr are most consistent with
lens systems residing in the bulge, and the best fit primary lens
masses are consistent with main sequence stars fainter than the Sun.
The only exception is the heavier lens mass for event 96-BLG-3, which
has a best fit mass of $1.2\msun$ with an uncertainty of a factor of
two.

If the heavier 96-BLG-3 lens is a main sequence star, then we must
have $\m2 \leq 1.3\msun$ to be consistent with our measurement of the
unlensed flux observed at the position of the source star.  High
resolution imaging of the source might be able to tighten this
constraint by resolving some of the unlensed flux into separate stars.
Similar constraints apply for the other stars, but are significantly
weaker because the implied masses are much lower.  However, these
constraints do not apply for stellar remnant lenses, which may make up
of order $20$\% of the lens population \cite{gould-remnant99}.

\section{Conclusions}
\label{sec-con}

After a survey of the MACHO database, it is very apparent that
microlensing by binary lens systems has been detected, and at a rate
that is roughly consistent with theoretical predictions for known
stellar populations.  However, a rigorous search has not yet been
implemented, and we cannot set hard limits on the binary microlensing
rate, nor on the characteristics of binary systems in the lensing
population.  This includes the incidence of planets around lensing
stars.  However, follow-up efforts such as MPS \cite{mps-98blg35} and
PLANET \cite{planet-ogle9814} are undertaking dense lightcurve
sampling, and are beginning to set meaningful limits on planetary
companions on an event-by-event basis.

It is also apparent that there are difficult degeneracies between
binary microlensing fits which cannot be resolved with the sparse
sampling ($\sim$ once per night) of the microlensing survey
telescopes.  This is especially important in short duration and/or low
level deviations, such as caustic crossings or planetary 'spikes'.  In
both cases important information is contained in a small fraction of
the lightcurve.  It is thus important that microlensing follow-up
continues with dense (tens of observations per night), precise ($\sim
1 \%$) sampling of event lightcurves.  The limitations of undersampled
datasets are apparent in the analysis of 97-BLG-24, where we are sure
of a binary (possibly even planetary) signal, but are not able to
uniquely characterize it.

Three of our bulge events (95-BLG-12, 96-BLG-3 and 98-BLG-12) are
strongly blended in all MACHO and GMAN follow-up passbands, indicating
the lensing objects may be luminous at a detectable level.  High
resolution observations of these sources may eventually reveal the
appearance of a ``new'' source, as the lens proper motion separates it
from the lensed source at the rate of $\sim$ milli-arcseconds/yr.
This can be accomplished with the HST, or adaptive optics imaging
available on systems such as Gemini.  On the other hand, our binary
events on giant sources tend to show little blending.  Event
98-BLG-16, a main-sequence source which is highly blended in the MACHO
photometry, is resolved into separate objects in GMAN follow-up data,
and fit blend fractions indicate the lensing objects are in fact dark.

We have recovered a distribution of mass ratios for 16 of our Galactic
bulge microlensing candidates, and a distribution of lens proper
motions for 4 of these events where we have, to some degree, resolved
a caustic crossing.  The mass ratio distribution is consistent with
the relatively flat distribution seen in solar-type spectroscopic
binary systems.  The lens proper motions, when combined with the
likelihood analysis in Sec.~\ref{subsec-ppm}, imply a population of
binary lenses residing in the Galactic bulge at a distance of $7 \pm 1
\kpc$.  The lens masses generally appear sub-solar.

Finally, we would like to caution against over-interpretation of this
dataset, in particular the Magellanic Cloud subset, which is
admittedly incomplete.  In fact, there are undoubtedly a number of
unquantified biases in our ability to discriminate between the
single-lens and binary-lens case, although actual event detection is
more a function of the significance of the deviation from baseline
than lightcurve morphology.  However, observations of on-going binary
lensing events towards the Magellanic Clouds, such as 98-SMC-1, may on
an event-by-event basis allow us to examine the role the lensing
populations play in the larger context of Galactic dark matter.

\acknowledgements
\section*{Acknowledgments}

We are very grateful for the skilled support given our project by
S.~Chan, S.~Sabine, and the technical staff at the Mt. Stromlo
Observatory.  We especially thank J.D.~Reynolds for the network
software that has made this effort successful.  We would like to thank
the many staff and observers at CTIO and UTSO who have helped to make
the GMAN effort successful.

Work performed at LLNL is supported by the DOE under contract
W7405-ENG-48.  Work performed by the Center for Particle Astrophysics
personnel is supported in part by the Office of Science and Technology
Centers of NSF under cooperative agreement AST-8809616.  Work
performed at MSSSO is supported by the Bilateral Science and
Technology Program of the Australian Department of Industry,
Technology and Regional Development.  Astronomy at Wise Observatory is
supported by a grant from the Israel Science Foundation.  KG
acknowledges support from DOE grant DEFG03-90-ER 40546 and a Cottrell
Scholar award.  DM is also supported by Fondecyt 1990440.  CWS thanks
the Packard Foundation for their generous support.  WJS is supported
by a PPARC Advanced Fellowship.

\clearpage


\clearpage

\clearpage

\clearpage

\clearpage


\clearpage

\clearpage


\clearpage

\clearpage


\clearpage

\clearpage


\clearpage

\clearpage


\clearpage

\clearpage


\clearpage

\clearpage


\clearpage

\clearpage


\clearpage

\clearpage


\clearpage

\clearpage


\clearpage

\clearpage


\clearpage

\clearpage

\clearpage


\clearpage

\clearpage


\clearpage


\clearpage

\clearpage


\clearpage

\clearpage

\clearpage


\clearpage

\clearpage


\clearpage

\clearpage


\clearpage

\clearpage


\clearpage

\clearpage


\clearpage

\clearpage


\clearpage


\clearpage


\clearpage


\clearpage


\begin{deluxetable}{llll|cc|cc}
\tablecaption{Candidate Lensed Source Stars \label{tab-sources} }
\tablewidth{0pt}
\tablehead{
        \colhead {Event } &
        \colhead {MACHO Id } &
        \colhead {RA (J2000) } &
        \colhead {DEC (J2000) } &
        \colhead {$V$ \tablenotemark{a} } &
        \colhead {$V-R$ } &
        \colhead {$V$ \tablenotemark{b} } &
        \colhead {$V-R$ } 
}
\startdata
LMC-1 (-----) & 79.5628.1547  & 05 14 44.5  & -68 48 00.1
   & 19.73  & 0.54  & 19.75  & 0.55 \nl

LMC-1 (-~-~-) & \nodata  & \nodata  & \nodata
   & 19.73  & 0.54  & 19.78  & 0.55 \nl

LMC-9         & 80.6468.2746  & 05 20 20.2  & -69 15 11.8
   & 19.62  & 0.28  & 21.43  & 0.55 \nl

LMC-10       & 18.3324.1765  & 05 01 16.0 &  -69 07 33.1
   & 19.49  & 0.13  & 21.57  & 0.13 \nl

98-SMC-1 & 208.15683.4237  & 00 45 35.2  & -72 52 34.1
   & 21.62  & 0.20  & 22.28  & 0.09 \nl

\hline

OGLE-7 / 119-A & 119.20226.2119  & 18 03 35.7  & -29 42 01.2
   & 19.16  & 0.84  & 20.21  & 0.90 \nl

403-C & 403.47793.2961  & 17 55 57.9  & -29 26 12.1
   & 18.87  & 1.45  & 19.62  & 1.41 \nl

94-BLG-4 & 118.18141.731  & 17 58 36.7  & -30 02 19.2
   & 17.92  & 1.06  & 17.87  & 1.07 \nl

95-BLG-12 & 120.21263.1213  & 18 06 04.7  & -29 52 38.1
   & 18.88  & 0.92  & 20.39  & 0.81 \nl

96-BLG-3 & 119.19444.2055  & 18 01 45.5  & -29 49 46.8
   & 20.00  & 0.78  & 21.48  & 0.73 \nl

96-BLG-4 & 105.21417.101  & 18 06 11.9  & -28 16 52.7
   & 16.25  & 1.07  & 16.21  & 1.06 \nl

97-BLG-1 & 113.18674.756  & 17 59 53.3  & -29 09 07.8
   & 17.41  & 0.95  & 17.46  & 0.92 \nl

97-BLG-24 (-----) & 101.20650.1216  & 18 04 20.2  & -27 24 45.2
   & 18.01  & 0.67  & 19.96  & 0.65 \nl

97-BLG-24 (-~-~-) & \nodata  & \nodata  & \nodata
   & 18.01  & 0.67  & 20.7  & 0.64 \nl

97-BLG-28 & 108.18951.593  & 18 00 33.7  & -28 01 10.4
   & 18.10  & 1.18  & 18.19  & 1.25 \nl

97-BLG-41 & 402.47862.1576  & 17 56 20.6  & -28 47 41.9
   & 19.54  & 1.29  & \nodata  & \nodata \nl

98-BLG-12 & 179.21577.1740  & 18 06 31.7  & -26 16 01.5
   & 18.86  & 0.96  & 20.83  & 1.10 \nl

98-BLG-14 (-----) & 401.48408.649  & 17 59 08.9  & -28 24 54.6
   & 17.70  & 1.20  & 18.42  & 1.19 \nl

98-BLG-14 (-~-~-) & \nodata  & \nodata  & \nodata
   & 17.70  & 1.20  & 17.61  & 1.19 \nl

98-BLG-16 & 402.47863.110  & 17 56 18.1  & -28 46 04.9
   & 16.66  & 1.19  & 20.28  & 1.29 \nl

98-BLG-42 & 101.21045.2528 & 18 05 12.6  & -27 05 47.1
   & 19.33  & 0.85  & 20.20  & 0.86 \nl

97-BLG-d2 & 108.19073.2291 & 18 00 39.5 & -28 34 43.8
   & \nodata  & \nodata  & 20.14  & 0.87 \nl

108-E     & 108.19333.1878  & 18 01 21.1  & -28 32 39.4
   & 17.41  & 1.04  & 17.52  & 1.21 \nl

176-A     & 176.19219.978  & 18 01 04.4  & -27 30 41.3
   & 18.74  & 1.27  & 19.11  & 1.30 \nl
\enddata
\tablenotetext{a} { Standard magnitudes and color of the MACHO object
  that received the lensed flux.}
\tablenotetext{b} { Standard magnitudes and color of the actual lensed
  source star, as determined from the blend fraction in the binary lens
  fit.}
\tablenotetext{} { Information on the lensed MACHO objects and
de-blended source stars.  Events LMC-1, 97-BLG-24, and 98-BLG-14 are
each presented with 2 fits of similar significance, but different
event parameters.  We do not present a binary microlensing fit for
event 97-BLG-41, and we only include an estimate of the MACHO object's
baseline flux.  Event 97-BLG-d2 was found through difference image
analysis (DIA), which uniquely identifies the lensed source.}
\end{deluxetable} 

\clearpage

\begin{landscape}
\begin{deluxetable}{lcccccccc}
\tablecaption{Binary microlensing event parameters \label{tab-params} }
\tablewidth{0pt}
\scriptsize
\tablehead{
        \colhead {Event } &
        \colhead {$\chi^2/d.o.f.$} &
        \colhead {$\that$ } &
        \colhead {$\t0$ \tablenotemark{a} } &
        \colhead {$\umin$ } &
        \colhead {$a$ } &
        \colhead {$\theta$ (rad) } &
        \colhead {${\rm M}_1$ / ${\rm M}_2$ } &
        \colhead { $\tstar$ (days) } 
}
\startdata
LMC-1 (-----)
  & 2794.9/2179    
  & 35.54
  & 433.58
  & 0.150
  & -0.430
  & -0.488
  & 0.861
  & 0.112 \nl
LMC-1 (-~-~-)
  & 2799.4/2179      
  & 34.55
  & 433.58
  & -0.127
  & 1.217
  & 4.104
  & 108.890
  & 0.122    \nl
LMC-9
  & 1476.5/848   
  & 143.12 
  & 979.59
  & -0.054
  & 1.657
  & 0.086
  & 1.627
  & 0.651    \nl
LMC-10
  & 1672.9/1240    
  & 151.05
  & 585.74
  & 0.102
  & 0.823
  & -3.375
  & 0.034
  & \nodata    \nl
98-SMC-1
  & 1771.9/1583       
  & 147.58
  & 2354.93
  & 0.046
  & 0.664
  & -0.180
  & 0.388
  & 0.116    \nl

\hline

OGLE-7 / 119-A
  & 1997.6/1407   
  & 169.04
  & 550.04
  & 0.077
  & 1.045
  & -0.940
  & 1.212
  & 0.212    \nl
403-C
  & 318.7/342   
  & 21.37
  & 1688.75
  & -0.089
  & 1.227
  & 3.082
  & 0.556
  & \nodata    \nl
94-BLG-4
  & 923.5/1418   
  & 10.66
  & 884.13
  & 0.028
  & -1.068
  & 0.302
  & 18.084
  & 0.096    \nl
95-BLG-12
  & 1413.4/1466    
  & 307.41
  & 1264.07
  & 0.106
  & 0.421
  & -0.664
  & 2.148
  & \nodata    \nl
96-BLG-3
  & 1934.8/1542   
  & 182.45
  & 1545.66
  & 0.043
  & -0.436
  & 3.530
  & 0.392
  & 0.086    \nl
96-BLG-4
  & 1286.0/1605    
  & 162.44
  & 1815.29
  & 0.378
  & 7.454
  & -0.011
  & 1.140
  & \nodata    \nl
97-BLG-1
  & 1237.7/1305    
  & 68.52
  & 1884.02
  & 0.234
  & 0.931
  & 2.583
  & 0.418
  & 0.527    \nl
97-BLG-24 (-----)
  & 1230.6/943   
  & 30.77
  & 1968.38
  & 0.319
  & 2.077
  & -3.967
  & 4.236
  & 0.050    \nl
97-BLG-24 (-~-~-)
  & 1235.0/943    
  & 45.49
  & 1970.82
  & 0.0003
  & 1.752
  & 0.988
  & 0.035
  & 0.197    \nl
97-BLG-28
  & 2734.9/1404   
  & 52.76
  & 1991.87
  & 0.225
  & 0.707
  & 1.437
  & 0.210
  & 0.760    \nl
98-BLG-12
  & 440.5/507   
  & 239.92
  & 2321.95
  & 0.163
  & 0.793
  & 11.346
  & 0.681
  & \nodata    \nl
98-BLG-14 (-----)
  & 809.4/714  
  & 107.65
  & 2322.32
  & 0.214
  & 0.541
  & 3.531
  & 4.504
  & \nodata    \nl
98-BLG-14 (-~-~-)
  & 814.4/714  
  & 74.05
  & 2323.56
  & 0.396
  & 1.213
  & 4.429
  & 0.0857
  & \nodata    \nl


98-BLG-16
  & 461.2/616     
  & 69.50
  & 2316.57
  & 0.140
  & 0.758
  & 3.699
  & 1.476
  & 0.163    \nl
98-BLG-42
  & 890.4/395      
  & 49.19
  & 2422.41
  & -0.028
  & 1.260
  & 3.328
  & 3.065
  & 0.109    \nl
97-BLG-d2   
  & 314.1/584    
  & 92.65
  & 1971.07
  & -0.326
  & 0.965
  & 1.766
  & 0.390
  & \nodata    \nl
108-E
  & 934.4/597    
  & 71.30
  & 1927.66
  & 0.747
  & 0.637
  & -2.609
  & 1.514
  & 0.557    \nl
176-A
  & 684.0/578  
  & 60.02
  & 1230.06
  & 0.108
  & 0.965
  & 0.204
  & 1.514
  & 0.068    \nl
\enddata
\tablenotetext{a} { (JD $-$ 2448623.50). }
\tablenotetext{} { List of binary microlensing event parameters, as
described in Sec.~\ref{sec-bin}, for each of our candidates.  Also
included are the binary microlensing fit $\chi^2$, and the degrees of
freedom (equal to the number of data points minus 1 for each
constraint listed here, and minus 2 for each passband with
observations in Tab.~\ref{tab-bparams}). Events LMC-1, 97-BLG-24, and
98-BLG-14 are each presented with 2 fits of similar significance, but
different event parameters.}
\end{deluxetable} 

\clearpage

\begin{deluxetable}{lccccccccc}
\tablecaption{Binary microlensing event blend parameters \label{tab-bparams} }
\tablewidth{0pt}
\scriptsize
\tablehead{
        \colhead {Event } &
        \colhead { $f_{MACHO_R}$ } &
        \colhead { $f_{MACHO_B}$ } &
        \colhead { $f_{CTIO_R}$ } &
        \colhead { $f_{CTIO_B}$ } &
        \colhead { $f_{MSO30_R}$ } &
        \colhead { $f_{MSO74_R}$ } &
        \colhead { $f_{UTSO_R}$ } &
        \colhead { $f_{UTSO_V}$ } &
        \colhead { $f_{WISE_R}$ } 
}
\startdata
LMC-1 (-----)
  & 0.99
  & 0.98
  & \nodata
  & \nodata
  & \nodata
  & \nodata
  & \nodata
  & \nodata
  & \nodata \nl
LMC-1 (-~-~-)
  & 0.96
  & 0.95
  & \nodata
  & \nodata
  & \nodata
  & \nodata
  & \nodata
  & \nodata
  & \nodata \nl
LMC-9
  & 0.26
  & 0.17
  & \nodata
  & \nodata
  & \nodata
  & \nodata
  & \nodata
  & \nodata
  & \nodata \nl
LMC-10
  & 0.15
  & 0.15
  & \nodata
  & \nodata
  & \nodata
  & \nodata
  & \nodata
  & \nodata
  & \nodata \nl
98-SMC-1
  & 0.47
  & 0.56
  & 0.79
  & 1.07
  & \nodata
  & \nodata
  & \nodata
  & \nodata
  & \nodata \nl

\hline

OGLE-7 / 119-A
  & 0.41
  & 0.37
  & \nodata
  & \nodata
  & \nodata
  & \nodata
  & \nodata
  & \nodata
  & \nodata \nl
403-C
  & 0.48
  & 0.50
  & \nodata
  & \nodata
  & \nodata
  & \nodata
  & \nodata
  & \nodata
  & \nodata \nl
94-BLG-4
  & 1.06
  & 1.04
  & \nodata
  & \nodata
  & \nodata
  & \nodata
  & \nodata
  & \nodata
  & \nodata \nl
95-BLG-12
  & 0.22
  & 0.26
  & 0.26
  & \nodata
  & \nodata
  & \nodata
  & 0.27
  & 0.26
  & 0.31 \nl
96-BLG-3
  & 0.23
  & 0.25
  & 0.35
  & \nodata
  & \nodata
  & \nodata
  & 0.33
  & \nodata
  & \nodata \nl
96-BLG-4
  & 1.03
  & 1.03
  & 1.05
  & \nodata
  & \nodata
  & \nodata
  & 1.01 
  & \nodata
  & \nodata \nl
97-BLG-1
  & 0.93
  & 0.97
  & 0.87
  & \nodata
  & \nodata
  & \nodata
  & \nodata
  & \nodata
  & \nodata \nl
97-BLG-24 (-----)
  & 0.16
  & 0.17
  & \nodata
  & \nodata
  & 0.15
  & \nodata
  & \nodata
  & \nodata
  & 0.21 \nl
97-BLG-24 (-~-~-)
  & 0.08
  & 0.08
  & \nodata
  & \nodata
  & 0.08
  & \nodata
  & \nodata
  & \nodata
  & 0.11 \nl
97-BLG-28
  & 1.00
  & 0.90
  & 0.97
  & \nodata
  & \nodata
  & \nodata
  & \nodata
  & \nodata
  & \nodata \nl
98-BLG-12
  & 0.17
  & 0.14
  & 0.17
  & \nodata
  & \nodata
  & 0.16
  & \nodata
  & \nodata
  & \nodata \nl
98-BLG-14 (-----)
  & 0.51
  & 0.52
  & 0.52
  & \nodata
  & \nodata
  & 0.48
  & \nodata
  & \nodata
  & \nodata \nl
98-BLG-14 (-~-~-)
  & 1.08
  & 1.10
  & 1.07
  & \nodata
  & \nodata
  & 0.99
  & \nodata
  & \nodata
  & \nodata \nl


98-BLG-16
  & 0.04
  & 0.03
  & 1.10
  & \nodata
  & \nodata
  & 1.07
  & \nodata
  & \nodata
  & 0.04 \nl
98-BLG-42
  & 0.46
  & 0.45
  & \nodata
  & \nodata
  & 0.73
  & 0.74
  & \nodata
  & \nodata
  & \nodata \nl
97-BLG-d2   
  & 0.30
  & 0.14
  & \nodata
  & \nodata
  & \nodata
  & \nodata
  & \nodata
  & \nodata
  & \nodata \nl
108-E
  & 1.10
  & 0.86
  & \nodata
  & \nodata
  & \nodata
  & \nodata
  & \nodata
  & \nodata
  & \nodata \nl
176-A
  & 0.74
  & 0.70
  & \nodata
  & \nodata
  & \nodata
  & \nodata
  & \nodata
  & \nodata
  & \nodata \nl
\enddata
\tablenotetext{} { List of blending parameters for each of our
candidate binary microlensing events.  The parameter $f$ represents
the fraction of the object's flux which was lensed in each respective
passband.  Events LMC-1, 97-BLG-24, and 98-BLG-14 are each presented
with 2 fits of similar significance, but different event parameters.}
\end{deluxetable} 
\end{landscape}

\clearpage
\begin{landscape}
\begin{deluxetable}{lcccccccc}
\tablecaption{Intrinsic Source Properties \label{tab-size} }
\tablewidth{0pt}
\scriptsize
\tablehead{
        \colhead {Event } &
        \colhead {$\#$ RR Lyrae} &
        \colhead {$<\Evr>$ } &
        \colhead {$<\Av>$ } &
        \colhead {$(V-R)_0$ } &
        \colhead {$V_0$ } &
        \colhead {$\teff$ (K) } &
        \colhead {$\mbol$ } &
        \colhead {$\theta_*$ ($\mu$as) }
}
\startdata
119-A
  & 3
  & 0.47 (9)
  & 1.9  (3)
  & 0.43 (9)
  & 18.4 (3)
  & 5350 (500)
  & 18.2 (4)
  & 1.06 (29) \nl
96-BLG-3
  & 6
  & 0.42 (5)
  & 1.7  (2)
  & 0.31 (6)
  & 19.1 (2)
  & 6200 (500)
  & 19.1 (2)
  & 0.53 (11) \nl
97-BLG-28
  & 9
  & 0.67 (4)
  & 2.7  (2)
  & 0.58 (6)
  & 15.5 (2)
  & 4500 (200)
  & 15.0 (2)
  & 6.58 (90) \nl
98-BLG-42
  & 10
  & 0.53 (5)
  & 2.1  (2)
  & 0.33 (7)
  & 18.1 (2)
  & 6050 (500)
  & 18.1 (2)
  & 0.89 (20) \nl
\enddata
\tablenotetext{} { Estimates of the reddening and extinction, as
determined from RR Lyrae within $10'$, to each of the 4 sources where
we have a well constrained binary lens fit and a reliable measure of
$\tstar$.  The intrinsic source color $(V-R)_0$ and brightness $V_0$
are used to determine $\teff$ and $\mbol$, and from these we find
the source angular radius $\theta_*$.  }
\end{deluxetable} 
\end{landscape}

\clearpage

\begin{deluxetable}{lccccc}
\tablecaption{Properties of the Lensing Systems \label{tab-ppm} }
\tablewidth{0pt}
\scriptsize
\tablehead{
        \colhead {Event } &
        \colhead {$\mu$ (mas/yr)} &
        \colhead {$\dlens$ (kpc)} &
        \colhead {$\mtot / \msun$ } &
        \colhead {${\rm M}_1 / \msun$ } &
        \colhead {${\rm M}_2 / \msun$ }
}
\startdata
119-A
  & 1.84  (58)
  & $7.0{+0.8\atop -0.9}$ 
  & $0.89{+1.25\atop -0.42}$
  & $0.49{+0.69\atop -0.23}$
  & $0.40{+0.57\atop -0.19}$ \nl \nl
96-BLG-3
  & 2.26  (47)
  & $7.1{+0.8\atop -0.9}$ 
  & $1.61{+2.25\atop -0.76}$
  & $0.45{+0.63\atop -0.21}$
  & $1.15{+1.62\atop -0.55}$ \nl \nl
97-BLG-28
  & 3.16  (44)
  & $7.0{+0.8\atop -1.0}$
  & $0.26{+0.36\atop -0.13}$
  & $0.05{+0.06\atop -0.02}$
  & $0.22{+0.30\atop -0.11}$ \nl \nl
98-BLG-42
  & 2.98  (78)
  & $7.0{+0.8\atop -1.0}$
  & $0.19{+0.26\atop -0.09}$
  & $0.14{+0.20\atop -0.07}$
  & $0.05{+0.06\atop -0.02}$ \nl \nl
\enddata
\tablenotetext{} { The lens proper motion measurements, $\mu =
\theta_* / \tstar$, have been used in the likelihood analysis detailed
in Sec.~\ref{subsec-ppm} to produce the most likely lens distance and
mass estimates given here.  The errors quoted are $1 \sigma$.  We note
that the errors in ${\rm M}_1$ and ${\rm M}_2$ are completely
correlated.  }
\end{deluxetable} 

\end{document}